\documentclass[%
 reprint,longbibliography,preprintnumbers,
nofootinbib,
 amsmath,amssymb,
 aps,
pre,
]{revtex4-1}
\pdfoutput=1
\usepackage[utf8]{inputenc}
\usepackage{flushend}
\usepackage{dcolumn}
\usepackage{bm}
\usepackage{balance}

\usepackage[normalem]{ulem}

\usepackage[colorlinks = true,
            linkcolor = blue,
            urlcolor  = blue,
            citecolor =green,
            anchorcolor = blue]{hyperref}
\usepackage{verbatim}
\usepackage{color,ulem}
\usepackage[english]{babel}

\usepackage[utf8]{inputenc}
\input Starburst.fd
\newcommand*\initfamily{\usefont{U}{Starburst}{xl}{n}}\initfamily

\newcommand{\beq}{\begin{eqnarray}}
\newcommand{\eeq}{\end{eqnarray}}
\usepackage{amsmath}
\usepackage{tikz}
\usetikzlibrary{decorations.pathmorphing}
\usetikzlibrary{shapes.misc}
\tikzset{cross/.style={cross out, draw=black, minimum size=8*(#1-\pgflinewidth), inner sep=0pt, outer sep=0pt},
cross/.default={1pt}}
\usetikzlibrary{patterns,math}
\begin{document}

\title{Complete mathematical theory of the jamming transition: A 
perspective}

\author{\textbf{Alessio Zaccone}$^{1}$}%
 \email{alessio.zaccone@unimi.it}
 
 \vspace{1cm}
 
\affiliation{$^{1}$Department of Physics ``A. Pontremoli'', University of Milan, via Celoria 16,
20133 Milan, Italy.}

\begin{abstract}
The jamming transition of frictionless athermal particles is a paradigm to understand the mechanics of amorphous materials at the atomic scale. Concepts related to the jamming transition and the mechanical response of jammed packings have cross-fertilized into other areas such as atomistic descriptions of the elasticity and plasticity of glasses.
In this perspective article, the microscopic mathematical theory of the jamming transition is reviewed from first-principles. The starting point of the derivation is a microscopically-reversible particle-bath Hamiltonian from which the governing equation of motion for the grains under an external deformation is derived. From this equation of motion, microscopic expressions are obtained for both the shear modulus and the viscosity as a function of the distance from the jamming transition (respectively, above and below the transition). Regarding the vanishing of the shear modulus at the unjamming transition, this theory, as originally demonstrated in [Zaccone \& Scossa-Romano, Phys. Rev. B 83, 184205 (2011)], is currently the only quantitative microscopic theory in parameter-free agreement with numerical simulations of [O'Hern et al. Phys. Rev. E 68, 011306 (2003)] for jammed packings. The divergence of the viscosity upon approaching the jamming transition from below is derived here, for the first time, from the same microscopic Hamiltonian. The quantitative microscopic prediction of the diverging viscosity is shown to be in fair agreement with numerical results of sheared 2D soft disks from [Olsson \& Teitel, Phys. Rev. Lett. 99, 178001 (2007)].
\end{abstract}

\maketitle
\section{Introduction: the jamming paradigm}
Understanding the mechanical properties of disordered systems based on the microstructure and dynamics of the building blocks thereof is a key topic in soft matter and materials physics \cite{zaccone2023,Vitelli,kob_book,Torquato_book}.
In the early days of soft matter physics in the 1990s, emulsions, foams and other soft granular systems emerged as a playground to develop and test new theoretical concepts for disordered condensed matter systems and complex fluids \cite{deGennes_granular,Liu_emulsions,Durian,Behringer,Torquato_book,Weaire,Corwin2005,Corey_PRL}.

\begin{figure}
\includegraphics[width = 0.8 \linewidth]{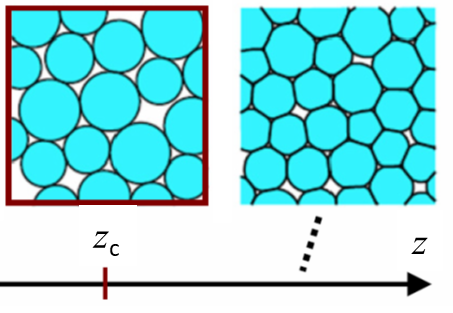}
	\caption{Schematic of the jamming transition of soft frictionless spheres. As more spheres are thrown into the box, their average coordination number $z$ reaches a critical value $z_c$ at which the spheres become jammed. This coincides with the emergence of shear rigidity \cite{Scossa}. Upon forcing more spheres into the box, their softness and deformability allows for a further increase of $z$ above the $z_c$ value. Reproduced with permission from J. Phys.: Condens. Matter 22, 033101 (2010). Copyright 2010 Institute of Physics.}
		\label{fig1}
\end{figure}

The advantage of considering these systems is that one can simplify a complex many-body problem of force-transmission and mechanical stability by ignoring all the complications deriving from the chemistry and even the Brownian motion of the building blocks. This led to the so-called jamming paradigm, and to the jamming transition of soft (frictionless) sphere packings \cite{Liu1998,vanHecke,Reichardt_rev}.
In a nutshell, soft athermal spheres are thrown into a box until each sphere touches a certain average number, $z$, of nearest-neighbours. At a critical value $z_c$ of $z$, the spheres are no longer able to readjust their positions and one has a "jammed" state \cite{OHern,Hideyuki}. This critical condition coincides with the emergence of rigidity, i.e. of a finite shear modulus $G$. As more spheres are jammed into the system, the coordination number $z$ gets larger than $z_c$ due to the softness of the sphere-sphere interaction, which implies a certain degree of deformability of the spheres (as typically is the case e.g. in emulsions \cite{Jasna2009}). The situation is schematically 
depicted in Fig. \ref{fig1}.
It has been recognized by Torquato and co-workers \cite{Truskett,Torquato_review} that jammed states are not unique: distinct jammed states can be generated, which differ by the degree of ordering as quantified by common metrics \cite{Truskett}. Under the strict-jamming constraint defined as "any collectively jammed configuration
that disallows all uniform volume-nonincreasing strains
of the system boundary" \cite{Torquato_2018}, this led to the definition of the Maximally Random Jammed (MRJ) state as the isostatic state $z=2d$ which is maximally disordered, rigid, and lacking any medium and long-range ordering. Less stringent jamming constraints include the \emph{local jamming} constraint in which a tagged particle is locally trapped by its nearest-neighbours such that it cannot be translated without moving the nearest-neighbours, and \emph{collective jamming}, where "no subset of particles can be
simultaneously displaced so that its members move out
of contact with one another and with the remainder set" \cite{Torquato_2018}.
Furthermore, the MRJ states have been recognized to be hyperuniform, where hyperuniformity is defined as the limiting behaviour of the static structure factor $S(\mathbf{k})$ vanishing at $\mathbf{k}\rightarrow 0$ in a power-law fashion, $S(\mathbf{k}) \sim |\mathbf{k}|^\alpha$, where $\alpha >0$ \cite{Torq_hyper}.

It is often said that the jamming transition presents a mixed first-order and second-order character. This is because some thermodynamic properties appear to change sharply at the transition \cite{Yuliang}, which would indicate a discontinuous character of the transition. In the spirit of Ehrenfest's classification, one could also consider derivatives of the free energy $F$ with respect to a thermodynamic variable or an external field. Two such quantities are the bulk modulus and the shear modulus, respectively. Since the latter goes to zero continuously at $z_c$ coming from above, and the bulk modulus jumps discontinuously, one would infer a continuous (second-order type) phase transition. It is, however, quite clear that traditional classifications such as the Ehrenfest classifications of phase transitions, which are based on equilibrium statistical mechanics, may lose their meaning for systems such as jammed packings which are athermal and far from thermodynamic equilibrium. Recent approaches have been directed towards understand jamming (and gelation \cite{Joep}) transitions in terms of dynamical nonequilibrium transitions \cite{Chaikin_Manna}. This is also relevant for the jamming transition of "sticky" or short-range attractive particles, which has been shown to belong to a different universality class with respect to repulsive jamming \cite{Lois}.
One should not forget, however, an important attempt to map the physics of granular systems, such as jammed packings, to that of equilibrium systems, using a volume ensemble of equiprobable jammed states in analogy to thermal equilibrium statistical mechanics. This approach is known as Edwards' extended statistical mechanics, for which the reader is referred to a comprehensive review \cite{MakseRMP}.

In general, as we are going to demonstrate below, methods based on the free energy, are less effective towards providing a quantitative mechanistic description of jammed systems. This is because, on one hand, they make the original problem much more difficult than it actually is (e.g. the problem of estimating the configurational entropy for random packings is practically unsolvable for realistically sized 3D systems \cite{Martiniani}). Also, on the other hand, the consideration of the $T\Delta S$ is superfluous at $T=0$, where the key non-zero quantity is the potential energy. The latter point will be exemplified in the following sections, where the success of the Hamiltonian theory in describing the observable properties of jamming systems will be illustrated.

In this paper, we first provide a brief summary of previous approaches to the jamming transition. 
Then, we shall give a self-contained presentation of the mathematical theory of the jamming transition from first-principles, i.e. from a microscopically reversible Hamiltonian. The theory is able to produce predictions for the two key experimental observables, i.e. the shear modulus (above jamming) and the viscosity (below jamming). 
We shall finally discuss how this theory can be naturally extended from the simplified paradigm of jamming to thermal and even atomic glassy systems.


\section{Previous approaches}
\subsection{Parisi's replica theory}
In equilibrium statistical mechanics, in order to evaluate the physical properties of a system, one has to take ensemble averages of quantities derived from the free energy $F = -k_{B}T \ln Z$. If the system is out of equilibrium and non-ergodic, the ensemble average can only be performed over different realizations (replicas) of the disorder, and not as the standard canonical ensemble. If the disordered structure of the system is frozen-in on the time-scale of observation, one speaks of "quenched average".
In order to average the logarithm over many replicas one resorts to the so-called replica trick for the partition function $Z$:
\begin{equation}
\ln Z = \lim_{n \rightarrow 0} \frac{Z^{n}-1}{n}\label{replica}
\end{equation}
which simplifies the calculation of $\overline{\ln Z}$, and reduces it to the much more manageable problem of calculating the disorder average $\overline{Z^{n}}$, where $n$ is an integer. A key assumption is that of \emph{self-averaging}\index{self-averaging}, whereby the average over a large realization of the system is indistinguishable from the average over many (smaller) realizations (replicas) of the disorder.
Introducing replicas allows one to perform this macroscopic average over different disorder realizations. 

The replica trick has been famously used by G. Parisi to obtain an exact solution for the free energy of the Sherrington-Kirkpatrick model of spin glasses. It was shown by Parisi that the mean-field solution breaks the invariance with respect to replica permutations, and thus one says that replica symmetry breaking (RSB) occurs. Even though taking the limit in Eq. \eqref{replica} has a highly non-trivial physical meaning, Parisi's theory was instrumental to elucidate the spin glass transition. Real systems where the spin glass state is observed are certain metal alloys, e.g. dilute manganese in copper or iron in gold. However, the spin glass transition of the real-world systems, which exhibit a sharp cusp of the susceptibility as predicted by Parisi's theory, is of the order of 30 K, i.e. so low that practical applications of these materials are limited. Afterwards, materials with higher spin glass transition temperatures (even up to 266 K) were discovered, which, however, do not exhibit any sharp cusp of the susceptibility (rather, a very broad peak) \cite{Kostya_2011}, and are not described by the replica theory. These materials appear to be supporting a different spin glass transition scenario, which is not governed by any "ideal" phase transition and is instead compatible with a dynamical crossover dominated by relaxation phenomena \cite{Kostya_2011,Kostya_book}.

Parisi's theory has proved to be fertile in the abstract sciences, in particular in maths (where probabilists such as Guerra and Talagrand took on the challenge to develop appropriate rigorous methods which recover Parisi's ansatz and results) and in computer science, where some of its conceptual tools have been used to study certain neural networks, as well as optimization problems.

Parisi's replica theory has been subsequently applied to the glass transition of hard spheres, where it turns out to be exact for infinite-dimensional systems $d \rightarrow \infty$ \cite{Parisi_RMP}.
The high-dimensional limit of the packing problem is, however, quite different from low-dimensional systems. For example, the densest packings in high-dimensions are disordered, as shown by a well-founded conjecture \cite{Torq_Still}, as opposed to what happens in low dimensions $d=2$ and $d=3$, where the densest packings are ordered (e.g. in $d=3$ the densest packing is the Face Cubic Crystal, FCC, as is well known \cite{Hales1}).
We should also mention that the replica symmetry breaking approach to evaluate the free energy of deformation has been fruitfully used in the mathematical theory of elasticity of polymer networks (e.g. rubber), for which the interested reader is referred to~\cite{Goldbart}.
It has also been used for a theoretical calculation of elastic moduli of model thermal (Lennard-Jones) glasses \cite{mezard,yoshino}. Interestingly, this calculation is largely based on the nonaffine deformation framework \cite{Lutsko} that we shall discuss more extensively in the following.

Although Parisi's replica theory has been extensively studied in the context of jammed packings \cite{Charbonneau}, to date it has yet to provide a quantitative description of the observable macroscopic features of jammed systems, in particular, of the shear modulus (above jamming) and of the divergence of the viscosity (below jamming). Moderate success has been achieved in qualitatively reproducing certain features of the pair correlation function of athermal hard-spheres \cite{Berthier_Zamp}.

The replica method has been applied to the calculation of the vibrational density of states (VDOS) of jammed sphere packings \cite{parisi2014}. In the original formulation, the calculation used a mixture of infinite-dimensional and one-dimensional assumptions. The infinite-dimensional approximation is intrinsic to the replica method's mean-field character, whereas the one-dimensional approximation was used in the definition of the Hessian matrix (see the definition below), which was defined as a $N \times N$ matrix, where $N$ is the number of particles, instead of $d N \times d N$. As pointed out in \cite{Cicuta}, the spectrum of the random block (Hessian) matrix of dimension $d N \times d N$ is quite different from that of the random (non-block) matrix of dimension $N \times N$. Also, from a physical point of view, the position vectors of the particles, which enter the definition of the Hessian, are $d$-dimensional vectors in the embedding $d$-dimensional space where the jammed packing lives, which is obviously a physical aspect that cannot be neglected.

This latter shortcoming was later corrected by Parisi and co-workers in Ref. \cite{Benetti} based on the random block matrix description introduced by Cicuta et al. \cite{Cicuta} and now provides an adequate description of the vibrational spectrum. For more details about random block matrices in this context, see Refs. \cite{Pernici2019,Cicuta_2023}, and for a recent pedagogic overview of solution methods to obtain eigenvalue spectra of sparse random matrices see Ref. \cite{Vivo}.

\subsection{Phenomenological theories}
The jamming transition has also been given a description by means of the theory of critical phenomena. In particular, a Widom-like scaling ansatz has been proposed for the jamming transition in \cite{Sethna}. While this approach ignores the microscopic mechanisms underlying the jamming transition, it had the important merit to highlight that it is, indeed, the potential (elastic) energy the only part of the free energy which is relevant for a description of the jamming transition as a phase transition. In particular, it is the potential elastic energy the only part of the free energy which can be expressed in a scale-invariant form consistent with known exponent relations \cite{Sethna}.

\subsection{Zaccone \& Scossa-Romano's exact solution}
It was recognized early on by Cates and co-workers \cite{Cates} that particle displacements in slightly deformed granular packings are strongly nonaffine, i.e. there is an extra (typically random) displacement on top of the affine displacement dictated by the strain tensor of the externall applied deformation.
Builgind on this consideration, an exact analytical solution for the jamming transition was found in Ref. \cite{Scossa} using the nonaffine deformation formalism. This theory will be re-derived in the following sections by first-principles starting from a microscopically reversible Hamiltonian. In this theory, the zero-frequency shear modulus of jammed packings is evaluated analytically at the particle-level by computing the nonaffine (negative) correction to the elastic energy of deformation. This was done using the trick of analytically evaluating the eigenvalues and eigenvectors of the Hessian matrix from the orientationally-averaged "copy" of the Hessian itself. The orientationally-averaged copy of the Hessian is a matrix that commutes with the original Hessian, and, therefore, its eigenvectors and eigenvalues are also eigenvectors and eigenvalues of the system's Hessian matrix. Using them to construct the nonaffine correction to the shear modulus, an exact analytical formula for the shear modulus of 3D jammed packings was obtained \cite{Scossa}:
\begin{equation}
G= G_{A}-G_{NA}= \frac{1}{30}\frac{N}{V} \kappa R_0^2 (z-6)
\label{modulus_closed}
\end{equation}
where $\kappa$ denotes the spring constant (e.g. for harmonically repulsive soft spheres), and $R_0$ is the particle diameter. Here, $G_{A}$ and $-G_{NA}$ are, respectively, the affine and the nonaffine contributions to the elasticity.

This formula is in excellent quantitative agreement, without any adjustable parameters, with numerical simulations of harmonically-repulsive 3D jammed packings from Ref. \cite{OHern}. A similar formula was derived for 2D jammed packings, where the shear modulus vanishes at $z_c = 4$ \cite{Scossa}.

The above formula Eq. \eqref{modulus_closed} correctly predicts the jamming transition to occur at the isostatic point $z_c = 2d$, in $d$-dimensions. Furthermore, it is the only theoretical solution, to the best of our knowledge, that can reproduce the jamming transition in quantitative parameter-free agreement with numerical simulations \cite{OHern}.

The solution of Ref. \cite{Scossa} was subsequently extended to account for excluded-volume correlations in the orientational distribution of nearest-neighbours, which are responsible for the decrease of the nonaffine (negative) contribution under hydrostatic compression compared to shear \cite{JAP,Schlegel2016}. This theory of the bulk modulus provides a one-parameter description of the experimentally observed bulk modulus of jammed emulsions \cite{Schlegel2016}, and resolves the paradox of packing-derived vs network-derived elastic systems first discussed on the basis of numerical simulations in \cite{Ellenbroek}.

Finally, the above formula can be expressed also in terms of the particle volume fraction $\phi$, by relating the coordination number to $\phi$ via the radial distribution function \cite{KSZ}. This, in turn, leads to a square-root scaling $G \sim (\phi - \phi_c)^{1/2}$ with respect to the distance from jamming, which coincides with the random close packing (RCP) fraction $\phi_c$, as discussed further below in Section VIII. This type of scaling relations are well studied for both frictionless \cite{Hatano,PhysRevE.72.051306} (where also the exponent $0.6$ has been observed \cite{Reichardt_exp}) and frictional jammed packings \cite{Haya}.

\section{From the microscopic Hamiltonian to the equations of motion under deformation}
Our goal is to derive the macroscopic features of the jamming transition starting from a microscopically-reversible Hamiltonian. 
We adopt, to this aim, the classical Caldeira-Leggett (CL) particle-bath Hamiltonian \cite{Caldeira}, originally proposed by Zwanzig \cite{Zwanzig1973}.
We will work at the level of classical mechanics, as appropriate for jamming, although generalizations of the CL Hamiltonian exists also for quantum systems \cite{Weiss} and, more recently, also for relativistic particles \cite{Petrosyan_2022,Zadra,Zaccone_quark}.

In the CL Hamiltonian, the dynamics of a tagged particle (mass $M$, position $Q$ and momentum $P$) is coupled to all other particles (treated as fictitious harmonic oscillators with mass $M_m$, position $Q_m$ and momentum $P_m$, representing the normal modes of the system) as follows (cfr. Ref.~\cite{Zwanzig1973,Zwanzig}):
\begin{equation}
\begin{split}
\mathcal{H}&=\frac{P^2}{2M}+U(Q)\\
&+\frac{1}{2}\sum_{m}\left[\frac{P_{m}^2}{M_{m}}+M_{m}\omega_{m}^2\left(Q_{m}
-\frac{\gamma_{m}Q}{M_{m}\omega_{m}^2}\right)^{2}\right].\label{Hamiltonian}
\end{split}
\end{equation}
The above Hamiltonian is obtained by summing together the Hamiltonian of the tagged particle: $\mathcal{H}_P=\frac{P^2}{2M}+U(Q)$ and the Hamiltonian of the bath of oscillators: $H_{B}=\frac{1}{2}\sum_{m}\left[\frac{P_{m}^2}{M_{m}}+M_{m}\omega_{m}^2\left(Q_{m}
-\frac{\gamma_{m}Q}{M_{m}\omega_{m}^2}\right)^{2}\right]$, where the index $m$ runs over the $N$ oscillators forming the bath. The dynamics of the tagged particle and the dynamics of the oscillators' are coupled via two terms in the bath Hamiltonian (upon expanding the square in the last term): $- Q \sum_i{\gamma_m Q_m} + Q^2 \sum_m \frac{\gamma_m^2}{2M_m \omega_m^2}$. Hence, in the Caldeira–Leggett Hamiltonian, the bath is coupled to the position $Q$ of the tagged particle. The $\gamma_m$ are coefficients which depend on the details of the coupling, hence they depend on the long-range anharmonic interaction between the tagged particle and all other particles (atoms and molecules) with which it interacts in the system. The second term $Q^2 \sum_m \frac{\gamma_m^2}{2M_m \omega_m^2}$ is a counter-term which must be considered to make sure that dissipation is homogeneous across the whole system. As the bath couples to the position, neglecting this term would lead to a dynamics which is not translation-invariant, which would be unphysical. More details about this can be found in textbooks, e.g. Nitzan \cite{Nitzan} or Weiss \cite{Weiss}.
This formalism can be extended to include the presence of an externally applied time-dependent field (driven bath) \cite{Cui_GLE,Gamba}. 

The resulting generalized Langevin equation of motion for the displacement of the tagged particle can be derived by using the Euler-Lagrange equations (cfr. \cite{Zwanzig1973} or a textbook \cite{Zwanzig}, for the full derivation). One solves for the $m$-th oscillator coordinate and replaces it back into the Euler-Lagrange equation for the tagged particle. By collecting together all terms and factors which depend only on the initial conditions of the bath oscillator dynamics to form the "noise", and introducing the mass-scaled tagged-particle displacement $s=Q\sqrt{M}$, the final generalized Langevin equation (GLE) reads as:
\begin{equation}
\ddot{s}=-U'(s)-\int_{-\infty}^t \nu(t-t')\frac{ds}{dt'}dt' + F_{p}(t),
\label{2.4gle1}
\end{equation}
where $F_{p}(t)$ is the thermal stochastic noise with zero average, $U$ is a local interaction potential (e.g. from the interaction of the tagged particle with the nearest neighbours), and $\nu$ is the friction kernel resulting from many  conservative interactions with all the other particles in the system, imposed by the dynamical bi-linear coupling. This friction is not to be confused with the "solid friction" between touching spheres, but is rather an effective viscous ("wet" as opposed to "solid") frictional dissipation arising from sweeping under carpet the dynamical degrees of freedom of all the particles but the tagged particle, which we are following in its dynamics \cite{Zwanzig}. 
Obviously, for athermal particles $F_p \approx 0$ (also implying that the fluctuation-dissipation relation is broken, as expected for strongly nonequilibrium systems), and, therefore, this term will be dropped throughout in the following derivations.

We rewrite the Eq. \eqref{2.4gle1} for a tagged particle in $d$-dimensions, which moves with an affine velocity prescribed by the strain-rate tensor
$\dot{\mathbf{F}}$ (where the dot indicates a time derivative, while the circle indicates quantities measured in the undeformed rest frame):
\begin{equation}
\ddot{\mathbf{r}}_i=\mathbf{f}_i-\int_{-\infty}^t\nu(t-t')\left(\dot{\mathring{\mathbf{r}}}_i - \mathbf{u}\right) dt'
\end{equation}
where $\mathbf{f}_i=-\partial{U}/\partial{\mathbf{r}}_i$ generalises the $-U'(s)$ to the tagged particle $i$.
Furthermore, we used the Galilean transformations to express the particle velocity in the moving frame: $\dot{\mathbf{r}}_{i}=\dot{\mathring{\mathbf{r}}}_i - \mathbf{u}$ where $\mathbf{u}=\dot{\mathbf{F}}\mathbf{\mathring{r}}_{i}$ represents the local velocity of the moving frame. 
Since we are going to work at constant deformation rates, hence at constant velocity of the moving frame, the Galilean transformation valid for inertial frames is appropriate.
This notation is consistent with the use of the circle on the particle position variables to signify that they are measured with respect to the reference rest frame. In terms of the original (undeformed) rest frame $\lbrace\mathbf{\mathring{r}}_i\rbrace$, the equation of motion can be written, for the particle position averaged over several oscillations, as
\begin{equation}
\mathbf{F}\,\mathbf{\ddot{\mathring{r}}}_i  =\mathbf{f}_i-
\int_{-\infty}^{t}\nu(t-t')\cdot\frac{d\mathring{\mathbf{r}}_i}{dt'}dt'.
\label{2.4restframe}
\end{equation}


We work in the linear regime of small strain $\parallel\mathbf{F}-\mathbf{1}\parallel \ll 1$ by making a perturbative expansion in the small displacement $\{\mathbf{s}_i(t)=\mathbf{\mathring{r}}_i(t)-\mathbf{\mathring{r}}_i\}$ around a known rest frame $\mathbf{\mathring{r}}_i$.
That is, we take $\mathbf{F}=\mathbf{1}+\delta\mathbf{F}+...$ where $\delta\mathbf{F}\approx\mathbf{F}-\mathbf{1}$ is the small parameter. Replacing this back into Eq. \eqref{2.4restframe} gives \cite{Cui_viscoelastic}
\begin{align}
&(\mathbf{1}+\delta\mathbf{F}+...)\frac{d^2\mathbf{s}_i}{dt^2}=\delta\mathbf{f}_i+\nonumber\\
&-(\mathbf{1}+\delta\mathbf{F}+...)\int_{-\infty}^{t}\nu(t-t')\cdot\frac{d\mathbf{s}_i}{dt'}dt'.
\label{2.4expansion}
\end{align}
For the term $\delta\mathbf{f}_i$, imposing mechanical equilibrium again \cite{Slonczewski}, which is $\mathbf{f}_{i}=0$,  
implies:
\begin{equation}
    \delta \mathbf{f}_{i} = \frac{\partial \mathbf{f}_{i}}{\partial \mathbf{\mathring{r}}_j }\delta \mathbf{\mathring{r}}_j + \frac{\partial \mathbf{f}_{i}}{\partial \bm{\eta}}:\delta \bm{\eta}\label{two_ter}
\end{equation}
where $\bm{\eta}$ is the strain tensor and in the first term we recognise
\begin{equation}
    \frac{\partial \mathbf{f}_{i}}{\partial \mathbf{\mathring{r}}_j }\delta \mathbf{\mathring{r}}_j =-\mathbf{H}_{ij}\mathbf{s}_{j}.
\end{equation}
Furthermore, $\mathbf{H}_{ij}$ represents the Hessian matrix, defined as
\begin{equation}
\mathbf{H}_{ij}=\frac{\partial U}{\partial \mathring{\mathbf{r}}_{i} \partial \mathring{\mathbf{r}}_{j}}\bigg\rvert_{\gamma \rightarrow 0} = \frac{\partial U}{\partial \mathbf{r}_{i} \partial \mathbf{r}_{j}}\bigg\rvert_{\mathbf{r}\rightarrow \mathbf{r}_{0}}\label{ring_Hessian}
\end{equation}
since $\mathring{\mathbf{r}}(\gamma)\rvert_{\gamma \rightarrow 0}=\mathbf{r}_{0}$, where $U$ denotes the total energy of the system.
For the second term in Eq. \eqref{two_ter} we have:
\begin{equation}
    \mathbf{\Xi}_{i,\kappa\chi}=\frac{\partial \mathbf{f}_{i}}{\partial \bm{\eta}_{\kappa\chi}}
    \label{aff_force}
\end{equation}
and the limit $\bm{\eta}_{\kappa\chi} \rightarrow 0$ is implied. Here $\bm{\eta}_{\kappa\chi}$ are the components of the Cauchy-Green strain tensor defined as $\bm{\eta} =\frac{1}{2}\left(\mathbf{F}^{T}\mathbf{F}-\mathbf{1} \right)$. This is a second-rank tensor, and should not be confused with the fluid viscosity $\eta$ (a scalar).
Using standard lattice dynamics, the affine force Eq. \eqref{aff_force} can be evaluated for a shear deformation $\kappa\chi=xy$ as \cite{Lemaitre}
\begin{equation}
\mathbf{\Xi}_{i,xy} = -\sum_{j}\left( \kappa_{ij}r_{ij} - t_{ij}\right) n_{ij}^{x} n_{ij}^{y} \mathbf{n}_{ij},
\label{xi_lattice}
\end{equation}
where $\kappa_{ij}$ is the bond spring constant between particles $i$ and $j$, as standard in lattice dynamics, $t_{ij}$ is the bond tension (first derivative of the interaction energy evaluated at the interparticle distance), and $r_{ij}$ is the modulus of the interparticle distance. Here the centrosymmetry or lack thereof is contained in the sum: 
$\sum_{j}...n_{ij}^{x} n_{ij}^{y} \mathbf{n}_{ij}$
where $\mathbf{n}_{ij}$ is the unit vector from particle $i$ to at any other particle $j$, while $n_{ij}^{x}$ and $n_{ij}^y$ are its $x$ and $y$ components, respectively.
It is clear that this sum, and hence the force $\mathbf{\Xi}_{i,xy}$, will be zero for a centrosymmetric arrangement of the particles $j$ around the particle $i$, and hence there will be no nonaffine motions \cite{Lemaitre,zaccone2023}. In a non-centrosymmetric configuration of particles, as is the case in a jammed packing (but the same is true also for liquids and glasses), the above factor will rarely be zero, and therefore the force $\mathbf{\Xi}_{i,xy}\bm{\eta}_{xy}$ will almost always be present and significant and will affect the mechanical stability of the system as indeed observed experimentally \cite{amelia}.

With these identifications, we can write Eq. (\ref{2.4expansion}), to first order in strain, for the particle displacement $\mathbf{s}_i$ measured with respect to the particle's position in the undeformed state:
\begin{equation}
\frac{d^2\mathbf{s}_i}{dt^2}+\int_{-\infty}^{t}\nu(t-t')
\frac{d\mathbf{s}_i}{dt'}dt'+\mathbf{H}_{ij}\mathbf{s}_{j}=\mathbf{\Xi}_{i,\kappa\chi}\bm{\eta}_{xy}.
\label{2.4gle2}
\end{equation}

The above equation can be solved by Fourier transformation followed by a normal mode decomposition \cite{Lemaitre}. All terms in the above Eq. \eqref{2.4gle2} are vectors in $\mathbb{R}^{3}$ and the equation is in manifestly covariant form.

To enable further manipulation, we extend all tensors to $dN$-dimensional, and, without loss of generality, we select $d=3$.
After applying Fourier transformation to Eq. \eqref{2.4gle2}, we obtain
\begin{equation}
-\omega^2\,\tilde{\mathbf{s}}+i\tilde{\nu}(\omega)\omega\,\tilde{\mathbf{s}}
+\mathbf{H}\,\tilde{\mathbf{s}}
=\mathbf{\Xi}_{\kappa\chi}\tilde{\bm{\eta}}_{xy},
\end{equation}
where $\tilde{\nu}(\omega)$ is the Fourier transform of $\nu(t)$ etc (we use the tilde consistently throughout to denote Fourier-transformed quantities). In the above equation, all the terms are now vectors in $\mathbb{R}^{3N}$ space.
Next, we apply normal mode decomposition in $\mathbb{R}^{3N}$ using the $3N$-dimensional eigenvectors of the Hessian as the basis set for the decomposition. This is equivalent to diagonalising the Hessian matrix $\mathbf{H}$. 
Proceeding in the same way as in \cite{Cui_viscoelastic}, we have that the $m$-th mode of displacement can be written as \cite{Lemaitre,Cui_viscoelastic}:
\begin{equation}
-\omega^2\hat{\tilde{s}}_m(\omega)+i\tilde{\nu}(\omega)\omega\,\hat{\tilde{s}}_m(\omega)
+\omega_m^2\hat{\tilde{s}}_m(\omega)
=\hat{\Xi}_{m,\kappa\chi}(\omega)\tilde{\bm{\eta}}_{\kappa\chi}.
\label{2.4kerneldecom}
\end{equation}
It was shown in \cite{Lemaitre}, by means of MD simulations for an amorphous Lennard-Jones system, that $\hat{\Xi}_{m,\kappa\chi}=\mathbf{v}_m\cdot\mathbf{\Xi}_{\kappa\chi}$ is self-averaging, and one might thus introduce the smooth correlator function on eigenfrequency shells
\begin{equation}
\Gamma_{\mu\nu\kappa\chi}(\omega)=\langle\hat{\Xi}_{m,\mu\nu}\hat{\Xi}_{m,\kappa\chi}\rangle_{\omega_m\in\{\omega,\omega+d\omega\}}. \label{Gamma}
\end{equation}

\section{Linear response theory of jammed packings}
Following the general protocol of ~\cite{Lemaitre} to find the oscillatory stress for a dynamic nonaffine deformation, the stress is obtained, to first order in the strain amplitude, as a function of $\omega$ (note that the summation convention is implied for repeated indices):
\begin{align}
\tilde{\sigma}_{\mu\nu}(\omega)&=C^A_{\mu\nu\kappa\chi}\tilde{\bm{\eta}}_{\kappa\chi}(\omega)-\frac{1}{\mathring{V}}\sum_{m}\hat{\Xi}_{m,\mu\nu}\hat{\tilde{s}}_m(\omega) \notag\\
&=C^A_{\mu\nu\kappa\chi}\tilde{\bm{\eta}}_{\kappa\chi}(\omega)-\frac{1}{\mathring{V}}\sum_{m}\frac{\hat{\Xi}_{m,\mu\nu}\hat{\Xi}_{m,\kappa\chi}}{\omega_{m}^2-\omega^2
+i\tilde{\nu}(\omega)\omega}\tilde{\bm{\eta}}_{\kappa\chi}(\omega)\notag\\
&\equiv C_{\mu\nu\kappa\chi}(\omega)\tilde{\bm{\eta}}_{\kappa\chi}(\omega).\label{generalized}
\end{align}

In the thermodynamic limit, we can replace the discrete sum over degrees of freedom with an integral over vibrational frequencies up to a cut-off frequency $\omega_{D}$. Hence, we need to replace the discrete sum over the $3N$ degrees of freedom (eigenmodes) with an integral, $\sum_{m}...= \sum_{l=1}^{d}\sum_{p=1}^{N}...\rightarrow \int_{0}^{\omega_D}g(\omega_p)\sum_{l=1}^{d}...d\omega_p$, where $g(\omega_p)$ is the vibrational density of states (VDOS). We follow the standard convention for the normalization of the VDOS of solids as in textbooks, cfr. p. 82 in \cite{Cohen}:
\begin{equation}
\int_{0}^{\omega_D}g(\omega_p)d\omega_p=N \label{normaliz}
\end{equation}
with $N$ the total number of particles in the system.

Then, the complex elastic constants tensor can be written as:
\begin{equation}
C_{\mu\nu\kappa\chi}(\omega)=C^A_{\mu\nu\kappa\chi}-\frac{1}{V}\int_0^{\omega_{D}}\frac{g(\omega_p)\Gamma_{\mu\nu\kappa\chi}(\omega_p)}{\omega_p^2-\omega^2+i\tilde{\nu}(\omega)\omega}d\omega_p,
\label{2.4modulus}
\end{equation}
where the VDOS is normalized according to Eq. \eqref{normaliz} \cite{Cohen}.
The above Eq. \eqref{2.4modulus} is a crucial result obtained in \cite{Cui_viscoelastic} in the context of atomic-scale materials such as metallic glasses.

Specializing to shear deformations, $\mu\nu\kappa\chi=xyxy$, we obtain the following expressions for the complex shear modulus $G^{*}$:
\begin{equation}
G^{*}(\omega)=G_A-\frac{1}{V}\int_0^{\omega_{D}}\frac{g(\omega_p)\Gamma_{xyxy}(\omega_p)}{m\omega_p^2-m\omega^2+i\tilde{\nu}(\omega)\omega}d\omega_p.
\label{compl_shear_modulus}
\end{equation}
In the above expression the frequencies are now in physical units of rad/s or Hertz. The first term on the r.h.s. is the affine shear modulus $G_A$, which is independent of $\omega$, sometimes referred to as the Born modulus. The second term on the r.h.s. is the nonaffine contribution, $-G_{NA}$, and is a strong function of the oscillatory frequency $\omega$ \cite{PhysRevB.95.054203}. In the zero-frequency limit $\omega \rightarrow 0$, one recovers $G=G_A - G_{NA}$ as in Eq. \eqref{modulus_closed}. It is worth noting that the coincidence of the quasistatic shear modulus of Ref. \cite{Scossa} with that obtained from taking the zero-frequency of oscillatory shear is, in general not valid in the regime of nonlinear deformations as shown in Ref. \cite{Otsuki}.

From now on, we should also use the shorthand notation $\Gamma_{xyxy}(\omega_p) \equiv \Gamma(\omega_p)$.

By separating real and imaginary part of the above expression Eq. \eqref{compl_shear_modulus}, we then get to the storage and loss moduli as \cite{Milkus2}:
\begin{equation}
    \begin{split}
    G'(\omega)&=G_A - \frac{1}{V}\int_{0}^{\omega_{D}}\frac{m\,g(\omega_p)\,\Gamma(\omega_p)\,(\omega_{p}^{2}-\omega^{2})}{m^{2}(\omega_{p}^{2}-\omega^{2})^{2}+\tilde{\nu}(\omega)^{2}\omega^{2}}d\omega_p\\
    G''(\omega)&=\frac{1}{V}\int_{0}^{\omega_{D}}\frac{g(\omega_p)\,\Gamma(\omega_p)\,\tilde{\nu}(\omega)\,\omega}{m^{2}(\omega_{p}^{2}-\omega^{2})^{2}+\tilde{\nu}(\omega)^{2}\omega^{2}}d\omega_p.\label{visco-moduli}
    \end{split}
\end{equation}
It is easy to check that the storage modulus $G'(\omega)$ reduces to the Born modulus $G_A \equiv G_\infty$ in the infinite-frequency limit, $\omega \rightarrow \infty$.
From the point of view of practical computation, the VDOS $g(\omega_p)$ can be obtained numerically via direct diagonalization of the Hessian matrix $\mathbf{H}_{ij}$, since its eigenvalues are related to the eigenfrequencies via $\lambda_p=m\omega_{p}^{2}$. Similarly, the affine-force correlator $\Gamma(\omega_p)$ can also be computed from its definition by knowing the positions of all the particles, their interactions and forces (so that the affine force fields $\mathbf{\Xi}$ can be computed) as well as the eigenvectors of the Hessian $\mathbf{v}_p$.

The above equations for $G'(\omega)$ and $G''(\omega)$ provide the
foundation for the theoretical prediction of power-law rheology near jamming. For the rigorous predictions of power-law rheology near jamming, the reader is referred to Ref. \cite{Milkus2}, while for numerical results Refs. \cite{Tighe,Hara} should be consulted.

\section{Exact solution for the shear modulus of jammed packings}
The low-frequency behaviour of $\Gamma(\omega_p)$ was derived analytically in Ref. \cite{Scossa} for a system in $d$-dimensions:
\begin{equation}
\Gamma_{xyxy}(\lambda_p) = \kappa
R_0^2\, d\,\lambda_{p}\sum_{l=1}^{d}B_{l,xyxy},\label{correlator}
\end{equation}
 which gives $\langle \hat{\Xi}_{p,xy}^{2}\rangle \propto \lambda_{p}$, thus implying (from its definition above): 
\begin{equation}
\Gamma_{xyxy}(\omega_p) \propto \omega_{p}^{2},\label{scoss}
\end{equation}
a key result that was derived for the first time in \cite{Scossa}.
This analytical estimate appears to work reasonably well in the low-eigenfrequency part of the $\Gamma(\omega_p)$ spectrum of amorphous solids \cite{Palyulin,Lemaitre,Milkus2}. Recent independent numerical verifications of Eq. \eqref{scoss} have been provided by Szamel and Flenner \cite{flenner2024} and by Grießer and Pastewka \cite{Pastewka}.
Furthermore, $\sum_{l}B_{l,\mu\nu\kappa\chi}$ is a geometric coefficient that depends only on the geometry of macroscopic deformation and its exact values for a given deformation  field can be found tabulated in Eq. (14) of Ref. \cite{Scossa}. Using the exact geometric coefficients from Ref. \cite{Scossa}, one can thus evaluate Eq. \eqref{scoss} including the prefactors \cite{Scossa}:
\begin{equation}
\Gamma_{xyxy}(\omega_p) = \frac{3}{15} m\kappa R_{0}^{2} \omega_{p}^{2}.\label{scos}
\end{equation}

Upon replacing Eq. \eqref{scos} into the second term on the r.h.s. of the first equation (for $G'$) in Eq. \eqref{visco-moduli}, and upon taking the zero-frequency limit, $\omega \rightarrow 0$, we can thus easily identify:
\begin{equation}
    G_{NA}=\frac{3}{15}\frac{N}{V}\kappa R_{0}^{2}.\label{nonaf}
\end{equation}
It important to note that, in the above derivation, the details of the shape of the VDOS $g(\omega_p)$ are irrelevant for the calculation of the shear modulus, the only thing that matters being the normalization of the VDOS.

Next, we need to compute the affine shear modulus, or Born modulus or infinite-frequency shear modulus, $G_{A} \equiv \frac{1}{V} \frac{\partial^{2}U}{\partial \gamma^{2}}$. 
Recalling the Born-Huang formula for the affine shear elastic constant \cite{Born_Huang,zaccone2023}:
\begin{equation}
G_A = \frac{1}{V}\sum_{ij} r_{ij}^{2} \kappa_{ij}n^{x}_{ij}n^{y}_{ij}n^{x}_{ij}n^{y}_{ij},
\label{BH}
\end{equation}
where $n^{x}_{ij}$ denotes the $x$ Cartesian component of the unit vector $\mathbf{n}_{ij}$ connecting particle $i$ to particle $j$, $r_{ij}$ is the modulus of the distance between $i$ and $j$, and the sum runs over pairs $ij$ of particles in direct contact.
The above Born-Huang formula Eq. \eqref{BH} is derived under the two assumptions of (i) purely affine transformation of the distance between $i$ and $j$ under deformation: $\mathbf{r}_{ij}'= \mathbf{F} \cdot \mathbf{r}_{ij}$, and (ii) pairwise central-force contact forces.

Next, we need to evaluate the sum $\sum_{ij}...$ over nearest-neighbour particles pairs $ij $. We shall assume that, on average, the nearest-neighbours are oriented at random around the tagged particle at the center of the spherical frame (this is accurate for the shear modulus, while for the bulk modulus one has to include further correlations due to excluded-volume as explained in \cite{Schlegel2016}).
Consequently, from the random distribution assumption we have $\langle n_{ij}^{x}n_{ij}^{y}n_{ij}^{x}n_{ij}^{y}\rangle = \sum_{l}B_{l,xyxy}$, where $\langle...\rangle=\frac{1}{4\pi}\int_{\Omega}... \sin \theta_{ij} d \theta_{ij}d\phi_{ij}$ denotes the angular averaging of bond unit vectors over the solid angle $\Omega$. Upon neglecting residual bond stresses ($t_{ij}=0$ for all bonds), we obtain 
\begin{equation}
G_{A}=\frac{1}{30}\frac{N}{V}\kappa z R_{0}^2, 
\label{affine_closed}
\end{equation}
where $z$ is the coordination number, or nearest-neighbour number, which arises from the sum over all neighbouring pairs of particles.

\begin{figure}
\includegraphics[width = 0.98 \linewidth]{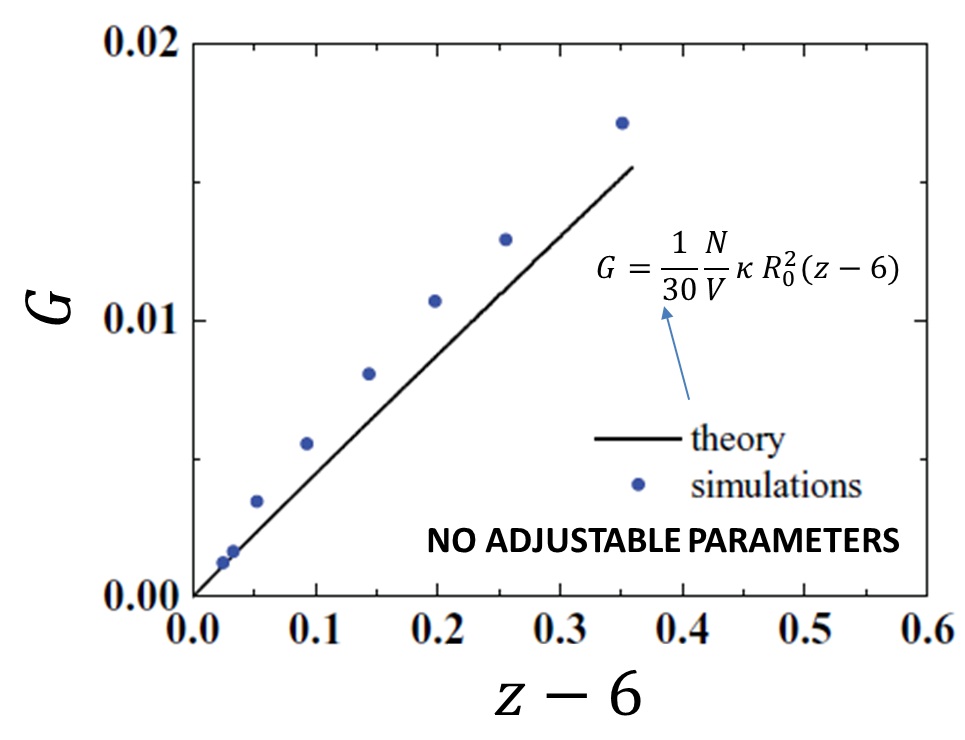}
	\caption{Quantitative parameter-free comparison between the exact formula for the shear modulus $G$ of harmonic jammed packings as a function of the coordination number $z$ measured with respect to the isostatic condition $z_c=6$ in $d=3$. The circles are simulations data  from O'Hern et al. \cite{OHern} while the solid line is the theory from Ref. \cite{Scossa}, reported in this article as Eq. \eqref{modulus_closed}. The comparison was originally shown in Ref. \cite{Scossa} and does not involve any free or adjustable parameter. Reproduced with permission from Phys. Rev. B 83, 184205 (2010). Copyright 2010 American Physical Society.}
		\label{fig2}
\end{figure}

We can now go back to the $\omega \rightarrow 0$ limit of the first one of Eqs. \eqref{visco-moduli} and replace Eq. \eqref{affine_closed} and Eq. \eqref{nonaf} therein, to finally obtain, for the static zero-frequency shear modulus:
\begin{equation}
G= G_{A}-G_{NA}= \frac{1}{30}\frac{N}{V} \kappa R_0^2 (z-6)
\nonumber
\end{equation}
which is Eq. \eqref{modulus_closed}, originally derived in Ref. \cite{Scossa} via a slightly different protocol, and via yet another route in Ref. \cite{Cui2019}.
This formula provides a quantitative parameter-free description of the jamming of soft sphere packings, as shown in Fig. \ref{fig2}, where the formula is compared with numerical simulations of harmonically-repulsive soft spheres of \cite{OHern}. No adjustable parameter is used in the comparison.

Importantly, for simple thermal liquids at thermodynamic equilibrium, it has been demonstrated in Ref. \cite{wittmer}, by using nonaffine response theory combined with equilibrium statistical mechanics, that the affine shear modulus $G_{A}$ is identical, with opposite sign, to the nonaffine part, thus resulting in $G'(\omega=0)=0$ for liquids at equilibrium. This is an important check for the correctness and generality of the nonaffine deformation theory \cite{zaccone2023}.

Finally, the above microscopic elasticity theory describes the ``hardening'', i.e. the increase of the shear modulus, as the jammed packing is increasingly compressed above the jamming point (compression hardening). Similarly, one can start from the jamming point and apply a shearing instead of compression protocol, which results in ``shearing hardening'' \cite{Fanlong}. The above description of the elasticity of jammed packings has been recently extended also to this protocol, which revealed the emergence of long-range anisotropic correlations responsible for the observed increase of stiffness \cite{Fanlong}.

\section{Viscosity divergence a``st jamming}
\subsection{Derivation of the viscosity formula}
A complete mathematical theory of the jamming transition should also be able to provide a prediction of the divergence of the viscosity of athermal hard spheres upon approaching the jamming transition from below, i.e. for $z < 6$.

We recall that the viscosity can be obtained from the loss viscoelastic modulus $G''$ using nonaffine response theory as (cfr. the introductory pages of Chapter 3 in \cite{zaccone2023}):
\begin{equation}
    \eta = \frac{G''}{\omega}.
\end{equation}

The nonaffine response theory developed from first principles in Sections III-IV above provides the following form for the loss modulus $G''$ (cfr. Eq. \eqref{visco-moduli}):
\begin{equation}
 G''(\omega)=\frac{1}{V}\int_{0}^{\omega_{D}}\frac{g(\omega_p)\,\Gamma(\omega_p)\,\tilde{\nu}(\omega)\,\omega}{m^{2}(\omega_{p}^{2}-\omega^{2})^{2}+\tilde{\nu}(\omega)^{2}\omega^{2}}d\omega_p. \label{loss_repeated}
 \end{equation}
For shear deformation ($\kappa\chi=xy$), 
Eq. \eqref{generalized} becomes:
\begin{equation}
    \sigma_{xy}(\omega)=G^{*}(\omega)\gamma(\omega).
\end{equation}
Since $G^{*}=G' + iG''$ and there is a factor $\omega$ in the above expression for $G''$ in Eq. \eqref{loss_repeated}, the theory correctly recovers, for the dissipative part of the stress, $\sigma_{xy}'$, the Newton law of laminar viscous flow (cfr. Newton's law of viscous liquids $\sigma' = \eta \dot{\gamma}$):
\begin{equation}
    \sigma_{xy}' = \eta \, i \omega \gamma = \eta \dot{\gamma}
\end{equation}
with, indeed, a zero-frequency shear viscosity given by \cite{PhysRevE.108.044101}:
\begin{equation}
    \eta = \frac{1}{V}\tilde{\nu}(0)\int_{0}^{\omega_{D}}\frac{g(\omega_p)\,\Gamma(\omega_p)}{m^{2}\omega_{p}^{4}}d\omega_p. \label{viscosity_nonaffine}
\end{equation}
This formula provides a direct, and unprecedented, connection between the viscosity $\eta$ and the vibrational density of states (VDOS) $g(\omega_p)$ and was derived in \cite{PhysRevE.108.044101}.
Furthermore, this formula shows that the viscosity goes to zero whenever the memory function or spectral function goes to zero at zero frequency, i.e. when $\tilde{\nu}(\omega \rightarrow 0)=0$. This is physically meaningful because a nearly dissipationless fluid with the non-kinetic part of the viscosity equal to zero can only arise when the correlations decay to zero completely in the long-time limit, which is not observed for real viscous fluids. This formula has been recently extended also to high-energy relativistic fluids in Ref. \cite{Zaccone_quark}.

We shall now apply this formula to weakly-sheared hard spheres below jamming by inserting a suitable mathematical description of the VDOS in the integral of the viscosity in Eq. \eqref{viscosity_nonaffine}. To this aim, we need to discuss the vibrational density of states (VDOS). Differently from the case of solid-like packings above jamming, where the shear modulus was found to be independent of the shape of the VDOS, thanks to the cancellation of $\Gamma(\omega_p)$ with $\omega_{p}^{2}$, in this case, as is clear from Eq. \eqref{viscosity_nonaffine}, this cancellation cannot happen. This is because, on one hand, as numerically verified by Saitoh et al. \cite{Saitoh}, the scaling  $\Gamma(\omega_p) \sim \omega_p^2$ as predicted in \cite{Scossa} still holds also below jamming, but, this time, we have $\omega_{p}^{4}$ instead of $\omega_{p}^{2}$, in the denominator in the integral. Hence, we can anticipate that the form of the viscosity and its scaling upon approaching jamming from below will depend on the details of the VDOS.

\subsection{Vibrational density of states (VDOS)}
The VDOS for athermal hard spheres under shear was computed numerically by Lerner et al. \cite{Lerner2012} in 2D and 3D and by Saitoh et al. \cite{Saitoh} in 2D.
The VDOS $g(\omega_{p})$ features a plateau with a lower edge and an upper edge, delimiting the support of the distribution, plus a low-energy solitary "single" mode, well separated from the other modes of the distribution.

It was argued by Parisi and co-workers \cite{Franz} that the VDOS spectrum of random hard sphere packings can be approximated by the Marchenko-Pastur distribution of random matrix theory \cite{Bai2010}. Essentially, this is also what the replica theory does in the mean-field $d \rightarrow \infty$ limit \cite{parisi2014}. However, the problem was correctly formulated in terms of finding the spectrum of a $dN \times dN$ block random matrix only in 2018 in Ref. \cite{Cicuta} and subsequently in Ref. \cite{Benetti}. See also recent mean-field solutions in various spatial dimensions implementing the same principles \cite{Harukuni_PRR,Harukuni,Mizuno_Berthier}.

In practice, the Hessian matrix can be identified as a block Laplacian random matrix. The blocks are $d \times d$ since this is the dimensionality of the unit vectors $\mathbf{n}_{ij}$ connecting particles $i$ and $j$, where the indices $i$ and $j$ run from $1$ to $N$.

Each row of the Hessian is made of a small and random number of non-zero entries, which makes it a sparse random matrix \cite{Vivo}. The off-diagonal entries $\mathbf{H}_{ij}$, with $i<j$, are identical independent random variables, whereas the diagonal entries are defined as $\mathbf{H}_{ij}=-\sum_{j\neq i}\mathbf{H}_{ij}$. The latter requirement is dictated by enforcing mechanical equilibrium on every particle $i$.\\

In the formulation of the Hessian matrix $\mathbf{H}_{ij}$, the unit vector $\mathbf{n}_{ij}$ provides the direction from particle $i$ to particle $j$. The general form of the Hessian matrix, in Cartesian components $\alpha,\beta=x,y,z$, reads as \cite{Scossa}:
\begin{equation}
H_{ij}^{\alpha\beta}= \delta_{ij}\sum_{s}\kappa c_{is} n_{is}^{\alpha}
n_{is}^{\beta} -(1-\delta_{ij}) \kappa c_{ij} n_{ij}^{\alpha}
n_{ij}^{\beta}
\label{hessian_lattice}
\end{equation}
where $\delta_{ij}$ is Kronecker's delta,  $c_{ij}$  is a random occupancy matrix with $c_{ij}=1$ if
$i$ and $j$ are nearest neighbours and $c_{ij}=0$ otherwise. $c_{ij}$ is a matrix where each row and each column have on average $z$ elements equal to 1 distributed randomly with the constraint that the matrix be symmetric. Furthermore, we assume $\kappa_{ij} \equiv \kappa$ is the same for all bonds $ij$ between nearest neighbours. The above notation also makes explicit the diagonal terms of the Hessian (the first term in Eq.~(\ref{hessian_lattice}) and the off-diagonal terms (the second contribution). 

In this description, every $d \times d$ off-diagonal block has probability $1-z/N$ of being a null matrix and a probability $z/N$ of being a rank one matrix, $X_{ij}=X_{j,i}=(X_{ij})^T=\mathbf{n}_{ij}\mathbf{n}_{ij}^{T}$ where $\mathbf{n}_{ij}$ is the $d$-dimensional (random) vector of unit length, chosen with uniform probability on the $d$-dimensional sphere, introduced previously. Furthermore, $\mathbf{n}_{ij}\mathbf{n}_{ij}^{T}$ is the usual matrix (or dyadic) product of a column vector times a row vector, which gives a rank-one matrix.

It was shown in Ref. \cite{Cicuta} that, in the limit $d \rightarrow \infty$, taken while keeping the ratio $z/d$ fixed, the above Hessian matrix has an eigenvalue spectrum given by the following Marchenko-Pastur (MP) distribution:
 \begin{equation}
\rho_{MP}(\lambda_{p}) = \frac{ \sqrt{(b-\lambda_{p})(\lambda_{p}-a)}}{4 \pi\,\lambda_{p}} \quad , \quad 0\leq a\leq \lambda_{p}\leq b \label{Marchenko}
\end{equation}
where $\lambda_{p}$ denotes the $p$-th eigenvalue of the Hessian, with the following definition of edge parameters:
\begin{equation}
a=\left(\sqrt{2}-\sqrt{\frac{z}{d}}\right)^2 \quad, \quad b=\left(\sqrt{2}-\sqrt{\frac{z}{d}}\right)^2.
\end{equation}
For a more formal mathematical proof that Eq. \eqref{Marchenko} is indeed the $d \rightarrow \infty$ limit of the spectrum of the above Hessian matrix Eq. \eqref{hessian_lattice}, the interested reader is referred to Ref. \cite{Pernici2019}.

Since the eigenvalue is related to the eigenfrequency via $\lambda_{p}=m \omega_{p}^{2}$, one obtains that the lower edge of the plateau in the VDOS is given by:
\begin{equation}
\omega_{*}=\left| \sqrt{2}-\sqrt{\frac{z}{d}}\right|
\end{equation}

As one can easily verify, the lower edge of the support $\omega_{*}$ admits a Taylor expansion about $z \rightarrow 2d^{-}$ (e.g. in $d=2$ approaching $z=4^{-}$ from below jamming) with the leading term $\sim (2d -z)$:
\begin{equation}
   \omega_{*}=\left(\sqrt{2}-\sqrt{\frac{z}{d}}\right)\approx 0.177\, (2d -z) + \text{h.o.t} \label{edge}
\end{equation}

The above Marchenko-Pastur distribution for the eigenvalues of the Hessian thus implies a Marchenko-Pastur distribution for the eigenfrequencies $\omega_{p}$, which has been shown to provide a fair description of the VDOS of random jammed solids also above jamming in Ref. \cite{Milkus_RMT}, where, \emph{mutatis mutandis}, $\omega_{*} \sim (z -2d)$.

This derivation from Ref. \cite{Cicuta} is, to our knowledge, the only simple analytical prediction of the VDOS of hard spheres, both below and above jamming, based on random matrix theory, which is able to capture the essential features of the Hessian spectrum as a block random matrix of dimension $d N \times dN$. Other approaches neglect the block random matrix character of the Hessian and are thus effectively one-dimensional \cite{parisi2014,Franz} or they do arrive at the same results of the above treatment at the expense of substantial computational complications \cite{Benetti}.

\begin{figure}
\includegraphics[width = 0.99 \linewidth]{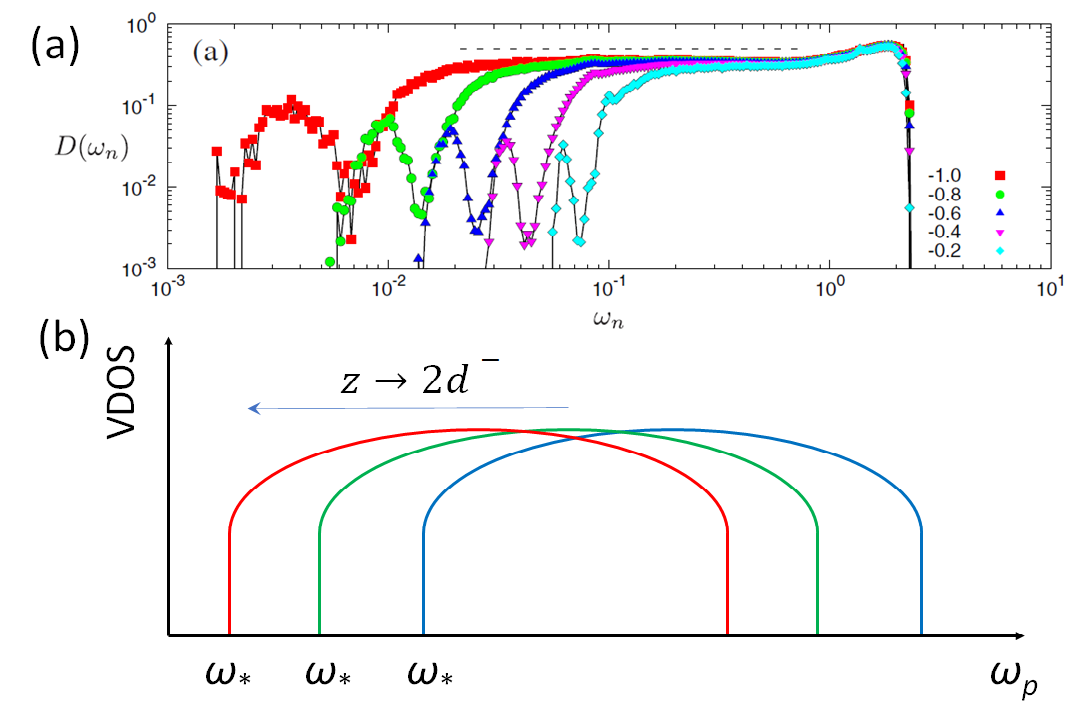}
	\caption{Panel (a) presents numerical calculations of the VDOS of sheared athermal hard disks upon approaching jamming from below. Reproduced with permission from Phys. Rev. Lett. 124, 118001 (2020). Copyright 2020 American Physical Society. The average coordination number $z$ (and the packing fraction $\phi$) increase from right to left (i.e. from cyan to red). Also shown is a small peak between zero frequency and the lower edge of the plateau. This is actually a single solitary mode (a Dirac delta) which appears as a peak due to averaging over several realizations, as explained in Ref. \cite{Saitoh}. Panel (b) shows a schematic cartoon of the Marchenko-Pastur spectrum \cite{Cicuta}, where the lower edge $\omega_{*}$ shrinks to zero as $\sim (2d -z)$ when $z$ increases towards the jamming point $z_c=2d$.}
		\label{fig3}
\end{figure}

Numerical results based on numerical diagonalization of the Hessian for sheared hard disks below jamming computed in Ref. \cite{Saitoh} are shown in Fig. \ref{fig3} (a). A schematic trend of the random matrix theory predictions \cite{Cicuta} is shown in Fig. \ref{fig3} (b).

\subsection{Divergence of the viscosity at jamming from below}
We can now substitute the above mathematical results for the VDOS $g(\omega_p)$ inside the viscosity formula, Eq. \eqref{viscosity_nonaffine}. 
In particular, based on the shape of the VDOS below jamming predicted by RMT \cite{Cicuta}, and verified numerically in Ref. \cite{Saitoh}, we shall put, recalling Eq. \eqref{edge}:
\begin{equation}
    g(\omega_p) \approx K, \,\,\,\ \forall \, \omega_p > \omega_*\approx 0.177\, (2d -z), \label{DOs_liq}
\end{equation}
where $K = const$.

As usual, we should also specify the form of the affine-force correlator, $\Gamma(\omega_p)$.
Saitoh et al. \cite{Saitoh} have numerically verified for weakly sheared hard spheres below jamming that $\Gamma(\omega_p)$ is given, also in this case, by the theory of Zaccone \& Scossa-Romano \cite{Scossa}, Eq. \eqref{scos}.
Putting all these results together (Eq. \eqref{viscosity_nonaffine} with \eqref{DOs_liq} and \eqref{scos}), we thus obtain the following mathematical prediction for the viscosity upon approaching the jamming transition from below:
\begin{equation}
 \eta = \frac{3}{15}\frac{N}{V}\tilde{\nu}(0) \kappa R_{0}^{2} K' \left(\frac{1}{\omega_{*}} - \frac{1}{\omega_{D}}\right).    \label{viscosity_div}  
\end{equation}
By substituting Eq. \eqref{edge} in the above expression for the viscosity, we thus obtain
\begin{equation}
 \eta = \frac{3}{15}\frac{N}{V}\tilde{\nu}(0) \kappa R_{0}^{2} K' \left\{\frac{1}{[0.177\, (2d -z)]} - \frac{1} {\omega_{D}}\right\},
   \label{viscosity_appr}
\end{equation}
where we have set $K' \equiv K/N$ to account for the normalization of the VDOS as discussed above.
This expression, in the limit $z \rightarrow 2d^{-}$ reduces to
\begin{equation}
 \eta = 1.13 \frac{N}{V}\tilde{\nu}(0) \kappa R_{0}^{2} K' (2d -z)^{-1}.
   \label{viscosity_approx}
\end{equation}
For the scaling exponent of $\omega_{*}$ vs $(2d-z)$, Saitoh et al. \cite{Saitoh} in their more recent numerical simulations found a slightly larger exponent $1.3$ (whereas earlier simulations \cite{Lerner2012} found an exponent $1$):
\begin{equation}
    \omega_{*} \sim (2d -z)^{-1.3}.
\end{equation}
Using Saitoh's result, we obtain the following formula for the viscosity 
\begin{equation}
 \eta = 1.13 \rho \tilde{\nu}(0) \kappa R_{0}^{2} K' (2d -z)^{-1.3},
   \label{viscosity_fin}
\end{equation}
where $\rho =N/S$ and $\rho =N/V$ is the number density in $d=2$ and $d=3$, respectively.
In the $d=2$ hard disks system studied by Saitoh et al. \cite{Saitoh}, it was found from numerics that
\begin{equation}
    \mid z- 2d \mid \sim \mid \rho - \rho_c \mid
\end{equation}
where $\rho$ is the number density and $\rho_c$ is the number density at jamming. Using this result, the theory we have developed above thus predicts the divergence of the viscosity upon approaching jamming from below as a function of density as:
\begin{equation}
 \eta \sim \mid \rho - \rho_c \mid^{-1.3}
   \label{viscosity_den}
\end{equation}
thus, with a critical exponent $1.3$ which is not too far from the exponent $1.6$ measured in numerical simulations by Olsson \& Teitel for sheared athermal disks in 2D \cite{Olsson,Olsson_2024}.
It could be interesting to extend this mathematical approach to the viscosity of bidisperse and polydisperse hard spheres, which are known to jam at larger packing fractions \cite{Torq_bin,Anzivino,Desmond}. For the case of bidisperse spheres, for example, the viscosity is indeed observed to diverge at a larger (MRJ) packing fraction \cite{Torq_visc}, in agreement with these theories. Also, a relatively unexplored case is that of shear-driven jamming for attractive particles for which mostly only experiments exist \cite{Trappe2001}, or with competing interactions \cite{PRL_solidification}. In the latter case, the kinetics is understood in terms of shear-assisted thermal activation \cite{Zaccone2009}, but a systematic study of critical exponents is lacking and the above approach could be also applicable in future work.

Finally, we should note that the memory function or friction kernel $\nu(t)$ can be evaluated on the basis of MD simulations for which different methods are available. As discussed in \cite{zaccone2023}, one of them is based on the fluctuation-dissipation theorem \cite{Schmid,Pelargonio}, whereas another method is based on time-correlation functions \cite{Kuehn_2016}. 
The above results may be useful for a microscopic understanding of the phase diagram of shear jamming \cite{Bulbul,Yuliang_rev}.

\section{Power-law creep above jamming}
We have seen, in the previous sections, that the nonaffine deformation theory, derived from a microscopically-reversible Hamiltonian, is able to quantitatively predict the critical behaviour of the shear modulus above jamming and to semi-quantitatively predict the divergence of the viscosity at jamming approached from below.
We shall now consider the viscoelastic creep behaviour upon approaching jamming from above, by leveraging the microscopic linear viscoelastic theory given by Eqs. \eqref{visco-moduli}.

A main tenet of linear response theory, the Boltzmann's superposition principle, stipulates that the stress at time $t$ resulting from the application of strain at an earlier time $t'$ is given by 
\begin{equation}
\sigma(t)=\int_{-\infty}^tG(t-t')\dot{\gamma}(t')dt
\label{A3.stress}
\end{equation}
where $G(t)$ is the time-dependent elastic modulus, also called the ``relaxation'' modulus or creep modulus.

In the experimental practice, $G(t)$ is measured via the material response to a step strain of amplitude $\gamma_{0}$: $\dot{\gamma}=\gamma_0 \delta(t')$ from which the above Boltzmann equation becomes 
\begin{equation}
\sigma(t)=G(t)\gamma_0.
\end{equation}
Upon inverting the stress-strain relation, one can make it explicit for the strain instead of the stress:
\begin{equation}
    \gamma(t)=\int_{-\infty}^t J(t-t')\dot{\sigma}(t')dt
\end{equation}
to obtain:
\begin{equation}
    \gamma(t)=J(t)\sigma_0
\end{equation}
which is the strain response to a step disturbance in the stress. Also, $J(t)$ is the compliance or compliant modulus, i.e. the ratio of strain to stress or just the inverse of the elastic modulus $G(t)$.
For crystalline materials, the creep modulus exhibits a $1/3$ power-law that was reported already in 1910 by Andrade: $ \gamma(t) \sim t^{1/3}$, hence the name "Andrade'' creep. The microscopic mechanism of the Andrade creep in crystals has been elucidated in terms of the (Peierls-Nabarro) thermally-activated motion of dislocations.

Assuming a Markovian viscous damping $\nu$, and using the microscopic nonaffine viscoelastic equations presented in Sec. IV, Eqs. \eqref{visco-moduli}, the creep modulus is retrieved via a simple Fourier transformation from the frequency to the time domain:
\begin{equation}
\begin{split}
&G(t)=G_A- \frac{1}{2\pi V} \int_{-\infty}^{\infty} \int_{0}^{\infty} \frac{g(\omega_p)
\Gamma(\omega_p) \exp(i \omega t)}{\omega_{p}^2 - \omega^2  + i \nu
\omega} d\omega_{p} d\omega\\
&= G_A - \frac{t e^{- \frac{\nu}{2}t}}{V} \int_{0}^{\infty} g(\omega_p)
\Gamma(\omega_p) \,\mathrm{sinc}\left(\frac{1}{2} \sqrt{4 \omega_{p}^2 - \nu^2} t\right)
d\omega_{p}.\label{relax_G}
\end{split}
\end{equation}
For large $\nu$ and large times, Eq. \eqref{relax_G} can be put in simpler form. First we approximate $\sqrt{\nu^2 - 4 \omega_{p}^2}
\approx \nu - 2\,\frac{\omega_{p}^2}{\nu}$, where we use $\omega_{p} \ll \nu$.
Then, we substitute this result in Eq. \eqref{relax_G} and we use the definition of $\sinh(x)$ to get
    \begin{equation}\label{Gtapprox}
    \begin{gathered}
    G(t)\,\approx\,\frac{2}{V}\, e^{- \frac{\nu}{2}t} \int_{0}^{\infty}
\frac{g(\omega_{p}) \Gamma(\omega_{p}) \sinh(\frac{\nu}{2} t - \frac{\omega_{p}^2}{\nu}
t)}{\omega_{p}^{2}(\nu - 2\frac{\omega_{p}^2}{\nu})} d\omega_{p}\\
    =\, \frac{1}{V}\, \int_{0}^{\infty} \frac{g(\omega_{p}) \Gamma(\omega_{p})(e^{-
\frac{\omega_{p}^2}{\nu} t} - e^{-\nu t + \frac{\omega_{p}^2}{\nu} t})}{\omega_{p}^{2}(\nu
- 2\frac{\omega_{p}^2}{\nu})} d\omega_{p}\\
    \approx\, \frac{1}{V}\,\frac{1}{\nu} \int_{0}^{\infty} \frac{g(\omega_{p})
\Gamma(\omega_{p})}{\omega_{p}^{2} } e^{- \frac{\omega_{p}^2}{\nu} t} d\omega_{p}.
    \end{gathered}
    \end{equation}
    In the last step we used $\nu \gg 2 \omega_{p}^2/\nu$ and $\nu t -
\omega_{p}^2 t/\nu \gg 1$. This corresponds to a set of Maxwell elements with
relaxation times $\tau_{p} = \nu/\omega_{p}^2$ \cite{Zaccone_2020}.

Let us now recall the standard connection
between the vibrational density of states (VDOS) and the eigenvalue spectrum $\rho(\lambda)$ of the Hessian
matrix, $g(\omega_p)d\omega_p = \rho(\lambda_p)d\lambda_p$, with $\omega_p^{2}=\lambda_p$ (in units of $m=1)$.
At the isostatic point of 3D jammed packings, $z_c=6$, the VDOS, instead of the Debye law $g(\omega_p)\sim \omega_{p}^{2}$, develops a plateau
of soft modes \cite{OHern}, $g(\omega_p)\sim const$. This limit corresponds to the
scaling $\rho(\lambda_p) \sim \lambda_{p}^{-1/2}$ in the eigenvalue distribution.
Mathematically, this scaling arises 
from the random-matrix eigenvalue spectrum, as it can be derived from the Marchenko-Pastur distribution
of random matrix theory $\rho_{MP}$ \cite{Cicuta} reported as Eq. \eqref{Marchenko} in Sec. VI.B. In the VDOS obtained numerically by digonalization of the Hessian matrix
, a
scaling $\rho(\lambda) \sim a + \lambda^{-1/2}$ is observed \cite{Milkus_RMT}, where $a$ is a constant. In the low-energy part of the spectrum, the scaling $\rho(\lambda_p) \sim
\lambda_{p}^{-1/2}$ dominates the asymptotic behavior. Recall now that
$\Gamma(\omega_p)\sim\omega_{p}^{2}$, according to Eq. \eqref{scoss} \cite{Scossa}, which directly implies
$\tilde{\Gamma}(\lambda_p)\sim\lambda_p$. Upon substituting these results in the last line
of Eq. \eqref{Gtapprox}, we obtain the following Laplace transform, which can be easily
evaluated:
    \begin{equation}
    \begin{split}
    G(t) & \sim \int_{0}^{\infty}
\frac{\rho(\lambda_p)\tilde{\Gamma}(\lambda_p)}{\lambda_p} e^{- \lambda_p t} d\lambda_p
\sim \int_{0}^{\infty} \frac{\lambda_{p}^{-1/2}\lambda_p}{\lambda_p} e^{- \lambda_p t}
d\lambda_p\\
    & \sim t^{-1/2},\label{power-law}
    \end{split}
        \end{equation}
a result which was derived in Ref. \cite{Milkus2}.
The last line arises upon performing the integral. 
Hence, the nonaffine deformation theory \cite{Scossa} together with the random matrix theory description of the eigenvalue spectrum of the Hessian, predict the power-creep $G(t) \sim t^{-1/2}$ with power-law $1/2$ that has been observed in numerical simulations by Tighe in 2011 \cite{Tighe}.

\section{Predicting the random close packing of spheres based on jamming}
\begin{figure}
\includegraphics[width = \linewidth]{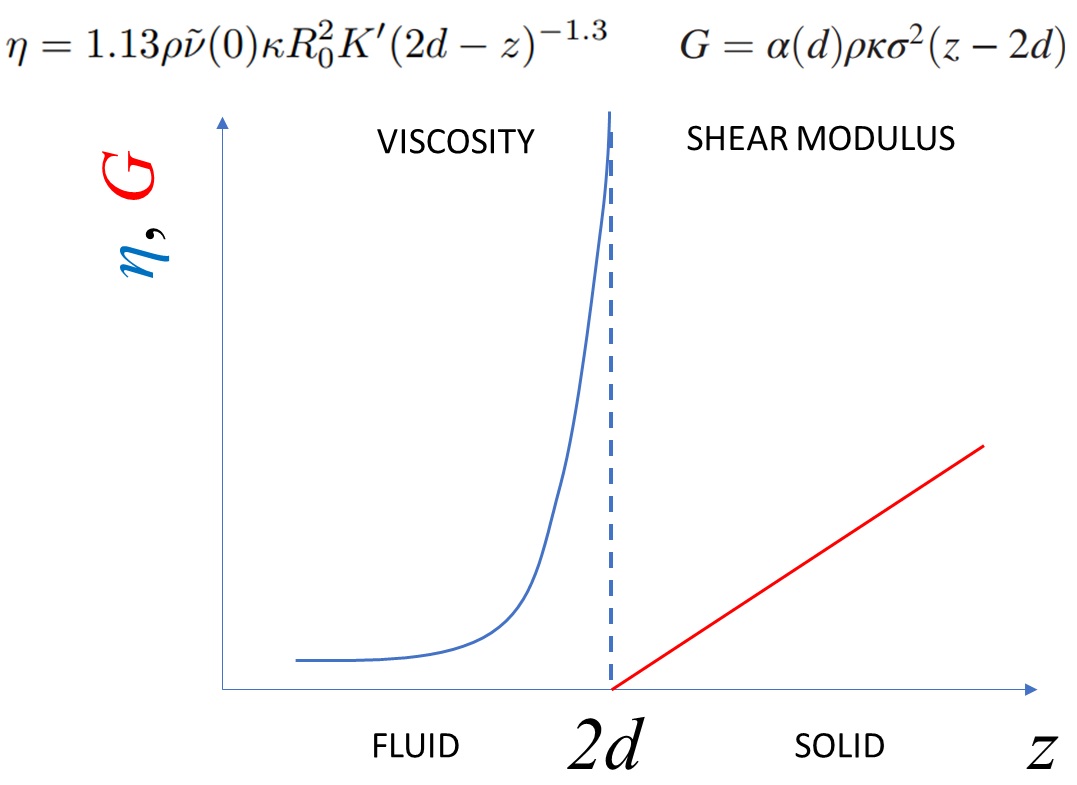}
	\caption{Schematic summary of the mathematical predictions of the nonaffine dynamic theory reviewed in this article. The viscosity diverges at the jamming point $z_c =2d$, approached from below, with an exponent $(2d -z)^{-1/3}$ in $d=2$, cfr. Eq. \eqref{viscosity_fin}. The static shear modulus rises from zero at $z_c=2d$ following Eq. \eqref{modulus_closed}\cite{Scossa} written for a generic $d$-dimensional system where the dimension-dependent prefactor $\alpha$ is given by $\alpha \big( d=2 \big)= 1/18$ in $d=2$ and by $\alpha \big( d=3 \big)= 1/30$ in $d=3.$}
		\label{fig4}
\end{figure}

The main predictions of the nonaffine deformation theory for the jamming transition are schematically summarized in Fig. \ref{fig4}.

The power-law scalings in the viscosity and in the shear modulus on the two sides of the transition are compatible with a second-order phase transition scenario reminiscent of gelation \cite{Durand} and of the metal-insulator transition in random resistor networks \cite{Efros}.

A special attention deserves the critical point $z_c = 2d$, or jamming point or point-J. 
Since Bernal's pioneering experiments on balls packings \cite{Bernal} it is known that the closest packing of equal-sized (monodisperse) spheres occurs when the coordination number is about $6$ (or $2d$), when the solid friction between the spheres is negligible.

Unlike the case of other transitions or packing densities of regular (e.g. crystalline) structures, the maximum random packing density of spheres has remained an unsolved problem for a long time. In the absence of analytical solutions, many computational simulations studies have located the random close packing (RCP) fraction in a range $\phi_{RCP} \approx 0.63-0.66$ \cite{Torquato_review}.
Also, it has been importantly recognized that introducing even a small amount of structural order can lead to an increase of the observed $\phi_{RCP}$ \cite{Truskett}. Hence, while the critical coordination number is always $z_c = 2d$ sharp, significant uncertainties have surrounded the determination of the corresponding solid volume fraction, denoted as $\phi_J$ or $\phi_{RCP}$.

Recently, an analytical solution for the random close packing fraction has been derived by leveraging the $d$-dimensional onset of rigidity encoded in the shear modulus expression \cite{Scossa}:
\begin{equation} 
G = \alpha (d) \rho \kappa \sigma^{2} \big( z - 2d \big), \label{dimen}
\end{equation}
where $z$ is the coordination number, $\rho = N/V$ is the number density (obviously, $\rho=N/S$ in $d=2$), $\kappa$ is the spring constant of the nearest-neighbor interaction, and $\alpha$ is a constant prefactor given by $\alpha \big( d=2 \big)= 1/18$ in $d=2$ and by $\alpha \big( d=3 \big)= 1/30$ in $d=3.$

Based on the mathematical theory of jamming presented in the previous Sections, it is clear that a system of athermal hard spheres is fluid below jamming ($z<2d$) and is solid above jamming ($z>2d$). This fact, together with the rigidity theory of Zaccone \& Scossa-Romano, Eq. \eqref{dimen}, provides a clear-cut quantitative criterion to define the jamming transition, and hence can be used to provide a robust estimate of $\phi_{RCP}$, since the latter must coincide with the emergence of rigidity in the system (if that, by absurdum, was not the case, particles may still rearrange in order to find an even denser configuration until rigidity sets in preventing further motions) \cite{Zaccone_2022}. Upon recalling the relation between the radial distribution function (rdf) $g(r)$ and $z$, one arrives at a relationship between the contact value of the rdf, $g(\sigma)$ (a function of the system's density $\phi$), the critical contact number $2d$ and the corresponding packing fraction $\phi_{RCP}$. Here, the assumption is used that RCP can be obtained via an equilibrium-like crowding process \cite{Kamien,Anzivino,Ken}, which allows one to use analytical approximations for the HS equation of state in order to estimate $g(\sigma)$ and its dependence on $\phi$.
This relation, however, involves also an unknown dimensionful coefficient $g_0$, which is the prefactor of the Dirac delta used in describing the contact part of the rdf \cite{Torquato_2018}. This coefficient $g_0$ can be determined using an effective boundary condition on the contact value in the hard-sphere (HS) phase diagram. For example, a point in the HS phase diagram where both $z$ and $\phi$ are known exactly is the face centered cubic (FCC) packing, corresponding to a FCC arrangement of hard spheres. Using this known point in the phase diagram, one can determine $g_0$ and eventually get the following estimate for the random close packing fraction in $d=3$:
\begin{widetext}
\begin{equation}
\phi_{RCP} =\frac{2\sqrt{648+\pi[\,\pi(54-24\sqrt2\pi+5\pi^2)-108\sqrt2\,]}}{36\sqrt2+\pi(\sqrt2\pi - 36)}+\frac{2(36\sqrt2-48\pi)}{36\sqrt2+\pi(\sqrt2\pi-36)}-3=0.658963
\label{rcp_1}
\end{equation}
as derived in Ref. \cite{Zaccone_2022}
As suggested by C. Likos \cite{Likos} in a commentary on Ref. \cite{Zaccone_2022}, another possible choice to determine $g_0$ is to use the body center cubic crystal (BCC) instead of the FCC. For the BCC, one also knows, exactly, both values of $z$, which is 8, and of the corresponding $\phi$, which is $\pi \sqrt{3}/8$. Using this choice, one therefore obtains the following exact prediction for the RCP fraction:
    \begin{equation}
\phi_{RCP} = \frac{2 \sqrt{3} \left(768+\sqrt{16384+\pi  \left(1024 \sqrt{3}+3 \pi  \left(-192-128 \sqrt{3} \pi +39 \pi ^2\right)\right)}\right)-2304 \pi }{256 \sqrt{3}+3 \pi  \left(\sqrt{3} \pi -112\right)}-7 = 0.64332 \label{rcp_2}
\end{equation}
\end{widetext}
which retrieves the most quoted value in numerical studies for frictionless monodisperse sphere packings in 3D, $\phi_{RCP}\approx 0.64$ \cite{OHern}.
These two estimates, Eq. \eqref{rcp_1} and Eq. \eqref{rcp_2}, constrain the values of the RCP volume fraction in $d=3$ to a relatively narrow range. In retrospect, the value $0.64332$ obtained using the BCC boundary condition might be closer to empirical observations in view of its being the "known" point in the HS diagram which is closest to the RCP.

Similarly, an estimate can be obtained for $\phi_{RCP}$ also in $d=2$, giving the value 0.88644, which is not far from the numerical estimate of Ref. \cite{Makse_2010}, although also in this case quite a spread of values are reported in numerical studies \cite{QUICK,OHern,Makse_2010,Brouwers}, and the situation is made more complicated by the vicinity of the 2D melting point, with its peculiar Berezinskii-Kosterlitz-Thouless phenomenology. Also, an important caveat in 2D is that, for monodisperse disks, RCP values as high as 0.88 are often associated with polycrystalline ordering. Indeed, there is an important conceptual and quantitative difference (including packing fractions)
between 2D RCP versus 2D MRJ states of monodispersed circular disks that shas been highlighted in Ref. \cite{Torq_2D} and should be noted in this context.

Interestingly, however, the most recent results on 2D random disk packings with particle size polydispersity obtained using the Monte-Carlo swap algorithm with irreversible collective particle swaps, provide values which are quantitatively consistent with the analytical theory of Refs. \cite{Anzivino,Zaccone_2022}.

The one presented above from Refs. \cite{Zaccone_2022,Likos} remains, to date, the only simple analytical estimate of RCP in $d=3$. It has been successfully extended to provide analogous analytical expressions for the RCP of unequally sized spheres, e.g. binary mixtures of hard spheres and polydisperse packings in Ref. \cite{Anzivino}. The analytical relations allow one to find the closest packing fractions as a function of polydispersity, in the case of continuous size distributions, or of the ratio between the two sizes and the mixture composition in the case of binary mixtures. The analytical relations of Refs.\cite{Zaccone_2022,Anzivino} have found widespread application in various contexts from physical chemistry \cite{Blondot} to planetary science and astronomy, where the lunar regolith has been argued to follow the analytical prediction of Ref. \cite{Anzivino} in recent work \cite{regolith}. In particular, the theoretical expressions of Ref. \cite{Anzivino} correctly predicts the increase of packing fraction with increasing the size polydispersity, with a plateau saturating at some large value below $\phi=1$, as observed in countless simulations and experiments, including, most recently, silica crush curve experiments \cite{Malamud_2024}.

Finally, we should mention the effect of quenched disorder \cite{Reichardt_pin}, in particular the effect of particle pinning \cite{Graves_1}. In numerical simulations of jammed packings, it is observed that even with a small fraction $n_f$ of pinned particles, the average contact number needed to jam the system gets reduced. So, for example, in 3D, the system of frictionless particles will jam when the average coordination is $z_{c}(n_f)$, which is a decreasing function of $n_f$.
This, in turn, leads to jamming at lower packing fractions. This could be interpreted by recalling that the vanishing of the shear modulus at $z_c=2d$ is a direct consequence of the sum over $dN$ degrees of freedom in the nonaffine correction to the shear modulus. With a fraction of particles being pinned, the total number of degrees of freedom available for nonaffine relaxation of forces will, thus, be less than $dN$. We would also expect the lower edge of the VDOS in the unjammed state, $\omega^{*}$, to vanish at a lower value than $2d$, to reflect the lowered number of degrees of freedom in the system due to pinning. 
It could be interesting, in future work, to put these considerations in a more quantitative form. This may also be useful for applications, because jamming with pinning or quenched disorder is a model to understand the phenomenon of clogging in industrial and environmental applications \cite{Reichardt_clog}.

\section{Bulk modulus}
A special consideration also goes to the bulk modulus of jammed packings. Ellenbroek et al. \cite{Ellenbroek} were the first to observe, in numerical simulations, that the shear modulus of random networks (with no excluded volume) and of jammed packings behave the same, whereas their bulk modulus behaves very differently. In particular, they showed that the bulk modulus of networks behaves like the shear modulus with the distance from jamming, i.e. $K \sim (z-2d)$, while in jammed packing it behaves in a nearly affine manner, i.e. $K \sim z$. This apparent paradox raised by numerical simulations has been resolved by the nonaffine deformation theory for $d=2$ in Ref. \cite{JAP} and for $d=3$ in Ref. \cite{Schlegel2016}. It was shown that, under a shear deformation, a description of the angular distribution of particle contact orientations as a random process in the solid angle suffices to describe the shear modulus. This is because non-random correlations due to excluded volume between nearby particles cancel each other out under the strongly anisotropic shear field. Instead, under a hydrostatic (hence, isotropic) field such as compression, the excluded-volume correlations do not cancel and act as to reduce the nonaffine correction, $-K_{NA}$, to the bulk modulus, thus leaving $K=K_A - K_{NA} \approx K_A \sim z$. The mathematical details of this proof can be found in the original papers \cite{JAP,Schlegel2016}
 or in Chapter 2 of \cite{zaccone2023}.
Within the context and scope of affine theories, progress has been recently made in extending predictions of the compression behaviour of jammed packings to the nonlinear compaction regime and to shape-changing grains in Refs. \cite{Azema1,Azema2}.
 
\section{Application to atomic-scale materials}
The physics of jamming provides an ideal ground to understand the microscopic elasticity of amorphous solids at a deeper mechanistic level. This understanding has then been translated to real-world materials, where the constituent building blocks are not athermal grains but atoms subjected to a plethora of interactions with other atoms (as well as with electrons) and to thermal motion. 
The nonaffine theory presented above, after having been proved successful on the example of jammed packings, has been extended to real materials. An analytical formula has been derived for the low-frequency static shear modulus of polymer glasses, in Ref. \cite{ZacconePRL}, which provides a good description of the temperature-dependent shear modulus of polystyrene glass upon approaching the glass transition from below. The same framework also yields a molecular-level expression for the glass transition temperature of polymer glass, which also predicts the correlation between glass transition temperature and colloidal stability of polymer colloids \cite{Anzivino2023}. This formula was based on a previous extension of the Zaccone \& Scossa-Romano formula Eq. \eqref{modulus_closed} to particles with covalent bonds represented by three-body bond-bending interactions \cite{MPLB}. If only covalent bonds are present in the system, this changes the critical scaling, from $G \sim (z - 6)$ to $G \sim (z - 2.4)$ \cite{MPLB}, in agreement with early simulations of covalent networks \cite{Thorpe}.
If, besides covalent bonds, there are also van der Waals (central-force) attractive interactions, the critical scaling is changed further to $G \sim (z - 4)$, in 3D, as derived in Ref. \cite{ZacconePRL}.
By adding thermal expansion to the picture, an analytical formula was derived for the low-frequency static shear modulus of polymer glasses, in Ref. \cite{ZacconePRL}, as a function of temperature which is able to predict the sharp drop, by several orders of magnitude, of the shear modulus of polymer glasses upon approaching the glass transition from below.

Another early application of the nonaffine deformation theory to polymer materials, has been to coarse-grained (Kremer-Grest) models of polymer glasses \cite{Palyulin} and to "jammed" athermal polymer models \cite{Hoy}. In \cite{Palyulin} the nonaffine equations for the viscoelastic moduli Eqs. \eqref{visco-moduli} (computed using the VDOS from direct diagonalization of the Hessian matrix) have been shown to provide an excellent parameter-free prediction of the $G'(\omega)$ and $G''(\omega)$ measured in molecular simulations in LAMMPS. The comparison is parameter-free, because the same value of viscous damping $\nu$ was used both in Eqs. \eqref{visco-moduli} and in the Langevin thermostat employed in the MD simulations. Subsequently, the same approach has been applied to atomistic simulations of real polymer glasses, in Ref. \cite{Elder}, and, in particular, to two thermoset networks with differing polar side groups, dicyclopentadiene (DCPD) and 5-norbornene-2-methanol (NBOH). In this case, the atomic-scale damping parameter $\nu$ was determined by matching the theoretical predictions of Eqs. \eqref{visco-moduli} with the MD simulations at high frequency where the MD simulations are trustworthy. This value of $\nu$ has then been used to produce predictions at much lower frequencies, where the MD simulation is no longer able to provide reliable data. Furthermore, the atomic vibrations responsible for the nonaffine softening were identified as a route towards chemical design of super-hard materials \cite{Elder}.
The application to atomistic polymer networks has been further extended to the case of a cross-linked epoxy system of diglycidyl ether of bisphenol A (DGEBA) and poly(oxypropylene) diamine in Ref. \cite{vaibhav}. Using the same protocol to determine the only unknown parameter $\nu$ from the match with MD at high frequency, it has been shown that the nonaffine deformation theory is able to make predictions of the viscoelastic moduli across many orders of magnitude in frequency, and to be able to predict the experimentally measured values of the modulus at temperatures below the glass transition. 

Finally, the nonaffine deformation theory has been also applied to hard atomic-scale materials such as $\alpha$-quartz. This is a paradigmatic crystalline solid which is non-centrosymmetric, and thus one expects the nonaffine correction $-G_{NA}$ to be particularly important. This was indeed the case in the calculations of Ref. \cite{quartz}, where the nonaffine lattice dynamic theory was extended to long-range electrostatic interactions by means of the Ewald method. The theoretical predictions were found in excellent agreement with experimental data from the literature.

\section{Conclusion and outlook}
In conclusion, we have provided a bird-eye perspective of the mathematical approach to the jamming transition, which has cross-fertilized into other fields to offer a robust methodology to compute the mechanics of real amorphous materials at the atomic scale.

As emphasized in \cite{Sethna}, the key starting quantity is not the free energy but rather the potential energy of the system, as one would naively expect for athermal systems with elastic interactions between the constituents. In this sense, considering the free energy instead of the potential energy is a rather unfruitful complication, which prevents one from obtaining quantitative results in simple analytical form (although, of course, it has been an important field of fundamental research with ramifications in the abstract sciences). Hence, starting from a microscopically reversible particle-bath Hamiltonian, by further imposing mechanical equilibrium at each step in the deformation for all particles in the system, and by assuming a random angular distribution of nearest-neighbours in the solid angle, an exact analytical formula for the shear modulus of frictionless jammed packings was derived in Ref. \cite{Scossa}, which still stands as the only quantitative analytical prediction of the quintessential property of jamming, the emergence of shear rigidity, in parameter-free agreement with the numerical simulations of Ref. \cite{OHern}. The same theory also produces formulae for the viscoelastic moduli and for the viscosity below jamming. It has been shown in this paper, that an analytical prediction of the viscosity of jamming systems can be obtained starting from the same microscopically reversible Hamiltonian used to derive the shear modulus of jammed packings above jamming. The prediction relies on the mathematical description of the vibrational density of states (VDOS). While the latter is \emph{de facto} irrelevant for the prediction of the shear modulus, it is, instead, crucial for the prediction of the viscosity. A digression on the VDOS showed that its salient features are captured by random matrix theory (RMT) applied to the Hessian matrix, which is a block random matrix, including the dependence of the lower edge on the distance to jamming \cite{Cicuta}. A power-law divergence of the viscosity with an exponent 1.3 is predicted, which is not far from the numerical result 1.6 of Ref. \cite{Olsson}.
Furthermore, the nonaffine deformation theory provides correct predictions also for the power-law creep modulus near jamming \cite{Milkus2} and for the bulk modulus \cite{JAP,Schlegel2016}. 

The application of these concepts and the associated framework to atomic-scale materials has been briefly summarized. In particular, active current ares of research, which certainly will remain very active in the next years, are directed towards identifying mathematically well-defined topological defects in amorphous solids. The goal is to provide a deeper understanding of the plastic deformation of amorphous solids \cite{Falk,Arratia} in terms of topological defects in the same way as crystal plasticity is understood and described in terms of dislocations. 
In 2021, well-defined topological defects associated with a continuous Burgers vector were discovered in simulations of glasses \cite{Baggioli} and theorized in \cite{Landry}. Following this discovery, the existence of discretized topological defects in the eigenvector field of glasses has been further reported for 2D glasses in \cite{Wu2023,desmarchelier} and for 3D glasses in \cite{nexus}, with growing evidence about their key role in triggering plastic events and plastic failure. Much studied are also the screening properties of dipoles of topological defects \cite{moshe,procaccia}. In this context, it will be of great interest to apply the new topological defects to the oscillatory yielding of glassy and jammed systems \cite{Denisov2015,Ido,Aime2023}, where the onset of irreversibility and chaos at the plastic transition is thought to coincide with an absorbing phase transition \cite{Ido2}.

In the context of amorphous and granular materials, applying the above mathematical concepts to systems of particles with non-spherical shape \cite{Corey,Donev,hoy2,moussa} still presents challenges for a fully analytical (non-numerical) description, and is an important task for future mathematical work in this area, especially for the rational design of materials with tailored mechanical properties \cite{Jaeger_shape,Truskett_inverse}. The same applies to frictional particles (with "solid" friction in the contacts) \cite{Schwartz_2,Yujie,Dauchot,Henkes}, force chains \cite{Daniels,Azema}, wet granular media \cite{bonn} and nanocomposites \cite{bonn_composite,Kumar}. For example, the theory of Zaccone and Scossa-Romano \cite{Scossa} has been extended to frictional jamming by means of a numerical evaluation of the eigenmodes and of the Hessian in Ref. \cite{Hayakawa1}, for the linear elastic regime, and in Ref. \cite{Hayakawa2} for the finite deformation regime.

Another, technologically-relevant, problem where these concepts may prove fertile is that of flow-induced clogging instabilities caused by jamming of suspended particles, where a complex interplay between fluid mechanics, interparticle interactions and particle-scale mechanics characterize the solidification process \cite{Hua,Sauret}.

Further exciting research directions include the application of the above nonaffine deformation and rigidity concepts to biological systems, from classical topics such as the actin network mechanical properties \cite{frey} to the recent proposal that embryonic tissue phase transitions could be due to an underlying rigidity percolation transition similar to jamming as described above \cite{Heisenberg} and motility-driven jamming and rigidity transitions in biological tissues \cite{Manning,Schwartz}. Although the concept of jamming as described in this article presents significant differences with the epithelial (un)jamming in biophysics and pathology \cite{Fredberg}, this is nevertheless an analogy which deserves to be further explored in future studies. This can be done, for instance, via implementation of active motion \cite{Hannezo} in the above generalized Langevin dynamics Eq. \eqref{2.4gle1} 
\cite{Janssen}.

Finally, there is a growing interest in developing synthetic materials that mimic biological systems and involve jamming transition(s). For example, hybrid  soft composites mimic tissue mechanics \cite{Fang2020,Janmey}. These fibrous networks contain densely-packed particles near jamming \cite{Shivers,Zhao2024}. In these composite, hybrid, materials, the nonaffine displacements of the densely packed particles were found to play a crucial role in determining the global mechanics \cite{Shivers}. Mathematical concepts such as those reviewed in this article may prove useful to understand the mechanics of these emerging biomimetic materials, especially for the rational tuning of their mechanical performance in biomedical applications.

\section*{Data availability}
Data sharing is not applicable to this article as no new data were created or analyzed in this study.

\subsection*{Acknowledgments} 
The author is indebted to Giovanni Cicuta, Salvatore Torquato, Hisao Hayakawa, Émilien Azéma, and Edouard Hannezo for reading the manuscript and for providing useful feedback.
The author gratefully acknowledges funding from the European Union through Horizon Europe ERC Grant number: 101043968 ``Multimech'', and from US Army Research Office through contract nr. W911NF-22-2-0256. 

\bibliographystyle{apsrev4-1}

\bibliography{refs}

\begin{thebibliography}{176}%
\makeatletter
\providecommand \@ifxundefined [1]{%
 \@ifx{#1\undefined}
}%
\providecommand \@ifnum [1]{%
 \ifnum #1\expandafter \@firstoftwo
 \else \expandafter \@secondoftwo
 \fi
}%
\providecommand \@ifx [1]{%
 \ifx #1\expandafter \@firstoftwo
 \else \expandafter \@secondoftwo
 \fi
}%
\providecommand \natexlab [1]{#1}%
\providecommand \enquote  [1]{``#1''}%
\providecommand \bibnamefont  [1]{#1}%
\providecommand \bibfnamefont [1]{#1}%
\providecommand \citenamefont [1]{#1}%
\providecommand \href@noop [0]{\@secondoftwo}%
\providecommand \href [0]{\begingroup \@sanitize@url \@href}%
\providecommand \@href[1]{\@@startlink{#1}\@@href}%
\providecommand \@@href[1]{\endgroup#1\@@endlink}%
\providecommand \@sanitize@url [0]{\catcode `\\12\catcode `\$12\catcode `\&12\catcode `\#12\catcode `\^12\catcode `\_12\catcode `\%12\relax}%
\providecommand \@@startlink[1]{}%
\providecommand \@@endlink[0]{}%
\providecommand \url  [0]{\begingroup\@sanitize@url \@url }%
\providecommand \@url [1]{\endgroup\@href {#1}{\urlprefix }}%
\providecommand \urlprefix  [0]{URL }%
\providecommand \Eprint [0]{\href }%
\providecommand \doibase [0]{http://dx.doi.org/}%
\providecommand \selectlanguage [0]{\@gobble}%
\providecommand \bibinfo  [0]{\@secondoftwo}%
\providecommand \bibfield  [0]{\@secondoftwo}%
\providecommand \translation [1]{[#1]}%
\providecommand \BibitemOpen [0]{}%
\providecommand \bibitemStop [0]{}%
\providecommand \bibitemNoStop [0]{.\EOS\space}%
\providecommand \EOS [0]{\spacefactor3000\relax}%
\providecommand \BibitemShut  [1]{\csname bibitem#1\endcsname}%
\let\auto@bib@innerbib\@empty
\bibitem [{\citenamefont {Zaccone}(2023{\natexlab{a}})}]{zaccone2023}%
  \BibitemOpen
  \bibfield  {author} {\bibinfo {author} {\bibfnamefont {A.}~\bibnamefont {Zaccone}},\ }\href@noop {} {\emph {\bibinfo {title} {Theory of Disordered Solids}}}\ (\bibinfo  {publisher} {Springer},\ \bibinfo {address} {Cham},\ \bibinfo {year} {2023})\BibitemShut {NoStop}%
\bibitem [{\citenamefont {van Saarloos}\ \emph {et~al.}(2024)\citenamefont {van Saarloos}, \citenamefont {Vitelli},\ and\ \citenamefont {Zeravcic}}]{Vitelli}%
  \BibitemOpen
  \bibfield  {author} {\bibinfo {author} {\bibfnamefont {W.}~\bibnamefont {van Saarloos}}, \bibinfo {author} {\bibfnamefont {V.}~\bibnamefont {Vitelli}}, \ and\ \bibinfo {author} {\bibfnamefont {Z.}~\bibnamefont {Zeravcic}},\ }\href@noop {} {\emph {\bibinfo {title} {Soft Matter: Concepts, Phenomena, and Applications}}}\ (\bibinfo  {publisher} {Princeton University Press},\ \bibinfo {address} {Princeton},\ \bibinfo {year} {2024})\BibitemShut {NoStop}%
\bibitem [{\citenamefont {Binder}\ and\ \citenamefont {Kob}(2011)}]{kob_book}%
  \BibitemOpen
  \bibfield  {author} {\bibinfo {author} {\bibfnamefont {K.}~\bibnamefont {Binder}}\ and\ \bibinfo {author} {\bibfnamefont {W.}~\bibnamefont {Kob}},\ }\href@noop {} {\emph {\bibinfo {title} {Glassy Materials and Disordered Solids}}}\ (\bibinfo  {publisher} {World Scientific, Singapore},\ \bibinfo {year} {2011})\BibitemShut {NoStop}%
\bibitem [{\citenamefont {Torquato}(2002)}]{Torquato_book}%
  \BibitemOpen
  \bibfield  {author} {\bibinfo {author} {\bibfnamefont {S.}~\bibnamefont {Torquato}},\ }\href@noop {} {\emph {\bibinfo {title} {Random Heterogeneous Materials: microstructure and macroscopic properties}}}\ (\bibinfo  {publisher} {Springer, New York},\ \bibinfo {year} {2002})\BibitemShut {NoStop}%
\bibitem [{\citenamefont {de~Gennes}(1999)}]{deGennes_granular}%
  \BibitemOpen
  \bibfield  {author} {\bibinfo {author} {\bibfnamefont {P.~G.}\ \bibnamefont {de~Gennes}},\ }\href {\doibase 10.1103/RevModPhys.71.S374} {\bibfield  {journal} {\bibinfo  {journal} {Rev. Mod. Phys.}\ }\textbf {\bibinfo {volume} {71}},\ \bibinfo {pages} {S374} (\bibinfo {year} {1999})}\BibitemShut {NoStop}%
\bibitem [{\citenamefont {Liu}\ \emph {et~al.}(1996)\citenamefont {Liu}, \citenamefont {Ramaswamy}, \citenamefont {Mason}, \citenamefont {Gang},\ and\ \citenamefont {Weitz}}]{Liu_emulsions}%
  \BibitemOpen
  \bibfield  {author} {\bibinfo {author} {\bibfnamefont {A.~J.}\ \bibnamefont {Liu}}, \bibinfo {author} {\bibfnamefont {S.}~\bibnamefont {Ramaswamy}}, \bibinfo {author} {\bibfnamefont {T.~G.}\ \bibnamefont {Mason}}, \bibinfo {author} {\bibfnamefont {H.}~\bibnamefont {Gang}}, \ and\ \bibinfo {author} {\bibfnamefont {D.~A.}\ \bibnamefont {Weitz}},\ }\href {\doibase 10.1103/PhysRevLett.76.3017} {\bibfield  {journal} {\bibinfo  {journal} {Phys. Rev. Lett.}\ }\textbf {\bibinfo {volume} {76}},\ \bibinfo {pages} {3017} (\bibinfo {year} {1996})}\BibitemShut {NoStop}%
\bibitem [{\citenamefont {Durian}(1995)}]{Durian}%
  \BibitemOpen
  \bibfield  {author} {\bibinfo {author} {\bibfnamefont {D.~J.}\ \bibnamefont {Durian}},\ }\href {\doibase 10.1103/PhysRevLett.75.4780} {\bibfield  {journal} {\bibinfo  {journal} {Phys. Rev. Lett.}\ }\textbf {\bibinfo {volume} {75}},\ \bibinfo {pages} {4780} (\bibinfo {year} {1995})}\BibitemShut {NoStop}%
\bibitem [{\citenamefont {Jaeger}\ \emph {et~al.}(1996)\citenamefont {Jaeger}, \citenamefont {Nagel},\ and\ \citenamefont {Behringer}}]{Behringer}%
  \BibitemOpen
  \bibfield  {author} {\bibinfo {author} {\bibfnamefont {H.~M.}\ \bibnamefont {Jaeger}}, \bibinfo {author} {\bibfnamefont {S.~R.}\ \bibnamefont {Nagel}}, \ and\ \bibinfo {author} {\bibfnamefont {R.~P.}\ \bibnamefont {Behringer}},\ }\href {\doibase 10.1103/RevModPhys.68.1259} {\bibfield  {journal} {\bibinfo  {journal} {Rev. Mod. Phys.}\ }\textbf {\bibinfo {volume} {68}},\ \bibinfo {pages} {1259} (\bibinfo {year} {1996})}\BibitemShut {NoStop}%
\bibitem [{\citenamefont {Weaire}\ and\ \citenamefont {Hutzler}(2000)}]{Weaire}%
  \BibitemOpen
  \bibfield  {author} {\bibinfo {author} {\bibfnamefont {D.}~\bibnamefont {Weaire}}\ and\ \bibinfo {author} {\bibfnamefont {S.}~\bibnamefont {Hutzler}},\ }\href {\doibase 10.1093/oso/9780198505518.001.0001} {\emph {\bibinfo {title} {{The Physics of Foams}}}}\ (\bibinfo  {publisher} {Oxford University Press},\ \bibinfo {year} {2000})\BibitemShut {NoStop}%
\bibitem [{\citenamefont {Corwin}\ \emph {et~al.}(2005)\citenamefont {Corwin}, \citenamefont {Jaeger},\ and\ \citenamefont {Nagel}}]{Corwin2005}%
  \BibitemOpen
  \bibfield  {author} {\bibinfo {author} {\bibfnamefont {E.~I.}\ \bibnamefont {Corwin}}, \bibinfo {author} {\bibfnamefont {H.~M.}\ \bibnamefont {Jaeger}}, \ and\ \bibinfo {author} {\bibfnamefont {S.~R.}\ \bibnamefont {Nagel}},\ }\href {\doibase 10.1038/nature03698} {\bibfield  {journal} {\bibinfo  {journal} {Nature}\ }\textbf {\bibinfo {volume} {435}},\ \bibinfo {pages} {1075} (\bibinfo {year} {2005})}\BibitemShut {NoStop}%
\bibitem [{\citenamefont {O'Hern}\ \emph {et~al.}(2002)\citenamefont {O'Hern}, \citenamefont {Langer}, \citenamefont {Liu},\ and\ \citenamefont {Nagel}}]{Corey_PRL}%
  \BibitemOpen
  \bibfield  {author} {\bibinfo {author} {\bibfnamefont {C.~S.}\ \bibnamefont {O'Hern}}, \bibinfo {author} {\bibfnamefont {S.~A.}\ \bibnamefont {Langer}}, \bibinfo {author} {\bibfnamefont {A.~J.}\ \bibnamefont {Liu}}, \ and\ \bibinfo {author} {\bibfnamefont {S.~R.}\ \bibnamefont {Nagel}},\ }\href {\doibase 10.1103/PhysRevLett.88.075507} {\bibfield  {journal} {\bibinfo  {journal} {Phys. Rev. Lett.}\ }\textbf {\bibinfo {volume} {88}},\ \bibinfo {pages} {075507} (\bibinfo {year} {2002})}\BibitemShut {NoStop}%
\bibitem [{\citenamefont {Liu}\ and\ \citenamefont {Nagel}(1998)}]{Liu1998}%
  \BibitemOpen
  \bibfield  {author} {\bibinfo {author} {\bibfnamefont {A.~J.}\ \bibnamefont {Liu}}\ and\ \bibinfo {author} {\bibfnamefont {S.~R.}\ \bibnamefont {Nagel}},\ }\href {\doibase 10.1038/23819} {\bibfield  {journal} {\bibinfo  {journal} {Nature}\ }\textbf {\bibinfo {volume} {396}},\ \bibinfo {pages} {21} (\bibinfo {year} {1998})}\BibitemShut {NoStop}%
\bibitem [{\citenamefont {van Hecke}(2009)}]{vanHecke}%
  \BibitemOpen
  \bibfield  {author} {\bibinfo {author} {\bibfnamefont {M.}~\bibnamefont {van Hecke}},\ }\href {\doibase 10.1088/0953-8984/22/3/033101} {\bibfield  {journal} {\bibinfo  {journal} {Journal of Physics: Condensed Matter}\ }\textbf {\bibinfo {volume} {22}},\ \bibinfo {pages} {033101} (\bibinfo {year} {2009})}\BibitemShut {NoStop}%
\bibitem [{\citenamefont {Reichhardt}\ and\ \citenamefont {Reichhardt}(2014)}]{Reichardt_rev}%
  \BibitemOpen
  \bibfield  {author} {\bibinfo {author} {\bibfnamefont {C.}~\bibnamefont {Reichhardt}}\ and\ \bibinfo {author} {\bibfnamefont {C.~J.~O.}\ \bibnamefont {Reichhardt}},\ }\href {\doibase 10.1039/C3SM53154F} {\bibfield  {journal} {\bibinfo  {journal} {Soft Matter}\ }\textbf {\bibinfo {volume} {10}},\ \bibinfo {pages} {2932} (\bibinfo {year} {2014})}\BibitemShut {NoStop}%
\bibitem [{\citenamefont {O'Hern}\ \emph {et~al.}(2003)\citenamefont {O'Hern}, \citenamefont {Silbert}, \citenamefont {Liu},\ and\ \citenamefont {Nagel}}]{OHern}%
  \BibitemOpen
  \bibfield  {author} {\bibinfo {author} {\bibfnamefont {C.~S.}\ \bibnamefont {O'Hern}}, \bibinfo {author} {\bibfnamefont {L.~E.}\ \bibnamefont {Silbert}}, \bibinfo {author} {\bibfnamefont {A.~J.}\ \bibnamefont {Liu}}, \ and\ \bibinfo {author} {\bibfnamefont {S.~R.}\ \bibnamefont {Nagel}},\ }\href@noop {} {\bibfield  {journal} {\bibinfo  {journal} {Phys. Rev. E}\ }\textbf {\bibinfo {volume} {68}},\ \bibinfo {pages} {011306} (\bibinfo {year} {2003})}\BibitemShut {NoStop}%
\bibitem [{\citenamefont {Mizuno}\ \emph {et~al.}(2016)\citenamefont {Mizuno}, \citenamefont {Silbert},\ and\ \citenamefont {Sperl}}]{Hideyuki}%
  \BibitemOpen
  \bibfield  {author} {\bibinfo {author} {\bibfnamefont {H.}~\bibnamefont {Mizuno}}, \bibinfo {author} {\bibfnamefont {L.~E.}\ \bibnamefont {Silbert}}, \ and\ \bibinfo {author} {\bibfnamefont {M.}~\bibnamefont {Sperl}},\ }\href {\doibase 10.1103/PhysRevLett.116.068302} {\bibfield  {journal} {\bibinfo  {journal} {Phys. Rev. Lett.}\ }\textbf {\bibinfo {volume} {116}},\ \bibinfo {pages} {068302} (\bibinfo {year} {2016})}\BibitemShut {NoStop}%
\bibitem [{\citenamefont {Clusel}\ \emph {et~al.}(2009)\citenamefont {Clusel}, \citenamefont {Corwin}, \citenamefont {Siemens},\ and\ \citenamefont {Bruji{\'{c}}}}]{Jasna2009}%
  \BibitemOpen
  \bibfield  {author} {\bibinfo {author} {\bibfnamefont {M.}~\bibnamefont {Clusel}}, \bibinfo {author} {\bibfnamefont {E.~I.}\ \bibnamefont {Corwin}}, \bibinfo {author} {\bibfnamefont {A.~O.~N.}\ \bibnamefont {Siemens}}, \ and\ \bibinfo {author} {\bibfnamefont {J.}~\bibnamefont {Bruji{\'{c}}}},\ }\href {\doibase 10.1038/nature08158} {\bibfield  {journal} {\bibinfo  {journal} {Nature}\ }\textbf {\bibinfo {volume} {460}},\ \bibinfo {pages} {611} (\bibinfo {year} {2009})}\BibitemShut {NoStop}%
\bibitem [{\citenamefont {Torquato}\ \emph {et~al.}(2000)\citenamefont {Torquato}, \citenamefont {Truskett},\ and\ \citenamefont {Debenedetti}}]{Truskett}%
  \BibitemOpen
  \bibfield  {author} {\bibinfo {author} {\bibfnamefont {S.}~\bibnamefont {Torquato}}, \bibinfo {author} {\bibfnamefont {T.~M.}\ \bibnamefont {Truskett}}, \ and\ \bibinfo {author} {\bibfnamefont {P.~G.}\ \bibnamefont {Debenedetti}},\ }\href {\doibase 10.1103/PhysRevLett.84.2064} {\bibfield  {journal} {\bibinfo  {journal} {Phys. Rev. Lett.}\ }\textbf {\bibinfo {volume} {84}},\ \bibinfo {pages} {2064} (\bibinfo {year} {2000})}\BibitemShut {NoStop}%
\bibitem [{\citenamefont {Torquato}\ and\ \citenamefont {Stillinger}(2010)}]{Torquato_review}%
  \BibitemOpen
  \bibfield  {author} {\bibinfo {author} {\bibfnamefont {S.}~\bibnamefont {Torquato}}\ and\ \bibinfo {author} {\bibfnamefont {F.~H.}\ \bibnamefont {Stillinger}},\ }\href {https://link.aps.org/doi/10.1103/RevModPhys.82.2633} {\bibfield  {journal} {\bibinfo  {journal} {Rev. Mod. Phys.}\ }\textbf {\bibinfo {volume} {82}},\ \bibinfo {pages} {2633} (\bibinfo {year} {2010})}\BibitemShut {NoStop}%
\bibitem [{\citenamefont {Torquato}(2018)}]{Torquato_2018}%
  \BibitemOpen
  \bibfield  {author} {\bibinfo {author} {\bibfnamefont {S.}~\bibnamefont {Torquato}},\ }\href@noop {} {\bibfield  {journal} {\bibinfo  {journal} {The Journal of Chemical Physics}\ }\textbf {\bibinfo {volume} {149}},\ \bibinfo {pages} {020901} (\bibinfo {year} {2018})}\BibitemShut {NoStop}%
\bibitem [{\citenamefont {Maher}\ \emph {et~al.}(2023)\citenamefont {Maher}, \citenamefont {Jiao},\ and\ \citenamefont {Torquato}}]{Torq_hyper}%
  \BibitemOpen
  \bibfield  {author} {\bibinfo {author} {\bibfnamefont {C.~E.}\ \bibnamefont {Maher}}, \bibinfo {author} {\bibfnamefont {Y.}~\bibnamefont {Jiao}}, \ and\ \bibinfo {author} {\bibfnamefont {S.}~\bibnamefont {Torquato}},\ }\href {\doibase 10.1103/PhysRevE.108.064602} {\bibfield  {journal} {\bibinfo  {journal} {Phys. Rev. E}\ }\textbf {\bibinfo {volume} {108}},\ \bibinfo {pages} {064602} (\bibinfo {year} {2023})}\BibitemShut {NoStop}%
\bibitem [{\citenamefont {Zaccone}\ and\ \citenamefont {Scossa-Romano}(2011)}]{Scossa}%
  \BibitemOpen
  \bibfield  {author} {\bibinfo {author} {\bibfnamefont {A.}~\bibnamefont {Zaccone}}\ and\ \bibinfo {author} {\bibfnamefont {E.}~\bibnamefont {Scossa-Romano}},\ }\href {\doibase 10.1103/PhysRevB.83.184205} {\bibfield  {journal} {\bibinfo  {journal} {Phys. Rev. B}\ }\textbf {\bibinfo {volume} {83}},\ \bibinfo {pages} {184205} (\bibinfo {year} {2011})}\BibitemShut {NoStop}%
\bibitem [{\citenamefont {Jin}\ and\ \citenamefont {Makse}(2010)}]{Yuliang}%
  \BibitemOpen
  \bibfield  {author} {\bibinfo {author} {\bibfnamefont {Y.}~\bibnamefont {Jin}}\ and\ \bibinfo {author} {\bibfnamefont {H.~A.}\ \bibnamefont {Makse}},\ }\href {\doibase https://doi.org/10.1016/j.physa.2010.08.010} {\bibfield  {journal} {\bibinfo  {journal} {Physica A: Statistical Mechanics and its Applications}\ }\textbf {\bibinfo {volume} {389}},\ \bibinfo {pages} {5362} (\bibinfo {year} {2010})}\BibitemShut {NoStop}%
\bibitem [{\citenamefont {Rouwhorst}\ \emph {et~al.}(2020)\citenamefont {Rouwhorst}, \citenamefont {Ness}, \citenamefont {Stoyanov}, \citenamefont {Zaccone},\ and\ \citenamefont {Schall}}]{Joep}%
  \BibitemOpen
  \bibfield  {author} {\bibinfo {author} {\bibfnamefont {J.}~\bibnamefont {Rouwhorst}}, \bibinfo {author} {\bibfnamefont {C.}~\bibnamefont {Ness}}, \bibinfo {author} {\bibfnamefont {S.}~\bibnamefont {Stoyanov}}, \bibinfo {author} {\bibfnamefont {A.}~\bibnamefont {Zaccone}}, \ and\ \bibinfo {author} {\bibfnamefont {P.}~\bibnamefont {Schall}},\ }\href@noop {} {\bibfield  {journal} {\bibinfo  {journal} {Nature Communications}\ }\textbf {\bibinfo {volume} {11}},\ \bibinfo {pages} {3558} (\bibinfo {year} {2020})}\BibitemShut {NoStop}%
\bibitem [{\citenamefont {Wilken}\ \emph {et~al.}(2021)\citenamefont {Wilken}, \citenamefont {Guerra}, \citenamefont {Levine},\ and\ \citenamefont {Chaikin}}]{Chaikin_Manna}%
  \BibitemOpen
  \bibfield  {author} {\bibinfo {author} {\bibfnamefont {S.}~\bibnamefont {Wilken}}, \bibinfo {author} {\bibfnamefont {R.~E.}\ \bibnamefont {Guerra}}, \bibinfo {author} {\bibfnamefont {D.}~\bibnamefont {Levine}}, \ and\ \bibinfo {author} {\bibfnamefont {P.~M.}\ \bibnamefont {Chaikin}},\ }\href {\doibase 10.1103/PhysRevLett.127.038002} {\bibfield  {journal} {\bibinfo  {journal} {Phys. Rev. Lett.}\ }\textbf {\bibinfo {volume} {127}},\ \bibinfo {pages} {038002} (\bibinfo {year} {2021})}\BibitemShut {NoStop}%
\bibitem [{\citenamefont {Lois}\ \emph {et~al.}(2008)\citenamefont {Lois}, \citenamefont {Blawzdziewicz},\ and\ \citenamefont {O'Hern}}]{Lois}%
  \BibitemOpen
  \bibfield  {author} {\bibinfo {author} {\bibfnamefont {G.}~\bibnamefont {Lois}}, \bibinfo {author} {\bibfnamefont {J.}~\bibnamefont {Blawzdziewicz}}, \ and\ \bibinfo {author} {\bibfnamefont {C.~S.}\ \bibnamefont {O'Hern}},\ }\href {\doibase 10.1103/PhysRevLett.100.028001} {\bibfield  {journal} {\bibinfo  {journal} {Phys. Rev. Lett.}\ }\textbf {\bibinfo {volume} {100}},\ \bibinfo {pages} {028001} (\bibinfo {year} {2008})}\BibitemShut {NoStop}%
\bibitem [{\citenamefont {Baule}\ \emph {et~al.}(2018)\citenamefont {Baule}, \citenamefont {Morone}, \citenamefont {Herrmann},\ and\ \citenamefont {Makse}}]{MakseRMP}%
  \BibitemOpen
  \bibfield  {author} {\bibinfo {author} {\bibfnamefont {A.}~\bibnamefont {Baule}}, \bibinfo {author} {\bibfnamefont {F.}~\bibnamefont {Morone}}, \bibinfo {author} {\bibfnamefont {H.~J.}\ \bibnamefont {Herrmann}}, \ and\ \bibinfo {author} {\bibfnamefont {H.~A.}\ \bibnamefont {Makse}},\ }\href {\doibase 10.1103/RevModPhys.90.015006} {\bibfield  {journal} {\bibinfo  {journal} {Rev. Mod. Phys.}\ }\textbf {\bibinfo {volume} {90}},\ \bibinfo {pages} {015006} (\bibinfo {year} {2018})}\BibitemShut {NoStop}%
\bibitem [{\citenamefont {Martiniani}\ \emph {et~al.}(2016)\citenamefont {Martiniani}, \citenamefont {Schrenk}, \citenamefont {Stevenson}, \citenamefont {Wales},\ and\ \citenamefont {Frenkel}}]{Martiniani}%
  \BibitemOpen
  \bibfield  {author} {\bibinfo {author} {\bibfnamefont {S.}~\bibnamefont {Martiniani}}, \bibinfo {author} {\bibfnamefont {K.~J.}\ \bibnamefont {Schrenk}}, \bibinfo {author} {\bibfnamefont {J.~D.}\ \bibnamefont {Stevenson}}, \bibinfo {author} {\bibfnamefont {D.~J.}\ \bibnamefont {Wales}}, \ and\ \bibinfo {author} {\bibfnamefont {D.}~\bibnamefont {Frenkel}},\ }\href {\doibase 10.1103/PhysRevE.93.012906} {\bibfield  {journal} {\bibinfo  {journal} {Phys. Rev. E}\ }\textbf {\bibinfo {volume} {93}},\ \bibinfo {pages} {012906} (\bibinfo {year} {2016})}\BibitemShut {NoStop}%
\bibitem [{\citenamefont {Trachenko}(2011)}]{Kostya_2011}%
  \BibitemOpen
  \bibfield  {author} {\bibinfo {author} {\bibfnamefont {K.}~\bibnamefont {Trachenko}},\ }\href {\doibase 10.1088/0953-8984/23/36/366003} {\bibfield  {journal} {\bibinfo  {journal} {Journal of Physics: Condensed Matter}\ }\textbf {\bibinfo {volume} {23}},\ \bibinfo {pages} {366003} (\bibinfo {year} {2011})}\BibitemShut {NoStop}%
\bibitem [{\citenamefont {Trachenko}(2023)}]{Kostya_book}%
  \BibitemOpen
  \bibfield  {author} {\bibinfo {author} {\bibfnamefont {K.}~\bibnamefont {Trachenko}},\ }\href@noop {} {\emph {\bibinfo {title} {Theory of Liquids}}}\ (\bibinfo  {publisher} {Cambridge University Press},\ \bibinfo {address} {Cambridge},\ \bibinfo {year} {2023})\BibitemShut {NoStop}%
\bibitem [{\citenamefont {Parisi}\ and\ \citenamefont {Zamponi}(2010)}]{Parisi_RMP}%
  \BibitemOpen
  \bibfield  {author} {\bibinfo {author} {\bibfnamefont {G.}~\bibnamefont {Parisi}}\ and\ \bibinfo {author} {\bibfnamefont {F.}~\bibnamefont {Zamponi}},\ }\href {\doibase 10.1103/RevModPhys.82.789} {\bibfield  {journal} {\bibinfo  {journal} {Rev. Mod. Phys.}\ }\textbf {\bibinfo {volume} {82}},\ \bibinfo {pages} {789} (\bibinfo {year} {2010})}\BibitemShut {NoStop}%
\bibitem [{\citenamefont {Stillinger}\ and\ \citenamefont {Torquato}(2006)}]{Torq_Still}%
  \BibitemOpen
  \bibfield  {author} {\bibinfo {author} {\bibfnamefont {F.~H.}\ \bibnamefont {Stillinger}}\ and\ \bibinfo {author} {\bibfnamefont {S.}~\bibnamefont {Torquato}},\ }\href@noop {} {\bibfield  {journal} {\bibinfo  {journal} {Experimental Mathematics}\ }\textbf {\bibinfo {volume} {15}},\ \bibinfo {pages} {307 } (\bibinfo {year} {2006})}\BibitemShut {NoStop}%
\bibitem [{\citenamefont {Hales}(2005)}]{Hales1}%
  \BibitemOpen
  \bibfield  {author} {\bibinfo {author} {\bibfnamefont {T.~C.}\ \bibnamefont {Hales}},\ }\href@noop {} {\bibfield  {journal} {\bibinfo  {journal} {Annals of Mathematics}\ }\textbf {\bibinfo {volume} {162}},\ \bibinfo {pages} {1065} (\bibinfo {year} {2005})}\BibitemShut {NoStop}%
\bibitem [{\citenamefont {Paul M.~Goldbart}\ and\ \citenamefont {Zippelius}(1996)}]{Goldbart}%
  \BibitemOpen
  \bibfield  {author} {\bibinfo {author} {\bibfnamefont {H.~E.~C.}\ \bibnamefont {Paul M.~Goldbart}}\ and\ \bibinfo {author} {\bibfnamefont {A.}~\bibnamefont {Zippelius}},\ }\href {\doibase 10.1080/00018739600101527} {\bibfield  {journal} {\bibinfo  {journal} {Advances in Physics}\ }\textbf {\bibinfo {volume} {45}},\ \bibinfo {pages} {393} (\bibinfo {year} {1996})},\ \Eprint {http://arxiv.org/abs/https://doi.org/10.1080/00018739600101527} {https://doi.org/10.1080/00018739600101527} \BibitemShut {NoStop}%
\bibitem [{\citenamefont {Yoshino}\ and\ \citenamefont {M\'ezard}(2010)}]{mezard}%
  \BibitemOpen
  \bibfield  {author} {\bibinfo {author} {\bibfnamefont {H.}~\bibnamefont {Yoshino}}\ and\ \bibinfo {author} {\bibfnamefont {M.}~\bibnamefont {M\'ezard}},\ }\href@noop {} {\bibfield  {journal} {\bibinfo  {journal} {Phys. Rev. Lett.}\ }\textbf {\bibinfo {volume} {105}},\ \bibinfo {pages} {015504} (\bibinfo {year} {2010})}\BibitemShut {NoStop}%
\bibitem [{\citenamefont {Yoshino}(2012)}]{yoshino}%
  \BibitemOpen
  \bibfield  {author} {\bibinfo {author} {\bibfnamefont {H.}~\bibnamefont {Yoshino}},\ }\href@noop {} {\bibfield  {journal} {\bibinfo  {journal} {The Journal of Chemical Physics}\ }\textbf {\bibinfo {volume} {136}},\ \bibinfo {pages} {214108} (\bibinfo {year} {2012})}\BibitemShut {NoStop}%
\bibitem [{\citenamefont {Lutsko}(1989)}]{Lutsko}%
  \BibitemOpen
  \bibfield  {author} {\bibinfo {author} {\bibfnamefont {J.~F.}\ \bibnamefont {Lutsko}},\ }\href {\doibase 10.1063/1.342716} {\bibfield  {journal} {\bibinfo  {journal} {Journal of Applied Physics}\ }\textbf {\bibinfo {volume} {65}},\ \bibinfo {pages} {2991} (\bibinfo {year} {1989})},\ \Eprint {http://arxiv.org/abs/https://doi.org/10.1063/1.342716} {https://doi.org/10.1063/1.342716} \BibitemShut {NoStop}%
\bibitem [{\citenamefont {Charbonneau}\ \emph {et~al.}(2017)\citenamefont {Charbonneau}, \citenamefont {Kurchan}, \citenamefont {Parisi}, \citenamefont {Urbani},\ and\ \citenamefont {Zamponi}}]{Charbonneau}%
  \BibitemOpen
  \bibfield  {author} {\bibinfo {author} {\bibfnamefont {P.}~\bibnamefont {Charbonneau}}, \bibinfo {author} {\bibfnamefont {J.}~\bibnamefont {Kurchan}}, \bibinfo {author} {\bibfnamefont {G.}~\bibnamefont {Parisi}}, \bibinfo {author} {\bibfnamefont {P.}~\bibnamefont {Urbani}}, \ and\ \bibinfo {author} {\bibfnamefont {F.}~\bibnamefont {Zamponi}},\ }\href@noop {} {\bibfield  {journal} {\bibinfo  {journal} {Annual Review of Condensed Matter Physics}\ }\textbf {\bibinfo {volume} {8}},\ \bibinfo {pages} {265} (\bibinfo {year} {2017})}\BibitemShut {NoStop}%
\bibitem [{\citenamefont {Berthier}\ \emph {et~al.}(2011)\citenamefont {Berthier}, \citenamefont {Jacquin},\ and\ \citenamefont {Zamponi}}]{Berthier_Zamp}%
  \BibitemOpen
  \bibfield  {author} {\bibinfo {author} {\bibfnamefont {L.}~\bibnamefont {Berthier}}, \bibinfo {author} {\bibfnamefont {H.}~\bibnamefont {Jacquin}}, \ and\ \bibinfo {author} {\bibfnamefont {F.}~\bibnamefont {Zamponi}},\ }\href {\doibase 10.1103/PhysRevE.84.051103} {\bibfield  {journal} {\bibinfo  {journal} {Phys. Rev. E}\ }\textbf {\bibinfo {volume} {84}},\ \bibinfo {pages} {051103} (\bibinfo {year} {2011})}\BibitemShut {NoStop}%
\bibitem [{\citenamefont {Parisi}(2014)}]{parisi2014}%
  \BibitemOpen
  \bibfield  {author} {\bibinfo {author} {\bibfnamefont {G.}~\bibnamefont {Parisi}},\ }\href {https://arxiv.org/abs/1401.4413} {\enquote {\bibinfo {title} {Soft modes in jammed hard spheres (i): Mean field theory of the isostatic transition},}\ } (\bibinfo {year} {2014}),\ \Eprint {http://arxiv.org/abs/1401.4413} {arXiv:1401.4413 [cond-mat.soft]} \BibitemShut {NoStop}%
\bibitem [{\citenamefont {Cicuta}\ \emph {et~al.}(2018)\citenamefont {Cicuta}, \citenamefont {Krausser}, \citenamefont {Milkus},\ and\ \citenamefont {Zaccone}}]{Cicuta}%
  \BibitemOpen
  \bibfield  {author} {\bibinfo {author} {\bibfnamefont {G.~M.}\ \bibnamefont {Cicuta}}, \bibinfo {author} {\bibfnamefont {J.}~\bibnamefont {Krausser}}, \bibinfo {author} {\bibfnamefont {R.}~\bibnamefont {Milkus}}, \ and\ \bibinfo {author} {\bibfnamefont {A.}~\bibnamefont {Zaccone}},\ }\href {\doibase 10.1103/PhysRevE.97.032113} {\bibfield  {journal} {\bibinfo  {journal} {Phys. Rev. E}\ }\textbf {\bibinfo {volume} {97}},\ \bibinfo {pages} {032113} (\bibinfo {year} {2018})}\BibitemShut {NoStop}%
\bibitem [{\citenamefont {Benetti}\ \emph {et~al.}(2018)\citenamefont {Benetti}, \citenamefont {Parisi}, \citenamefont {Pietracaprina},\ and\ \citenamefont {Sicuro}}]{Benetti}%
  \BibitemOpen
  \bibfield  {author} {\bibinfo {author} {\bibfnamefont {F.~P.~C.}\ \bibnamefont {Benetti}}, \bibinfo {author} {\bibfnamefont {G.}~\bibnamefont {Parisi}}, \bibinfo {author} {\bibfnamefont {F.}~\bibnamefont {Pietracaprina}}, \ and\ \bibinfo {author} {\bibfnamefont {G.}~\bibnamefont {Sicuro}},\ }\href {\doibase 10.1103/PhysRevE.97.062157} {\bibfield  {journal} {\bibinfo  {journal} {Phys. Rev. E}\ }\textbf {\bibinfo {volume} {97}},\ \bibinfo {pages} {062157} (\bibinfo {year} {2018})}\BibitemShut {NoStop}%
\bibitem [{\citenamefont {Pernici}\ and\ \citenamefont {Cicuta}(2019)}]{Pernici2019}%
  \BibitemOpen
  \bibfield  {author} {\bibinfo {author} {\bibfnamefont {M.}~\bibnamefont {Pernici}}\ and\ \bibinfo {author} {\bibfnamefont {G.~M.}\ \bibnamefont {Cicuta}},\ }\href {\doibase 10.1007/s10955-019-02260-0} {\bibfield  {journal} {\bibinfo  {journal} {Journal of Statistical Physics}\ }\textbf {\bibinfo {volume} {175}},\ \bibinfo {pages} {384} (\bibinfo {year} {2019})}\BibitemShut {NoStop}%
\bibitem [{\citenamefont {Cicuta}\ and\ \citenamefont {Pernici}(2023)}]{Cicuta_2023}%
  \BibitemOpen
  \bibfield  {author} {\bibinfo {author} {\bibfnamefont {G.~M.}\ \bibnamefont {Cicuta}}\ and\ \bibinfo {author} {\bibfnamefont {M.}~\bibnamefont {Pernici}},\ }\href {\doibase 10.1088/2632-072X/acc71a} {\bibfield  {journal} {\bibinfo  {journal} {Journal of Physics: Complexity}\ }\textbf {\bibinfo {volume} {4}},\ \bibinfo {pages} {025004} (\bibinfo {year} {2023})}\BibitemShut {NoStop}%
\bibitem [{\citenamefont {Susca}\ \emph {et~al.}(2021)\citenamefont {Susca}, \citenamefont {Vivo},\ and\ \citenamefont {Kühn}}]{Vivo}%
  \BibitemOpen
  \bibfield  {author} {\bibinfo {author} {\bibfnamefont {V.~A.~R.}\ \bibnamefont {Susca}}, \bibinfo {author} {\bibfnamefont {P.}~\bibnamefont {Vivo}}, \ and\ \bibinfo {author} {\bibfnamefont {R.}~\bibnamefont {Kühn}},\ }\href {\doibase 10.21468/SciPostPhysLectNotes.33} {\bibfield  {journal} {\bibinfo  {journal} {SciPost Phys. Lect. Notes}\ ,\ \bibinfo {pages} {33}} (\bibinfo {year} {2021})}\BibitemShut {NoStop}%
\bibitem [{\citenamefont {Goodrich}\ \emph {et~al.}(2016)\citenamefont {Goodrich}, \citenamefont {Liu},\ and\ \citenamefont {Sethna}}]{Sethna}%
  \BibitemOpen
  \bibfield  {author} {\bibinfo {author} {\bibfnamefont {C.~P.}\ \bibnamefont {Goodrich}}, \bibinfo {author} {\bibfnamefont {A.~J.}\ \bibnamefont {Liu}}, \ and\ \bibinfo {author} {\bibfnamefont {J.~P.}\ \bibnamefont {Sethna}},\ }\href {\doibase 10.1073/pnas.1601858113} {\bibfield  {journal} {\bibinfo  {journal} {Proceedings of the National Academy of Sciences}\ }\textbf {\bibinfo {volume} {113}},\ \bibinfo {pages} {9745} (\bibinfo {year} {2016})},\ \Eprint {http://arxiv.org/abs/https://www.pnas.org/doi/pdf/10.1073/pnas.1601858113} {https://www.pnas.org/doi/pdf/10.1073/pnas.1601858113} \BibitemShut {NoStop}%
\bibitem [{\citenamefont {Cates}\ \emph {et~al.}(1998)\citenamefont {Cates}, \citenamefont {Wittmer}, \citenamefont {Bouchaud},\ and\ \citenamefont {Claudin}}]{Cates}%
  \BibitemOpen
  \bibfield  {author} {\bibinfo {author} {\bibfnamefont {M.~E.}\ \bibnamefont {Cates}}, \bibinfo {author} {\bibfnamefont {J.~P.}\ \bibnamefont {Wittmer}}, \bibinfo {author} {\bibfnamefont {J.-P.}\ \bibnamefont {Bouchaud}}, \ and\ \bibinfo {author} {\bibfnamefont {P.}~\bibnamefont {Claudin}},\ }\href {\doibase 10.1103/PhysRevLett.81.1841} {\bibfield  {journal} {\bibinfo  {journal} {Phys. Rev. Lett.}\ }\textbf {\bibinfo {volume} {81}},\ \bibinfo {pages} {1841} (\bibinfo {year} {1998})}\BibitemShut {NoStop}%
\bibitem [{\citenamefont {Zaccone}\ and\ \citenamefont {Terentjev}(2014)}]{JAP}%
  \BibitemOpen
  \bibfield  {author} {\bibinfo {author} {\bibfnamefont {A.}~\bibnamefont {Zaccone}}\ and\ \bibinfo {author} {\bibfnamefont {E.~M.}\ \bibnamefont {Terentjev}},\ }\href@noop {} {\bibfield  {journal} {\bibinfo  {journal} {Journal of Applied Physics}\ }\textbf {\bibinfo {volume} {115}},\ \bibinfo {pages} {033510} (\bibinfo {year} {2014})}\BibitemShut {NoStop}%
\bibitem [{\citenamefont {Schlegel}\ \emph {et~al.}(2016)\citenamefont {Schlegel}, \citenamefont {Brujic}, \citenamefont {Terentjev},\ and\ \citenamefont {Zaccone}}]{Schlegel2016}%
  \BibitemOpen
  \bibfield  {author} {\bibinfo {author} {\bibfnamefont {M.}~\bibnamefont {Schlegel}}, \bibinfo {author} {\bibfnamefont {J.}~\bibnamefont {Brujic}}, \bibinfo {author} {\bibfnamefont {E.~M.}\ \bibnamefont {Terentjev}}, \ and\ \bibinfo {author} {\bibfnamefont {A.}~\bibnamefont {Zaccone}},\ }\href {\doibase 10.1038/srep18724} {\bibfield  {journal} {\bibinfo  {journal} {Scientific Reports}\ }\textbf {\bibinfo {volume} {6}},\ \bibinfo {pages} {18724} (\bibinfo {year} {2016})}\BibitemShut {NoStop}%
\bibitem [{\citenamefont {Ellenbroek}\ \emph {et~al.}(2009)\citenamefont {Ellenbroek}, \citenamefont {Zeravcic}, \citenamefont {van Saarloos},\ and\ \citenamefont {van Hecke}}]{Ellenbroek}%
  \BibitemOpen
  \bibfield  {author} {\bibinfo {author} {\bibfnamefont {W.~G.}\ \bibnamefont {Ellenbroek}}, \bibinfo {author} {\bibfnamefont {Z.}~\bibnamefont {Zeravcic}}, \bibinfo {author} {\bibfnamefont {W.}~\bibnamefont {van Saarloos}}, \ and\ \bibinfo {author} {\bibfnamefont {M.}~\bibnamefont {van Hecke}},\ }\href@noop {} {\bibfield  {journal} {\bibinfo  {journal} {{EPL} (Europhysics Letters)}\ }\textbf {\bibinfo {volume} {87}},\ \bibinfo {pages} {34004} (\bibinfo {year} {2009})}\BibitemShut {NoStop}%
\bibitem [{\citenamefont {Krausser}\ \emph {et~al.}(2015)\citenamefont {Krausser}, \citenamefont {Samwer},\ and\ \citenamefont {Zaccone}}]{KSZ}%
  \BibitemOpen
  \bibfield  {author} {\bibinfo {author} {\bibfnamefont {J.}~\bibnamefont {Krausser}}, \bibinfo {author} {\bibfnamefont {K.~H.}\ \bibnamefont {Samwer}}, \ and\ \bibinfo {author} {\bibfnamefont {A.}~\bibnamefont {Zaccone}},\ }\href@noop {} {\bibfield  {journal} {\bibinfo  {journal} {Proceedings of the National Academy of Sciences}\ }\textbf {\bibinfo {volume} {112}},\ \bibinfo {pages} {13762} (\bibinfo {year} {2015})}\BibitemShut {NoStop}%
\bibitem [{\citenamefont {Hatano}(2008)}]{Hatano}%
  \BibitemOpen
  \bibfield  {author} {\bibinfo {author} {\bibfnamefont {T.}~\bibnamefont {Hatano}},\ }\href {\doibase 10.1143/JPSJ.77.123002} {\bibfield  {journal} {\bibinfo  {journal} {Journal of the Physical Society of Japan}\ }\textbf {\bibinfo {volume} {77}},\ \bibinfo {pages} {123002} (\bibinfo {year} {2008})},\ \Eprint {http://arxiv.org/abs/https://doi.org/10.1143/JPSJ.77.123002} {https://doi.org/10.1143/JPSJ.77.123002} \BibitemShut {NoStop}%
\bibitem [{\citenamefont {Wyart}\ \emph {et~al.}(2005)\citenamefont {Wyart}, \citenamefont {Silbert}, \citenamefont {Nagel},\ and\ \citenamefont {Witten}}]{PhysRevE.72.051306}%
  \BibitemOpen
  \bibfield  {author} {\bibinfo {author} {\bibfnamefont {M.}~\bibnamefont {Wyart}}, \bibinfo {author} {\bibfnamefont {L.~E.}\ \bibnamefont {Silbert}}, \bibinfo {author} {\bibfnamefont {S.~R.}\ \bibnamefont {Nagel}}, \ and\ \bibinfo {author} {\bibfnamefont {T.~A.}\ \bibnamefont {Witten}},\ }\href {\doibase 10.1103/PhysRevE.72.051306} {\bibfield  {journal} {\bibinfo  {journal} {Phys. Rev. E}\ }\textbf {\bibinfo {volume} {72}},\ \bibinfo {pages} {051306} (\bibinfo {year} {2005})}\BibitemShut {NoStop}%
\bibitem [{\citenamefont {Drocco}\ \emph {et~al.}(2005)\citenamefont {Drocco}, \citenamefont {Hastings}, \citenamefont {Reichhardt},\ and\ \citenamefont {Reichhardt}}]{Reichardt_exp}%
  \BibitemOpen
  \bibfield  {author} {\bibinfo {author} {\bibfnamefont {J.~A.}\ \bibnamefont {Drocco}}, \bibinfo {author} {\bibfnamefont {M.~B.}\ \bibnamefont {Hastings}}, \bibinfo {author} {\bibfnamefont {C.~J.~O.}\ \bibnamefont {Reichhardt}}, \ and\ \bibinfo {author} {\bibfnamefont {C.}~\bibnamefont {Reichhardt}},\ }\href {\doibase 10.1103/PhysRevLett.95.088001} {\bibfield  {journal} {\bibinfo  {journal} {Phys. Rev. Lett.}\ }\textbf {\bibinfo {volume} {95}},\ \bibinfo {pages} {088001} (\bibinfo {year} {2005})}\BibitemShut {NoStop}%
\bibitem [{\citenamefont {Otsuki}\ and\ \citenamefont {Hayakawa}(2011)}]{Haya}%
  \BibitemOpen
  \bibfield  {author} {\bibinfo {author} {\bibfnamefont {M.}~\bibnamefont {Otsuki}}\ and\ \bibinfo {author} {\bibfnamefont {H.}~\bibnamefont {Hayakawa}},\ }\href {\doibase 10.1103/PhysRevE.83.051301} {\bibfield  {journal} {\bibinfo  {journal} {Phys. Rev. E}\ }\textbf {\bibinfo {volume} {83}},\ \bibinfo {pages} {051301} (\bibinfo {year} {2011})}\BibitemShut {NoStop}%
\bibitem [{\citenamefont {Caldeira}\ and\ \citenamefont {Leggett}(1983)}]{Caldeira}%
  \BibitemOpen
  \bibfield  {author} {\bibinfo {author} {\bibfnamefont {A.}~\bibnamefont {Caldeira}}\ and\ \bibinfo {author} {\bibfnamefont {A.}~\bibnamefont {Leggett}},\ }\href@noop {} {\bibfield  {journal} {\bibinfo  {journal} {Annals of Physics}\ }\textbf {\bibinfo {volume} {149}},\ \bibinfo {pages} {374} (\bibinfo {year} {1983})}\BibitemShut {NoStop}%
\bibitem [{\citenamefont {Zwanzig}(1973)}]{Zwanzig1973}%
  \BibitemOpen
  \bibfield  {author} {\bibinfo {author} {\bibfnamefont {R.}~\bibnamefont {Zwanzig}},\ }\href {\doibase 10.1007/BF01008729} {\bibfield  {journal} {\bibinfo  {journal} {Journal of Statistical Physics}\ }\textbf {\bibinfo {volume} {9}},\ \bibinfo {pages} {215} (\bibinfo {year} {1973})}\BibitemShut {NoStop}%
\bibitem [{\citenamefont {Weiss}(2012)}]{Weiss}%
  \BibitemOpen
  \bibfield  {author} {\bibinfo {author} {\bibfnamefont {U.}~\bibnamefont {Weiss}},\ }\href {\doibase 10.1142/8334} {\emph {\bibinfo {title} {Quantum Dissipative Systems}}},\ \bibinfo {edition} {4th}\ ed.\ (\bibinfo  {publisher} {WORLD SCIENTIFIC},\ \bibinfo {year} {2012})\BibitemShut {NoStop}%
\bibitem [{\citenamefont {Petrosyan}\ and\ \citenamefont {Zaccone}(2022)}]{Petrosyan_2022}%
  \BibitemOpen
  \bibfield  {author} {\bibinfo {author} {\bibfnamefont {A.}~\bibnamefont {Petrosyan}}\ and\ \bibinfo {author} {\bibfnamefont {A.}~\bibnamefont {Zaccone}},\ }\href {\doibase 10.1088/1751-8121/ac3a33} {\bibfield  {journal} {\bibinfo  {journal} {Journal of Physics A: Mathematical and Theoretical}\ }\textbf {\bibinfo {volume} {55}},\ \bibinfo {pages} {015001} (\bibinfo {year} {2022})}\BibitemShut {NoStop}%
\bibitem [{\citenamefont {Zadra}\ \emph {et~al.}(2023)\citenamefont {Zadra}, \citenamefont {Petrosyan},\ and\ \citenamefont {Zaccone}}]{Zadra}%
  \BibitemOpen
  \bibfield  {author} {\bibinfo {author} {\bibfnamefont {F.}~\bibnamefont {Zadra}}, \bibinfo {author} {\bibfnamefont {A.}~\bibnamefont {Petrosyan}}, \ and\ \bibinfo {author} {\bibfnamefont {A.}~\bibnamefont {Zaccone}},\ }\href {\doibase 10.1103/PhysRevD.108.096012} {\bibfield  {journal} {\bibinfo  {journal} {Phys. Rev. D}\ }\textbf {\bibinfo {volume} {108}},\ \bibinfo {pages} {096012} (\bibinfo {year} {2023})}\BibitemShut {NoStop}%
\bibitem [{\citenamefont {Zaccone}(2024)}]{Zaccone_quark}%
  \BibitemOpen
  \bibfield  {author} {\bibinfo {author} {\bibfnamefont {A.}~\bibnamefont {Zaccone}},\ }\href {\doibase https://doi.org/10.1016/j.nuclphysb.2024.116483} {\bibfield  {journal} {\bibinfo  {journal} {Nuclear Physics B}\ }\textbf {\bibinfo {volume} {1000}},\ \bibinfo {pages} {116483} (\bibinfo {year} {2024})}\BibitemShut {NoStop}%
\bibitem [{\citenamefont {Zwanzig}\ and\ \citenamefont {Mountain}(1965)}]{Zwanzig}%
  \BibitemOpen
  \bibfield  {author} {\bibinfo {author} {\bibfnamefont {R.}~\bibnamefont {Zwanzig}}\ and\ \bibinfo {author} {\bibfnamefont {R.~D.}\ \bibnamefont {Mountain}},\ }\href@noop {} {\bibfield  {journal} {\bibinfo  {journal} {The Journal of Chemical Physics}\ }\textbf {\bibinfo {volume} {43}},\ \bibinfo {pages} {4464} (\bibinfo {year} {1965})}\BibitemShut {NoStop}%
\bibitem [{\citenamefont {Nitzan}(2006)}]{Nitzan}%
  \BibitemOpen
  \bibfield  {author} {\bibinfo {author} {\bibfnamefont {A.}~\bibnamefont {Nitzan}},\ }\href {\doibase 10.1093/oso/9780198529798.001.0001} {\emph {\bibinfo {title} {Chemical Dynamics in Condensed Phases: Relaxation, Transfer and Reactions in Condensed Molecular Systems}}}\ (\bibinfo  {publisher} {Oxford University Press},\ \bibinfo {year} {2006})\BibitemShut {NoStop}%
\bibitem [{\citenamefont {Cui}\ and\ \citenamefont {Zaccone}(2018)}]{Cui_GLE}%
  \BibitemOpen
  \bibfield  {author} {\bibinfo {author} {\bibfnamefont {B.}~\bibnamefont {Cui}}\ and\ \bibinfo {author} {\bibfnamefont {A.}~\bibnamefont {Zaccone}},\ }\href {\doibase 10.1103/PhysRevE.97.060102} {\bibfield  {journal} {\bibinfo  {journal} {Phys. Rev. E}\ }\textbf {\bibinfo {volume} {97}},\ \bibinfo {pages} {060102} (\bibinfo {year} {2018})}\BibitemShut {NoStop}%
\bibitem [{\citenamefont {Gamba}\ \emph {et~al.}(2024)\citenamefont {Gamba}, \citenamefont {Cui},\ and\ \citenamefont {Zaccone}}]{Gamba}%
  \BibitemOpen
  \bibfield  {author} {\bibinfo {author} {\bibfnamefont {D.}~\bibnamefont {Gamba}}, \bibinfo {author} {\bibfnamefont {B.}~\bibnamefont {Cui}}, \ and\ \bibinfo {author} {\bibfnamefont {A.}~\bibnamefont {Zaccone}},\ }\href {\doibase 10.1103/PhysRevE.110.054137} {\bibfield  {journal} {\bibinfo  {journal} {Phys. Rev. E}\ }\textbf {\bibinfo {volume} {110}},\ \bibinfo {pages} {054137} (\bibinfo {year} {2024})}\BibitemShut {NoStop}%
\bibitem [{\citenamefont {Cui}\ \emph {et~al.}(2017)\citenamefont {Cui}, \citenamefont {Yang}, \citenamefont {Qiao}, \citenamefont {Jiang}, \citenamefont {Dai}, \citenamefont {Wang},\ and\ \citenamefont {Zaccone}}]{Cui_viscoelastic}%
  \BibitemOpen
  \bibfield  {author} {\bibinfo {author} {\bibfnamefont {B.}~\bibnamefont {Cui}}, \bibinfo {author} {\bibfnamefont {J.}~\bibnamefont {Yang}}, \bibinfo {author} {\bibfnamefont {J.}~\bibnamefont {Qiao}}, \bibinfo {author} {\bibfnamefont {M.}~\bibnamefont {Jiang}}, \bibinfo {author} {\bibfnamefont {L.}~\bibnamefont {Dai}}, \bibinfo {author} {\bibfnamefont {Y.-J.}\ \bibnamefont {Wang}}, \ and\ \bibinfo {author} {\bibfnamefont {A.}~\bibnamefont {Zaccone}},\ }\href@noop {} {\bibfield  {journal} {\bibinfo  {journal} {Phys. Rev. B}\ }\textbf {\bibinfo {volume} {96}},\ \bibinfo {pages} {094203} (\bibinfo {year} {2017})}\BibitemShut {NoStop}%
\bibitem [{\citenamefont {Slonczewski}\ and\ \citenamefont {Thomas}(1970)}]{Slonczewski}%
  \BibitemOpen
  \bibfield  {author} {\bibinfo {author} {\bibfnamefont {J.~C.}\ \bibnamefont {Slonczewski}}\ and\ \bibinfo {author} {\bibfnamefont {H.}~\bibnamefont {Thomas}},\ }\href {\doibase 10.1103/PhysRevB.1.3599} {\bibfield  {journal} {\bibinfo  {journal} {Phys. Rev. B}\ }\textbf {\bibinfo {volume} {1}},\ \bibinfo {pages} {3599} (\bibinfo {year} {1970})}\BibitemShut {NoStop}%
\bibitem [{\citenamefont {Lema{\^i}tre}\ and\ \citenamefont {Maloney}(2006)}]{Lemaitre}%
  \BibitemOpen
  \bibfield  {author} {\bibinfo {author} {\bibfnamefont {A.}~\bibnamefont {Lema{\^i}tre}}\ and\ \bibinfo {author} {\bibfnamefont {C.}~\bibnamefont {Maloney}},\ }\href@noop {} {\bibfield  {journal} {\bibinfo  {journal} {Journal of Statistical Physics}\ }\textbf {\bibinfo {volume} {123}},\ \bibinfo {pages} {415} (\bibinfo {year} {2006})}\BibitemShut {NoStop}%
\bibitem [{\citenamefont {Liu}\ \emph {et~al.}(2022)\citenamefont {Liu}, \citenamefont {Bøjesen}, \citenamefont {Tabor}, \citenamefont {Mudie}, \citenamefont {Zaccone}, \citenamefont {Harrowell},\ and\ \citenamefont {Petersen}}]{amelia}%
  \BibitemOpen
  \bibfield  {author} {\bibinfo {author} {\bibfnamefont {A.~C.~Y.}\ \bibnamefont {Liu}}, \bibinfo {author} {\bibfnamefont {E.~D.}\ \bibnamefont {Bøjesen}}, \bibinfo {author} {\bibfnamefont {R.~F.}\ \bibnamefont {Tabor}}, \bibinfo {author} {\bibfnamefont {S.~T.}\ \bibnamefont {Mudie}}, \bibinfo {author} {\bibfnamefont {A.}~\bibnamefont {Zaccone}}, \bibinfo {author} {\bibfnamefont {P.}~\bibnamefont {Harrowell}}, \ and\ \bibinfo {author} {\bibfnamefont {T.~C.}\ \bibnamefont {Petersen}},\ }\href {\doibase 10.1126/sciadv.abn0681} {\bibfield  {journal} {\bibinfo  {journal} {Science Advances}\ }\textbf {\bibinfo {volume} {8}},\ \bibinfo {pages} {eabn0681} (\bibinfo {year} {2022})},\ \Eprint {http://arxiv.org/abs/https://www.science.org/doi/pdf/10.1126/sciadv.abn0681} {https://www.science.org/doi/pdf/10.1126/sciadv.abn0681} \BibitemShut {NoStop}%
\bibitem [{\citenamefont {Cohen}\ and\ \citenamefont {Louie}(2016)}]{Cohen}%
  \BibitemOpen
  \bibfield  {author} {\bibinfo {author} {\bibfnamefont {M.~L.}\ \bibnamefont {Cohen}}\ and\ \bibinfo {author} {\bibfnamefont {S.~G.}\ \bibnamefont {Louie}},\ }\href@noop {} {\emph {\bibinfo {title} {Fundamentals of Condensed Matter Physics}}}\ (\bibinfo  {publisher} {Cambridge University Press},\ \bibinfo {year} {2016})\BibitemShut {NoStop}%
\bibitem [{\citenamefont {Damart}\ \emph {et~al.}(2017)\citenamefont {Damart}, \citenamefont {Tanguy},\ and\ \citenamefont {Rodney}}]{PhysRevB.95.054203}%
  \BibitemOpen
  \bibfield  {author} {\bibinfo {author} {\bibfnamefont {T.}~\bibnamefont {Damart}}, \bibinfo {author} {\bibfnamefont {A.}~\bibnamefont {Tanguy}}, \ and\ \bibinfo {author} {\bibfnamefont {D.}~\bibnamefont {Rodney}},\ }\href {\doibase 10.1103/PhysRevB.95.054203} {\bibfield  {journal} {\bibinfo  {journal} {Phys. Rev. B}\ }\textbf {\bibinfo {volume} {95}},\ \bibinfo {pages} {054203} (\bibinfo {year} {2017})}\BibitemShut {NoStop}%
\bibitem [{\citenamefont {Otsuki}\ and\ \citenamefont {Hayakawa}(2022)}]{Otsuki}%
  \BibitemOpen
  \bibfield  {author} {\bibinfo {author} {\bibfnamefont {M.}~\bibnamefont {Otsuki}}\ and\ \bibinfo {author} {\bibfnamefont {H.}~\bibnamefont {Hayakawa}},\ }\href {\doibase 10.1103/PhysRevLett.128.208002} {\bibfield  {journal} {\bibinfo  {journal} {Phys. Rev. Lett.}\ }\textbf {\bibinfo {volume} {128}},\ \bibinfo {pages} {208002} (\bibinfo {year} {2022})}\BibitemShut {NoStop}%
\bibitem [{\citenamefont {Milkus}\ and\ \citenamefont {Zaccone}(2017)}]{Milkus2}%
  \BibitemOpen
  \bibfield  {author} {\bibinfo {author} {\bibfnamefont {R.}~\bibnamefont {Milkus}}\ and\ \bibinfo {author} {\bibfnamefont {A.}~\bibnamefont {Zaccone}},\ }\href@noop {} {\bibfield  {journal} {\bibinfo  {journal} {Phys. Rev. E}\ }\textbf {\bibinfo {volume} {95}},\ \bibinfo {pages} {023001} (\bibinfo {year} {2017})}\BibitemShut {NoStop}%
\bibitem [{\citenamefont {Tighe}(2011)}]{Tighe}%
  \BibitemOpen
  \bibfield  {author} {\bibinfo {author} {\bibfnamefont {B.~P.}\ \bibnamefont {Tighe}},\ }\href {\doibase 10.1103/PhysRevLett.107.158303} {\bibfield  {journal} {\bibinfo  {journal} {Phys. Rev. Lett.}\ }\textbf {\bibinfo {volume} {107}},\ \bibinfo {pages} {158303} (\bibinfo {year} {2011})}\BibitemShut {NoStop}%
\bibitem [{\citenamefont {Hara}\ \emph {et~al.}(2023)\citenamefont {Hara}, \citenamefont {Mizuno},\ and\ \citenamefont {Ikeda}}]{Hara}%
  \BibitemOpen
  \bibfield  {author} {\bibinfo {author} {\bibfnamefont {Y.}~\bibnamefont {Hara}}, \bibinfo {author} {\bibfnamefont {H.}~\bibnamefont {Mizuno}}, \ and\ \bibinfo {author} {\bibfnamefont {A.}~\bibnamefont {Ikeda}},\ }\href {\doibase 10.1039/D3SM00566F} {\bibfield  {journal} {\bibinfo  {journal} {Soft Matter}\ }\textbf {\bibinfo {volume} {19}},\ \bibinfo {pages} {6046} (\bibinfo {year} {2023})}\BibitemShut {NoStop}%
\bibitem [{\citenamefont {Palyulin}\ \emph {et~al.}(2018)\citenamefont {Palyulin}, \citenamefont {Ness}, \citenamefont {Milkus}, \citenamefont {Elder}, \citenamefont {Sirk},\ and\ \citenamefont {Zaccone}}]{Palyulin}%
  \BibitemOpen
  \bibfield  {author} {\bibinfo {author} {\bibfnamefont {V.~V.}\ \bibnamefont {Palyulin}}, \bibinfo {author} {\bibfnamefont {C.}~\bibnamefont {Ness}}, \bibinfo {author} {\bibfnamefont {R.}~\bibnamefont {Milkus}}, \bibinfo {author} {\bibfnamefont {R.~M.}\ \bibnamefont {Elder}}, \bibinfo {author} {\bibfnamefont {T.~W.}\ \bibnamefont {Sirk}}, \ and\ \bibinfo {author} {\bibfnamefont {A.}~\bibnamefont {Zaccone}},\ }\href@noop {} {\bibfield  {journal} {\bibinfo  {journal} {Soft Matter}\ }\textbf {\bibinfo {volume} {14}},\ \bibinfo {pages} {8475} (\bibinfo {year} {2018})}\BibitemShut {NoStop}%
\bibitem [{\citenamefont {Flenner}\ and\ \citenamefont {Szamel}(2024)}]{flenner2024}%
  \BibitemOpen
  \bibfield  {author} {\bibinfo {author} {\bibfnamefont {E.}~\bibnamefont {Flenner}}\ and\ \bibinfo {author} {\bibfnamefont {G.}~\bibnamefont {Szamel}},\ }\href {https://arxiv.org/abs/2406.18667} {\enquote {\bibinfo {title} {The origin of sound damping in amorphous solids: Defects and beyond},}\ } (\bibinfo {year} {2024}),\ \Eprint {http://arxiv.org/abs/2406.18667} {arXiv:2406.18667 [cond-mat.dis-nn]} \BibitemShut {NoStop}%
\bibitem [{\citenamefont {Grie\ss{}er}\ and\ \citenamefont {Pastewka}(2024)}]{Pastewka}%
  \BibitemOpen
  \bibfield  {author} {\bibinfo {author} {\bibfnamefont {J.}~\bibnamefont {Grie\ss{}er}}\ and\ \bibinfo {author} {\bibfnamefont {L.}~\bibnamefont {Pastewka}},\ }\href {\doibase 10.1103/PhysRevE.110.025001} {\bibfield  {journal} {\bibinfo  {journal} {Phys. Rev. E}\ }\textbf {\bibinfo {volume} {110}},\ \bibinfo {pages} {025001} (\bibinfo {year} {2024})}\BibitemShut {NoStop}%
\bibitem [{\citenamefont {Born}\ and\ \citenamefont {Huang}(1954)}]{Born_Huang}%
  \BibitemOpen
  \bibfield  {author} {\bibinfo {author} {\bibfnamefont {M.}~\bibnamefont {Born}}\ and\ \bibinfo {author} {\bibfnamefont {K.}~\bibnamefont {Huang}},\ }\href@noop {} {\emph {\bibinfo {title} {Dynamical Theory of Crystal Lattices}}}\ (\bibinfo  {publisher} {Clarendon Press},\ \bibinfo {address} {Oxford},\ \bibinfo {year} {1954})\BibitemShut {NoStop}%
\bibitem [{\citenamefont {Cui}\ \emph {et~al.}(2019{\natexlab{a}})\citenamefont {Cui}, \citenamefont {Ruocco},\ and\ \citenamefont {Zaccone}}]{Cui2019}%
  \BibitemOpen
  \bibfield  {author} {\bibinfo {author} {\bibfnamefont {B.}~\bibnamefont {Cui}}, \bibinfo {author} {\bibfnamefont {G.}~\bibnamefont {Ruocco}}, \ and\ \bibinfo {author} {\bibfnamefont {A.}~\bibnamefont {Zaccone}},\ }\href {\doibase 10.1007/s10035-019-0916-4} {\bibfield  {journal} {\bibinfo  {journal} {Granular Matter}\ }\textbf {\bibinfo {volume} {21}},\ \bibinfo {pages} {69} (\bibinfo {year} {2019}{\natexlab{a}})}\BibitemShut {NoStop}%
\bibitem [{\citenamefont {Wittmer}\ \emph {et~al.}(2013)\citenamefont {Wittmer}, \citenamefont {Xu}, \citenamefont {Polińska}, \citenamefont {Weysser},\ and\ \citenamefont {Baschnagel}}]{wittmer}%
  \BibitemOpen
  \bibfield  {author} {\bibinfo {author} {\bibfnamefont {J.~P.}\ \bibnamefont {Wittmer}}, \bibinfo {author} {\bibfnamefont {H.}~\bibnamefont {Xu}}, \bibinfo {author} {\bibfnamefont {P.}~\bibnamefont {Polińska}}, \bibinfo {author} {\bibfnamefont {F.}~\bibnamefont {Weysser}}, \ and\ \bibinfo {author} {\bibfnamefont {J.}~\bibnamefont {Baschnagel}},\ }\href@noop {} {\bibfield  {journal} {\bibinfo  {journal} {The Journal of Chemical Physics}\ }\textbf {\bibinfo {volume} {138}},\ \bibinfo {pages} {12A533} (\bibinfo {year} {2013})}\BibitemShut {NoStop}%
\bibitem [{\citenamefont {Pan}\ \emph {et~al.}(2023{\natexlab{a}})\citenamefont {Pan}, \citenamefont {Meng},\ and\ \citenamefont {Jin}}]{Fanlong}%
  \BibitemOpen
  \bibfield  {author} {\bibinfo {author} {\bibfnamefont {D.}~\bibnamefont {Pan}}, \bibinfo {author} {\bibfnamefont {F.}~\bibnamefont {Meng}}, \ and\ \bibinfo {author} {\bibfnamefont {Y.}~\bibnamefont {Jin}},\ }\href {\doibase 10.1093/pnasnexus/pgad047} {\bibfield  {journal} {\bibinfo  {journal} {PNAS Nexus}\ }\textbf {\bibinfo {volume} {2}},\ \bibinfo {pages} {pgad047} (\bibinfo {year} {2023}{\natexlab{a}})},\ \Eprint {http://arxiv.org/abs/https://academic.oup.com/pnasnexus/article-pdf/2/3/pgad047/50969403/pgad047.pdf} {https://academic.oup.com/pnasnexus/article-pdf/2/3/pgad047/50969403/pgad047.pdf} \BibitemShut {NoStop}%
\bibitem [{\citenamefont {Zaccone}(2023{\natexlab{b}})}]{PhysRevE.108.044101}%
  \BibitemOpen
  \bibfield  {author} {\bibinfo {author} {\bibfnamefont {A.}~\bibnamefont {Zaccone}},\ }\href {\doibase 10.1103/PhysRevE.108.044101} {\bibfield  {journal} {\bibinfo  {journal} {Phys. Rev. E}\ }\textbf {\bibinfo {volume} {108}},\ \bibinfo {pages} {044101} (\bibinfo {year} {2023}{\natexlab{b}})}\BibitemShut {NoStop}%
\bibitem [{\citenamefont {Saitoh}\ \emph {et~al.}(2020)\citenamefont {Saitoh}, \citenamefont {Hatano}, \citenamefont {Ikeda},\ and\ \citenamefont {Tighe}}]{Saitoh}%
  \BibitemOpen
  \bibfield  {author} {\bibinfo {author} {\bibfnamefont {K.}~\bibnamefont {Saitoh}}, \bibinfo {author} {\bibfnamefont {T.}~\bibnamefont {Hatano}}, \bibinfo {author} {\bibfnamefont {A.}~\bibnamefont {Ikeda}}, \ and\ \bibinfo {author} {\bibfnamefont {B.~P.}\ \bibnamefont {Tighe}},\ }\href {\doibase 10.1103/PhysRevLett.124.118001} {\bibfield  {journal} {\bibinfo  {journal} {Phys. Rev. Lett.}\ }\textbf {\bibinfo {volume} {124}},\ \bibinfo {pages} {118001} (\bibinfo {year} {2020})}\BibitemShut {NoStop}%
\bibitem [{\citenamefont {Lerner}\ \emph {et~al.}(2012)\citenamefont {Lerner}, \citenamefont {Düring},\ and\ \citenamefont {Wyart}}]{Lerner2012}%
  \BibitemOpen
  \bibfield  {author} {\bibinfo {author} {\bibfnamefont {E.}~\bibnamefont {Lerner}}, \bibinfo {author} {\bibfnamefont {G.}~\bibnamefont {Düring}}, \ and\ \bibinfo {author} {\bibfnamefont {M.}~\bibnamefont {Wyart}},\ }\href {\doibase 10.1073/pnas.1120215109} {\bibfield  {journal} {\bibinfo  {journal} {Proceedings of the National Academy of Sciences}\ }\textbf {\bibinfo {volume} {109}},\ \bibinfo {pages} {4798} (\bibinfo {year} {2012})},\ \Eprint {http://arxiv.org/abs/https://www.pnas.org/doi/pdf/10.1073/pnas.1120215109} {https://www.pnas.org/doi/pdf/10.1073/pnas.1120215109} \BibitemShut {NoStop}%
\bibitem [{\citenamefont {Franz}\ \emph {et~al.}(2015)\citenamefont {Franz}, \citenamefont {Parisi}, \citenamefont {Urbani},\ and\ \citenamefont {Zamponi}}]{Franz}%
  \BibitemOpen
  \bibfield  {author} {\bibinfo {author} {\bibfnamefont {S.}~\bibnamefont {Franz}}, \bibinfo {author} {\bibfnamefont {G.}~\bibnamefont {Parisi}}, \bibinfo {author} {\bibfnamefont {P.}~\bibnamefont {Urbani}}, \ and\ \bibinfo {author} {\bibfnamefont {F.}~\bibnamefont {Zamponi}},\ }\href {\doibase 10.1073/pnas.1511134112} {\bibfield  {journal} {\bibinfo  {journal} {Proceedings of the National Academy of Sciences}\ }\textbf {\bibinfo {volume} {112}},\ \bibinfo {pages} {14539} (\bibinfo {year} {2015})},\ \Eprint {http://arxiv.org/abs/https://www.pnas.org/doi/pdf/10.1073/pnas.1511134112} {https://www.pnas.org/doi/pdf/10.1073/pnas.1511134112} \BibitemShut {NoStop}%
\bibitem [{\citenamefont {Bai}\ and\ \citenamefont {Silverstein}(2010)}]{Bai2010}%
  \BibitemOpen
  \bibfield  {author} {\bibinfo {author} {\bibfnamefont {Z.}~\bibnamefont {Bai}}\ and\ \bibinfo {author} {\bibfnamefont {J.~W.}\ \bibnamefont {Silverstein}},\ }\href {\doibase 10.1007/978-1-4419-0661-8_1} {\emph {\bibinfo {title} {Spectral Analysis of Large Dimensional Random Matrices}}}\ (\bibinfo  {publisher} {Springer New York},\ \bibinfo {address} {New York, NY},\ \bibinfo {year} {2010})\BibitemShut {NoStop}%
\bibitem [{\citenamefont {Ikeda}(2020)}]{Harukuni_PRR}%
  \BibitemOpen
  \bibfield  {author} {\bibinfo {author} {\bibfnamefont {H.}~\bibnamefont {Ikeda}},\ }\href {\doibase 10.1103/PhysRevResearch.2.033220} {\bibfield  {journal} {\bibinfo  {journal} {Phys. Rev. Res.}\ }\textbf {\bibinfo {volume} {2}},\ \bibinfo {pages} {033220} (\bibinfo {year} {2020})}\BibitemShut {NoStop}%
\bibitem [{\citenamefont {Ikeda}\ and\ \citenamefont {Shimada}(2022)}]{Harukuni}%
  \BibitemOpen
  \bibfield  {author} {\bibinfo {author} {\bibfnamefont {H.}~\bibnamefont {Ikeda}}\ and\ \bibinfo {author} {\bibfnamefont {M.}~\bibnamefont {Shimada}},\ }\href {\doibase 10.1103/PhysRevE.106.024904} {\bibfield  {journal} {\bibinfo  {journal} {Phys. Rev. E}\ }\textbf {\bibinfo {volume} {106}},\ \bibinfo {pages} {024904} (\bibinfo {year} {2022})}\BibitemShut {NoStop}%
\bibitem [{\citenamefont {Shimada}\ \emph {et~al.}(2020)\citenamefont {Shimada}, \citenamefont {Mizuno}, \citenamefont {Berthier},\ and\ \citenamefont {Ikeda}}]{Mizuno_Berthier}%
  \BibitemOpen
  \bibfield  {author} {\bibinfo {author} {\bibfnamefont {M.}~\bibnamefont {Shimada}}, \bibinfo {author} {\bibfnamefont {H.}~\bibnamefont {Mizuno}}, \bibinfo {author} {\bibfnamefont {L.}~\bibnamefont {Berthier}}, \ and\ \bibinfo {author} {\bibfnamefont {A.}~\bibnamefont {Ikeda}},\ }\href {\doibase 10.1103/PhysRevE.101.052906} {\bibfield  {journal} {\bibinfo  {journal} {Phys. Rev. E}\ }\textbf {\bibinfo {volume} {101}},\ \bibinfo {pages} {052906} (\bibinfo {year} {2020})}\BibitemShut {NoStop}%
\bibitem [{\citenamefont {Baggioli}\ \emph {et~al.}(2019)\citenamefont {Baggioli}, \citenamefont {Milkus},\ and\ \citenamefont {Zaccone}}]{Milkus_RMT}%
  \BibitemOpen
  \bibfield  {author} {\bibinfo {author} {\bibfnamefont {M.}~\bibnamefont {Baggioli}}, \bibinfo {author} {\bibfnamefont {R.}~\bibnamefont {Milkus}}, \ and\ \bibinfo {author} {\bibfnamefont {A.}~\bibnamefont {Zaccone}},\ }\href {\doibase 10.1103/PhysRevE.100.062131} {\bibfield  {journal} {\bibinfo  {journal} {Phys. Rev. E}\ }\textbf {\bibinfo {volume} {100}},\ \bibinfo {pages} {062131} (\bibinfo {year} {2019})}\BibitemShut {NoStop}%
\bibitem [{\citenamefont {Olsson}\ and\ \citenamefont {Teitel}(2007)}]{Olsson}%
  \BibitemOpen
  \bibfield  {author} {\bibinfo {author} {\bibfnamefont {P.}~\bibnamefont {Olsson}}\ and\ \bibinfo {author} {\bibfnamefont {S.}~\bibnamefont {Teitel}},\ }\href {\doibase 10.1103/PhysRevLett.99.178001} {\bibfield  {journal} {\bibinfo  {journal} {Phys. Rev. Lett.}\ }\textbf {\bibinfo {volume} {99}},\ \bibinfo {pages} {178001} (\bibinfo {year} {2007})}\BibitemShut {NoStop}%
\bibitem [{\citenamefont {Hedstr\"om}\ and\ \citenamefont {Olsson}(2024)}]{Olsson_2024}%
  \BibitemOpen
  \bibfield  {author} {\bibinfo {author} {\bibfnamefont {L.}~\bibnamefont {Hedstr\"om}}\ and\ \bibinfo {author} {\bibfnamefont {P.}~\bibnamefont {Olsson}},\ }\href {\doibase 10.1103/PhysRevE.109.064904} {\bibfield  {journal} {\bibinfo  {journal} {Phys. Rev. E}\ }\textbf {\bibinfo {volume} {109}},\ \bibinfo {pages} {064904} (\bibinfo {year} {2024})}\BibitemShut {NoStop}%
\bibitem [{\citenamefont {Hopkins}\ \emph {et~al.}(2013)\citenamefont {Hopkins}, \citenamefont {Stillinger},\ and\ \citenamefont {Torquato}}]{Torq_bin}%
  \BibitemOpen
  \bibfield  {author} {\bibinfo {author} {\bibfnamefont {A.~B.}\ \bibnamefont {Hopkins}}, \bibinfo {author} {\bibfnamefont {F.~H.}\ \bibnamefont {Stillinger}}, \ and\ \bibinfo {author} {\bibfnamefont {S.}~\bibnamefont {Torquato}},\ }\href {\doibase 10.1103/PhysRevE.88.022205} {\bibfield  {journal} {\bibinfo  {journal} {Phys. Rev. E}\ }\textbf {\bibinfo {volume} {88}},\ \bibinfo {pages} {022205} (\bibinfo {year} {2013})}\BibitemShut {NoStop}%
\bibitem [{\citenamefont {Anzivino}\ \emph {et~al.}(2023)\citenamefont {Anzivino}, \citenamefont {Casiulis}, \citenamefont {Zhang}, \citenamefont {Moussa}, \citenamefont {Martiniani},\ and\ \citenamefont {Zaccone}}]{Anzivino}%
  \BibitemOpen
  \bibfield  {author} {\bibinfo {author} {\bibfnamefont {C.}~\bibnamefont {Anzivino}}, \bibinfo {author} {\bibfnamefont {M.}~\bibnamefont {Casiulis}}, \bibinfo {author} {\bibfnamefont {T.}~\bibnamefont {Zhang}}, \bibinfo {author} {\bibfnamefont {A.~S.}\ \bibnamefont {Moussa}}, \bibinfo {author} {\bibfnamefont {S.}~\bibnamefont {Martiniani}}, \ and\ \bibinfo {author} {\bibfnamefont {A.}~\bibnamefont {Zaccone}},\ }\href {\doibase 10.1063/5.0137111} {\bibfield  {journal} {\bibinfo  {journal} {The Journal of Chemical Physics}\ }\textbf {\bibinfo {volume} {158}},\ \bibinfo {pages} {044901} (\bibinfo {year} {2023})},\ \Eprint {http://arxiv.org/abs/https://pubs.aip.org/aip/jcp/article-pdf/doi/10.1063/5.0137111/16750608/044901\_1\_online.pdf} {https://pubs.aip.org/aip/jcp/article-pdf/doi/10.1063/5.0137111/16750608/044901\_1\_online.pdf} \BibitemShut {NoStop}%
\bibitem [{\citenamefont {Desmond}\ and\ \citenamefont {Weeks}(2014)}]{Desmond}%
  \BibitemOpen
  \bibfield  {author} {\bibinfo {author} {\bibfnamefont {K.~W.}\ \bibnamefont {Desmond}}\ and\ \bibinfo {author} {\bibfnamefont {E.~R.}\ \bibnamefont {Weeks}},\ }\href {\doibase 10.1103/PhysRevE.90.022204} {\bibfield  {journal} {\bibinfo  {journal} {Phys. Rev. E}\ }\textbf {\bibinfo {volume} {90}},\ \bibinfo {pages} {022204} (\bibinfo {year} {2014})}\BibitemShut {NoStop}%
\bibitem [{\citenamefont {Spangenberg}\ \emph {et~al.}(2014)\citenamefont {Spangenberg}, \citenamefont {Scherer}, \citenamefont {Hopkins},\ and\ \citenamefont {Torquato}}]{Torq_visc}%
  \BibitemOpen
  \bibfield  {author} {\bibinfo {author} {\bibfnamefont {J.}~\bibnamefont {Spangenberg}}, \bibinfo {author} {\bibfnamefont {G.~W.}\ \bibnamefont {Scherer}}, \bibinfo {author} {\bibfnamefont {A.~B.}\ \bibnamefont {Hopkins}}, \ and\ \bibinfo {author} {\bibfnamefont {S.}~\bibnamefont {Torquato}},\ }\href {\doibase 10.1063/1.4901463} {\bibfield  {journal} {\bibinfo  {journal} {Journal of Applied Physics}\ }\textbf {\bibinfo {volume} {116}},\ \bibinfo {pages} {184902} (\bibinfo {year} {2014})},\ \Eprint {http://arxiv.org/abs/https://pubs.aip.org/aip/jap/article-pdf/doi/10.1063/1.4901463/13974233/184902\_1\_online.pdf} {https://pubs.aip.org/aip/jap/article-pdf/doi/10.1063/1.4901463/13974233/184902\_1\_online.pdf} \BibitemShut {NoStop}%
\bibitem [{\citenamefont {Trappe}\ \emph {et~al.}(2001)\citenamefont {Trappe}, \citenamefont {Prasad}, \citenamefont {Cipelletti}, \citenamefont {Segre},\ and\ \citenamefont {Weitz}}]{Trappe2001}%
  \BibitemOpen
  \bibfield  {author} {\bibinfo {author} {\bibfnamefont {V.}~\bibnamefont {Trappe}}, \bibinfo {author} {\bibfnamefont {V.}~\bibnamefont {Prasad}}, \bibinfo {author} {\bibfnamefont {L.}~\bibnamefont {Cipelletti}}, \bibinfo {author} {\bibfnamefont {P.~N.}\ \bibnamefont {Segre}}, \ and\ \bibinfo {author} {\bibfnamefont {D.~A.}\ \bibnamefont {Weitz}},\ }\href {\doibase 10.1038/35081021} {\bibfield  {journal} {\bibinfo  {journal} {Nature}\ }\textbf {\bibinfo {volume} {411}},\ \bibinfo {pages} {772} (\bibinfo {year} {2001})}\BibitemShut {NoStop}%
\bibitem [{\citenamefont {Zaccone}\ \emph {et~al.}(2011)\citenamefont {Zaccone}, \citenamefont {Gentili}, \citenamefont {Wu}, \citenamefont {Morbidelli},\ and\ \citenamefont {Del~Gado}}]{PRL_solidification}%
  \BibitemOpen
  \bibfield  {author} {\bibinfo {author} {\bibfnamefont {A.}~\bibnamefont {Zaccone}}, \bibinfo {author} {\bibfnamefont {D.}~\bibnamefont {Gentili}}, \bibinfo {author} {\bibfnamefont {H.}~\bibnamefont {Wu}}, \bibinfo {author} {\bibfnamefont {M.}~\bibnamefont {Morbidelli}}, \ and\ \bibinfo {author} {\bibfnamefont {E.}~\bibnamefont {Del~Gado}},\ }\href {\doibase 10.1103/PhysRevLett.106.138301} {\bibfield  {journal} {\bibinfo  {journal} {Phys. Rev. Lett.}\ }\textbf {\bibinfo {volume} {106}},\ \bibinfo {pages} {138301} (\bibinfo {year} {2011})}\BibitemShut {NoStop}%
\bibitem [{\citenamefont {Zaccone}\ \emph {et~al.}(2009)\citenamefont {Zaccone}, \citenamefont {Wu}, \citenamefont {Gentili},\ and\ \citenamefont {Morbidelli}}]{Zaccone2009}%
  \BibitemOpen
  \bibfield  {author} {\bibinfo {author} {\bibfnamefont {A.}~\bibnamefont {Zaccone}}, \bibinfo {author} {\bibfnamefont {H.}~\bibnamefont {Wu}}, \bibinfo {author} {\bibfnamefont {D.}~\bibnamefont {Gentili}}, \ and\ \bibinfo {author} {\bibfnamefont {M.}~\bibnamefont {Morbidelli}},\ }\href {\doibase 10.1103/PhysRevE.80.051404} {\bibfield  {journal} {\bibinfo  {journal} {Phys. Rev. E}\ }\textbf {\bibinfo {volume} {80}},\ \bibinfo {pages} {051404} (\bibinfo {year} {2009})}\BibitemShut {NoStop}%
\bibitem [{\citenamefont {Jung}\ \emph {et~al.}(2017)\citenamefont {Jung}, \citenamefont {Hanke},\ and\ \citenamefont {Schmid}}]{Schmid}%
  \BibitemOpen
  \bibfield  {author} {\bibinfo {author} {\bibfnamefont {G.}~\bibnamefont {Jung}}, \bibinfo {author} {\bibfnamefont {M.}~\bibnamefont {Hanke}}, \ and\ \bibinfo {author} {\bibfnamefont {F.}~\bibnamefont {Schmid}},\ }\href@noop {} {\bibfield  {journal} {\bibinfo  {journal} {J. Chem. Theory Comput.}\ }\textbf {\bibinfo {volume} {13}},\ \bibinfo {pages} {2481} (\bibinfo {year} {2017})}\BibitemShut {NoStop}%
\bibitem [{\citenamefont {Pelargonio}\ and\ \citenamefont {Zaccone}(2023)}]{Pelargonio}%
  \BibitemOpen
  \bibfield  {author} {\bibinfo {author} {\bibfnamefont {S.}~\bibnamefont {Pelargonio}}\ and\ \bibinfo {author} {\bibfnamefont {A.}~\bibnamefont {Zaccone}},\ }\href {\doibase 10.1103/PhysRevE.107.064102} {\bibfield  {journal} {\bibinfo  {journal} {Phys. Rev. E}\ }\textbf {\bibinfo {volume} {107}},\ \bibinfo {pages} {064102} (\bibinfo {year} {2023})}\BibitemShut {NoStop}%
\bibitem [{\citenamefont {Gottwald}\ \emph {et~al.}(2016)\citenamefont {Gottwald}, \citenamefont {Ivanov},\ and\ \citenamefont {Kühn}}]{Kuehn_2016}%
  \BibitemOpen
  \bibfield  {author} {\bibinfo {author} {\bibfnamefont {F.}~\bibnamefont {Gottwald}}, \bibinfo {author} {\bibfnamefont {S.~D.}\ \bibnamefont {Ivanov}}, \ and\ \bibinfo {author} {\bibfnamefont {O.}~\bibnamefont {Kühn}},\ }\href@noop {} {\bibfield  {journal} {\bibinfo  {journal} {The Journal of Chemical Physics}\ }\textbf {\bibinfo {volume} {144}},\ \bibinfo {pages} {164102} (\bibinfo {year} {2016})}\BibitemShut {NoStop}%
\bibitem [{\citenamefont {Babu}\ \emph {et~al.}(2021)\citenamefont {Babu}, \citenamefont {Pan}, \citenamefont {Jin}, \citenamefont {Chakraborty},\ and\ \citenamefont {Sastry}}]{Bulbul}%
  \BibitemOpen
  \bibfield  {author} {\bibinfo {author} {\bibfnamefont {V.}~\bibnamefont {Babu}}, \bibinfo {author} {\bibfnamefont {D.}~\bibnamefont {Pan}}, \bibinfo {author} {\bibfnamefont {Y.}~\bibnamefont {Jin}}, \bibinfo {author} {\bibfnamefont {B.}~\bibnamefont {Chakraborty}}, \ and\ \bibinfo {author} {\bibfnamefont {S.}~\bibnamefont {Sastry}},\ }\href {\doibase 10.1039/D0SM02186E} {\bibfield  {journal} {\bibinfo  {journal} {Soft Matter}\ }\textbf {\bibinfo {volume} {17}},\ \bibinfo {pages} {3121} (\bibinfo {year} {2021})}\BibitemShut {NoStop}%
\bibitem [{\citenamefont {Pan}\ \emph {et~al.}(2023{\natexlab{b}})\citenamefont {Pan}, \citenamefont {Wang}, \citenamefont {Yoshino}, \citenamefont {Zhang},\ and\ \citenamefont {Jin}}]{Yuliang_rev}%
  \BibitemOpen
  \bibfield  {author} {\bibinfo {author} {\bibfnamefont {D.}~\bibnamefont {Pan}}, \bibinfo {author} {\bibfnamefont {Y.}~\bibnamefont {Wang}}, \bibinfo {author} {\bibfnamefont {H.}~\bibnamefont {Yoshino}}, \bibinfo {author} {\bibfnamefont {J.}~\bibnamefont {Zhang}}, \ and\ \bibinfo {author} {\bibfnamefont {Y.}~\bibnamefont {Jin}},\ }\href {\doibase https://doi.org/10.1016/j.physrep.2023.10.002} {\bibfield  {journal} {\bibinfo  {journal} {Physics Reports}\ }\textbf {\bibinfo {volume} {1038}},\ \bibinfo {pages} {1} (\bibinfo {year} {2023}{\natexlab{b}})},\ \bibinfo {note} {a review on shear jamming}\BibitemShut {NoStop}%
\bibitem [{\citenamefont {Zaccone}(2020)}]{Zaccone_2020}%
  \BibitemOpen
  \bibfield  {author} {\bibinfo {author} {\bibfnamefont {A.}~\bibnamefont {Zaccone}},\ }\href {\doibase 10.1088/1361-648X/ab6e41} {\bibfield  {journal} {\bibinfo  {journal} {Journal of Physics: Condensed Matter}\ }\textbf {\bibinfo {volume} {32}},\ \bibinfo {pages} {203001} (\bibinfo {year} {2020})}\BibitemShut {NoStop}%
\bibitem [{\citenamefont {Durand}\ \emph {et~al.}(1987)\citenamefont {Durand}, \citenamefont {Delsanti}, \citenamefont {Adam},\ and\ \citenamefont {Luck}}]{Durand}%
  \BibitemOpen
  \bibfield  {author} {\bibinfo {author} {\bibfnamefont {D.}~\bibnamefont {Durand}}, \bibinfo {author} {\bibfnamefont {M.}~\bibnamefont {Delsanti}}, \bibinfo {author} {\bibfnamefont {M.}~\bibnamefont {Adam}}, \ and\ \bibinfo {author} {\bibfnamefont {J.~M.}\ \bibnamefont {Luck}},\ }\href {\doibase 10.1209/0295-5075/3/3/008} {\bibfield  {journal} {\bibinfo  {journal} {Europhysics Letters}\ }\textbf {\bibinfo {volume} {3}},\ \bibinfo {pages} {297} (\bibinfo {year} {1987})}\BibitemShut {NoStop}%
\bibitem [{\citenamefont {Efros}\ and\ \citenamefont {Shklovskii}(1976)}]{Efros}%
  \BibitemOpen
  \bibfield  {author} {\bibinfo {author} {\bibfnamefont {A.~L.}\ \bibnamefont {Efros}}\ and\ \bibinfo {author} {\bibfnamefont {B.~I.}\ \bibnamefont {Shklovskii}},\ }\href {\doibase https://doi.org/10.1002/pssb.2220760205} {\bibfield  {journal} {\bibinfo  {journal} {physica status solidi (b)}\ }\textbf {\bibinfo {volume} {76}},\ \bibinfo {pages} {475} (\bibinfo {year} {1976})}\BibitemShut {NoStop}%
\bibitem [{\citenamefont {Bernal}\ and\ \citenamefont {Mason}(1960)}]{Bernal}%
  \BibitemOpen
  \bibfield  {author} {\bibinfo {author} {\bibfnamefont {J.~D.}\ \bibnamefont {Bernal}}\ and\ \bibinfo {author} {\bibfnamefont {J.}~\bibnamefont {Mason}},\ }\href@noop {} {\bibfield  {journal} {\bibinfo  {journal} {Nature}\ }\textbf {\bibinfo {volume} {188}},\ \bibinfo {pages} {910} (\bibinfo {year} {1960})}\BibitemShut {NoStop}%
\bibitem [{\citenamefont {Zaccone}(2022)}]{Zaccone_2022}%
  \BibitemOpen
  \bibfield  {author} {\bibinfo {author} {\bibfnamefont {A.}~\bibnamefont {Zaccone}},\ }\href {\doibase 10.1103/PhysRevLett.128.028002} {\bibfield  {journal} {\bibinfo  {journal} {Phys. Rev. Lett.}\ }\textbf {\bibinfo {volume} {128}},\ \bibinfo {pages} {028002} (\bibinfo {year} {2022})}\BibitemShut {NoStop}%
\bibitem [{\citenamefont {Kamien}\ and\ \citenamefont {Liu}(2007)}]{Kamien}%
  \BibitemOpen
  \bibfield  {author} {\bibinfo {author} {\bibfnamefont {R.~D.}\ \bibnamefont {Kamien}}\ and\ \bibinfo {author} {\bibfnamefont {A.~J.}\ \bibnamefont {Liu}},\ }\href {\doibase 10.1103/PhysRevLett.99.155501} {\bibfield  {journal} {\bibinfo  {journal} {Phys. Rev. Lett.}\ }\textbf {\bibinfo {volume} {99}},\ \bibinfo {pages} {155501} (\bibinfo {year} {2007})}\BibitemShut {NoStop}%
\bibitem [{\citenamefont {Chaki}\ \emph {et~al.}(2024)\citenamefont {Chaki}, \citenamefont {Mei},\ and\ \citenamefont {Schweizer}}]{Ken}%
  \BibitemOpen
  \bibfield  {author} {\bibinfo {author} {\bibfnamefont {S.}~\bibnamefont {Chaki}}, \bibinfo {author} {\bibfnamefont {B.}~\bibnamefont {Mei}}, \ and\ \bibinfo {author} {\bibfnamefont {K.~S.}\ \bibnamefont {Schweizer}},\ }\href {\doibase 10.1103/PhysRevE.110.034606} {\bibfield  {journal} {\bibinfo  {journal} {Phys. Rev. E}\ }\textbf {\bibinfo {volume} {110}},\ \bibinfo {pages} {034606} (\bibinfo {year} {2024})}\BibitemShut {NoStop}%
\bibitem [{\citenamefont {Likos}(2022)}]{Likos}%
  \BibitemOpen
  \bibfield  {author} {\bibinfo {author} {\bibfnamefont {C.}~\bibnamefont {Likos}},\ }\href {\doibase 10.36471/JCCM-March-2022-02} {\bibfield  {journal} {\bibinfo  {journal} {Journal Club for Condensed Matter Physics}\ } (\bibinfo {year} {2022}),\ 10.36471/JCCM-March-2022-02}\BibitemShut {NoStop}%
\bibitem [{\citenamefont {Meyer}\ \emph {et~al.}(2010)\citenamefont {Meyer}, \citenamefont {Song}, \citenamefont {Jin}, \citenamefont {Wang},\ and\ \citenamefont {Makse}}]{Makse_2010}%
  \BibitemOpen
  \bibfield  {author} {\bibinfo {author} {\bibfnamefont {S.}~\bibnamefont {Meyer}}, \bibinfo {author} {\bibfnamefont {C.}~\bibnamefont {Song}}, \bibinfo {author} {\bibfnamefont {Y.}~\bibnamefont {Jin}}, \bibinfo {author} {\bibfnamefont {K.}~\bibnamefont {Wang}}, \ and\ \bibinfo {author} {\bibfnamefont {H.~A.}\ \bibnamefont {Makse}},\ }\href@noop {} {\bibfield  {journal} {\bibinfo  {journal} {Physica A: Statistical Mechanics and its Applications}\ }\textbf {\bibinfo {volume} {389}},\ \bibinfo {pages} {5137} (\bibinfo {year} {2010})}\BibitemShut {NoStop}%
\bibitem [{\citenamefont {Quickenden}\ and\ \citenamefont {Tan}(1974)}]{QUICK}%
  \BibitemOpen
  \bibfield  {author} {\bibinfo {author} {\bibfnamefont {T.~I.}\ \bibnamefont {Quickenden}}\ and\ \bibinfo {author} {\bibfnamefont {G.~K.}\ \bibnamefont {Tan}},\ }\href {\doibase https://doi.org/10.1016/0021-9797(74)90181-7} {\bibfield  {journal} {\bibinfo  {journal} {Journal of Colloid and Interface Science}\ }\textbf {\bibinfo {volume} {48}},\ \bibinfo {pages} {382} (\bibinfo {year} {1974})}\BibitemShut {NoStop}%
\bibitem [{\citenamefont {Brouwers}(2023)}]{Brouwers}%
  \BibitemOpen
  \bibfield  {author} {\bibinfo {author} {\bibfnamefont {H.~J.~H.}\ \bibnamefont {Brouwers}},\ }\href {\doibase 10.1039/D3SM01254A} {\bibfield  {journal} {\bibinfo  {journal} {Soft Matter}\ }\textbf {\bibinfo {volume} {19}},\ \bibinfo {pages} {8465} (\bibinfo {year} {2023})}\BibitemShut {NoStop}%
\bibitem [{\citenamefont {Atkinson}\ \emph {et~al.}(2014)\citenamefont {Atkinson}, \citenamefont {Stillinger},\ and\ \citenamefont {Torquato}}]{Torq_2D}%
  \BibitemOpen
  \bibfield  {author} {\bibinfo {author} {\bibfnamefont {S.}~\bibnamefont {Atkinson}}, \bibinfo {author} {\bibfnamefont {F.~H.}\ \bibnamefont {Stillinger}}, \ and\ \bibinfo {author} {\bibfnamefont {S.}~\bibnamefont {Torquato}},\ }\href {\doibase 10.1073/pnas.1408371112} {\bibfield  {journal} {\bibinfo  {journal} {Proceedings of the National Academy of Sciences}\ }\textbf {\bibinfo {volume} {111}},\ \bibinfo {pages} {18436} (\bibinfo {year} {2014})},\ \Eprint {http://arxiv.org/abs/https://www.pnas.org/doi/pdf/10.1073/pnas.1408371112} {https://www.pnas.org/doi/pdf/10.1073/pnas.1408371112} \BibitemShut {NoStop}%
\bibitem [{\citenamefont {Blondot}\ \emph {et~al.}(2024)\citenamefont {Blondot}, \citenamefont {Gérard}, \citenamefont {Quibeuf}, \citenamefont {Arnold}, \citenamefont {Delteil}, \citenamefont {Bogicevic}, \citenamefont {Pons}, \citenamefont {Lequeux}, \citenamefont {Buil},\ and\ \citenamefont {Hermier}}]{Blondot}%
  \BibitemOpen
  \bibfield  {author} {\bibinfo {author} {\bibfnamefont {V.}~\bibnamefont {Blondot}}, \bibinfo {author} {\bibfnamefont {D.}~\bibnamefont {Gérard}}, \bibinfo {author} {\bibfnamefont {G.}~\bibnamefont {Quibeuf}}, \bibinfo {author} {\bibfnamefont {C.}~\bibnamefont {Arnold}}, \bibinfo {author} {\bibfnamefont {A.}~\bibnamefont {Delteil}}, \bibinfo {author} {\bibfnamefont {A.}~\bibnamefont {Bogicevic}}, \bibinfo {author} {\bibfnamefont {T.}~\bibnamefont {Pons}}, \bibinfo {author} {\bibfnamefont {N.}~\bibnamefont {Lequeux}}, \bibinfo {author} {\bibfnamefont {S.}~\bibnamefont {Buil}}, \ and\ \bibinfo {author} {\bibfnamefont {J.-P.}\ \bibnamefont {Hermier}},\ }\href {\doibase 10.1088/1361-6528/ad3832} {\bibfield  {journal} {\bibinfo  {journal} {Nanotechnology}\ }\textbf {\bibinfo {volume} {35}},\ \bibinfo {pages} {365001} (\bibinfo {year} {2024})}\BibitemShut {NoStop}%
\bibitem [{\citenamefont {Bürger}\ \emph {et~al.}(2024)\citenamefont {Bürger}, \citenamefont {Hayne}, \citenamefont {Gundlach}, \citenamefont {Läuter}, \citenamefont {Kramer},\ and\ \citenamefont {Blum}}]{regolith}%
  \BibitemOpen
  \bibfield  {author} {\bibinfo {author} {\bibfnamefont {J.}~\bibnamefont {Bürger}}, \bibinfo {author} {\bibfnamefont {P.~O.}\ \bibnamefont {Hayne}}, \bibinfo {author} {\bibfnamefont {B.}~\bibnamefont {Gundlach}}, \bibinfo {author} {\bibfnamefont {M.}~\bibnamefont {Läuter}}, \bibinfo {author} {\bibfnamefont {T.}~\bibnamefont {Kramer}}, \ and\ \bibinfo {author} {\bibfnamefont {J.}~\bibnamefont {Blum}},\ }\href {\doibase https://doi.org/10.1029/2023JE008152} {\bibfield  {journal} {\bibinfo  {journal} {Journal of Geophysical Research: Planets}\ }\textbf {\bibinfo {volume} {129}},\ \bibinfo {pages} {e2023JE008152} (\bibinfo {year} {2024})},\ \bibinfo {note} {e2023JE008152 2023JE008152}\BibitemShut {NoStop}%
\bibitem [{\citenamefont {Malamud}\ \emph {et~al.}(2024)\citenamefont {Malamud}, \citenamefont {Schäfer}, \citenamefont {Sebastián}, \citenamefont {Timpe}, \citenamefont {Essink}, \citenamefont {Kreuzig}, \citenamefont {Meier}, \citenamefont {Blum}, \citenamefont {Perets},\ and\ \citenamefont {Burger}}]{Malamud_2024}%
  \BibitemOpen
  \bibfield  {author} {\bibinfo {author} {\bibfnamefont {U.}~\bibnamefont {Malamud}}, \bibinfo {author} {\bibfnamefont {C.~M.}\ \bibnamefont {Schäfer}}, \bibinfo {author} {\bibfnamefont {I.~L.~S.}\ \bibnamefont {Sebastián}}, \bibinfo {author} {\bibfnamefont {M.}~\bibnamefont {Timpe}}, \bibinfo {author} {\bibfnamefont {K.~A.}\ \bibnamefont {Essink}}, \bibinfo {author} {\bibfnamefont {C.}~\bibnamefont {Kreuzig}}, \bibinfo {author} {\bibfnamefont {G.}~\bibnamefont {Meier}}, \bibinfo {author} {\bibfnamefont {J.}~\bibnamefont {Blum}}, \bibinfo {author} {\bibfnamefont {H.~B.}\ \bibnamefont {Perets}}, \ and\ \bibinfo {author} {\bibfnamefont {C.}~\bibnamefont {Burger}},\ }\href {\doibase 10.3847/1538-4357/ad6c4a} {\bibfield  {journal} {\bibinfo  {journal} {The Astrophysical Journal}\ }\textbf {\bibinfo {volume} {974}},\ \bibinfo {pages} {76} (\bibinfo {year} {2024})}\BibitemShut {NoStop}%
\bibitem [{\citenamefont {Olson~Reichhardt}\ \emph {et~al.}(2012)\citenamefont {Olson~Reichhardt}, \citenamefont {Groopman}, \citenamefont {Nussinov},\ and\ \citenamefont {Reichhardt}}]{Reichardt_pin}%
  \BibitemOpen
  \bibfield  {author} {\bibinfo {author} {\bibfnamefont {C.~J.}\ \bibnamefont {Olson~Reichhardt}}, \bibinfo {author} {\bibfnamefont {E.}~\bibnamefont {Groopman}}, \bibinfo {author} {\bibfnamefont {Z.}~\bibnamefont {Nussinov}}, \ and\ \bibinfo {author} {\bibfnamefont {C.}~\bibnamefont {Reichhardt}},\ }\href {\doibase 10.1103/PhysRevE.86.061301} {\bibfield  {journal} {\bibinfo  {journal} {Phys. Rev. E}\ }\textbf {\bibinfo {volume} {86}},\ \bibinfo {pages} {061301} (\bibinfo {year} {2012})}\BibitemShut {NoStop}%
\bibitem [{\citenamefont {Graves}\ \emph {et~al.}(2016)\citenamefont {Graves}, \citenamefont {Nashed}, \citenamefont {Padgett}, \citenamefont {Goodrich}, \citenamefont {Liu},\ and\ \citenamefont {Sethna}}]{Graves_1}%
  \BibitemOpen
  \bibfield  {author} {\bibinfo {author} {\bibfnamefont {A.~L.}\ \bibnamefont {Graves}}, \bibinfo {author} {\bibfnamefont {S.}~\bibnamefont {Nashed}}, \bibinfo {author} {\bibfnamefont {E.}~\bibnamefont {Padgett}}, \bibinfo {author} {\bibfnamefont {C.~P.}\ \bibnamefont {Goodrich}}, \bibinfo {author} {\bibfnamefont {A.~J.}\ \bibnamefont {Liu}}, \ and\ \bibinfo {author} {\bibfnamefont {J.~P.}\ \bibnamefont {Sethna}},\ }\href {\doibase 10.1103/PhysRevLett.116.235501} {\bibfield  {journal} {\bibinfo  {journal} {Phys. Rev. Lett.}\ }\textbf {\bibinfo {volume} {116}},\ \bibinfo {pages} {235501} (\bibinfo {year} {2016})}\BibitemShut {NoStop}%
\bibitem [{\citenamefont {P{\'e}ter}\ \emph {et~al.}(2018)\citenamefont {P{\'e}ter}, \citenamefont {Lib{\'a}l}, \citenamefont {Reichhardt},\ and\ \citenamefont {Reichhardt}}]{Reichardt_clog}%
  \BibitemOpen
  \bibfield  {author} {\bibinfo {author} {\bibfnamefont {H.}~\bibnamefont {P{\'e}ter}}, \bibinfo {author} {\bibfnamefont {A.}~\bibnamefont {Lib{\'a}l}}, \bibinfo {author} {\bibfnamefont {C.}~\bibnamefont {Reichhardt}}, \ and\ \bibinfo {author} {\bibfnamefont {C.~J.~O.}\ \bibnamefont {Reichhardt}},\ }\href {\doibase 10.1038/s41598-018-28256-6} {\bibfield  {journal} {\bibinfo  {journal} {Scientific Reports}\ }\textbf {\bibinfo {volume} {8}},\ \bibinfo {pages} {10252} (\bibinfo {year} {2018})}\BibitemShut {NoStop}%
\bibitem [{\citenamefont {C\'ardenas-Barrantes}\ \emph {et~al.}(2021)\citenamefont {C\'ardenas-Barrantes}, \citenamefont {Cantor}, \citenamefont {Bar\'es}, \citenamefont {Renouf},\ and\ \citenamefont {Az\'ema}}]{Azema1}%
  \BibitemOpen
  \bibfield  {author} {\bibinfo {author} {\bibfnamefont {M.}~\bibnamefont {C\'ardenas-Barrantes}}, \bibinfo {author} {\bibfnamefont {D.}~\bibnamefont {Cantor}}, \bibinfo {author} {\bibfnamefont {J.}~\bibnamefont {Bar\'es}}, \bibinfo {author} {\bibfnamefont {M.}~\bibnamefont {Renouf}}, \ and\ \bibinfo {author} {\bibfnamefont {E.}~\bibnamefont {Az\'ema}},\ }\href {\doibase 10.1103/PhysRevE.103.062902} {\bibfield  {journal} {\bibinfo  {journal} {Phys. Rev. E}\ }\textbf {\bibinfo {volume} {103}},\ \bibinfo {pages} {062902} (\bibinfo {year} {2021})}\BibitemShut {NoStop}%
\bibitem [{\citenamefont {Cárdenas-Barrantes}\ \emph {et~al.}(2022)\citenamefont {Cárdenas-Barrantes}, \citenamefont {Cantor}, \citenamefont {Barés}, \citenamefont {Renouf},\ and\ \citenamefont {Azéma}}]{Azema2}%
  \BibitemOpen
  \bibfield  {author} {\bibinfo {author} {\bibfnamefont {M.}~\bibnamefont {Cárdenas-Barrantes}}, \bibinfo {author} {\bibfnamefont {D.}~\bibnamefont {Cantor}}, \bibinfo {author} {\bibfnamefont {J.}~\bibnamefont {Barés}}, \bibinfo {author} {\bibfnamefont {M.}~\bibnamefont {Renouf}}, \ and\ \bibinfo {author} {\bibfnamefont {E.}~\bibnamefont {Azéma}},\ }\href {\doibase 10.1039/D1SM01241J} {\bibfield  {journal} {\bibinfo  {journal} {Soft Matter}\ }\textbf {\bibinfo {volume} {18}},\ \bibinfo {pages} {312} (\bibinfo {year} {2022})}\BibitemShut {NoStop}%
\bibitem [{\citenamefont {Zaccone}\ and\ \citenamefont {Terentjev}(2013)}]{ZacconePRL}%
  \BibitemOpen
  \bibfield  {author} {\bibinfo {author} {\bibfnamefont {A.}~\bibnamefont {Zaccone}}\ and\ \bibinfo {author} {\bibfnamefont {E.~M.}\ \bibnamefont {Terentjev}},\ }\href@noop {} {\bibfield  {journal} {\bibinfo  {journal} {Phys. Rev. Lett.}\ }\textbf {\bibinfo {volume} {110}},\ \bibinfo {pages} {178002} (\bibinfo {year} {2013})}\BibitemShut {NoStop}%
\bibitem [{\citenamefont {Anzivino}\ and\ \citenamefont {Zaccone}(2023)}]{Anzivino2023}%
  \BibitemOpen
  \bibfield  {author} {\bibinfo {author} {\bibfnamefont {C.}~\bibnamefont {Anzivino}}\ and\ \bibinfo {author} {\bibfnamefont {A.}~\bibnamefont {Zaccone}},\ }\href {\doibase 10.1021/acs.jpclett.3c02007} {\bibfield  {journal} {\bibinfo  {journal} {The Journal of Physical Chemistry Letters}\ }\textbf {\bibinfo {volume} {14}},\ \bibinfo {pages} {8846} (\bibinfo {year} {2023})}\BibitemShut {NoStop}%
\bibitem [{\citenamefont {Zaccone}(2013)}]{MPLB}%
  \BibitemOpen
  \bibfield  {author} {\bibinfo {author} {\bibfnamefont {A.}~\bibnamefont {Zaccone}},\ }\href {\doibase 10.1142/S0217984913300020} {\bibfield  {journal} {\bibinfo  {journal} {Modern Physics Letters B}\ }\textbf {\bibinfo {volume} {27}},\ \bibinfo {pages} {1330002} (\bibinfo {year} {2013})},\ \Eprint {http://arxiv.org/abs/https://doi.org/10.1142/S0217984913300020} {https://doi.org/10.1142/S0217984913300020} \BibitemShut {NoStop}%
\bibitem [{\citenamefont {He}\ and\ \citenamefont {Thorpe}(1985)}]{Thorpe}%
  \BibitemOpen
  \bibfield  {author} {\bibinfo {author} {\bibfnamefont {H.}~\bibnamefont {He}}\ and\ \bibinfo {author} {\bibfnamefont {M.~F.}\ \bibnamefont {Thorpe}},\ }\href {\doibase 10.1103/PhysRevLett.54.2107} {\bibfield  {journal} {\bibinfo  {journal} {Phys. Rev. Lett.}\ }\textbf {\bibinfo {volume} {54}},\ \bibinfo {pages} {2107} (\bibinfo {year} {1985})}\BibitemShut {NoStop}%
\bibitem [{\citenamefont {Hoy}(2017)}]{Hoy}%
  \BibitemOpen
  \bibfield  {author} {\bibinfo {author} {\bibfnamefont {R.~S.}\ \bibnamefont {Hoy}},\ }\href {\doibase 10.1103/PhysRevLett.118.068002} {\bibfield  {journal} {\bibinfo  {journal} {Phys. Rev. Lett.}\ }\textbf {\bibinfo {volume} {118}},\ \bibinfo {pages} {068002} (\bibinfo {year} {2017})}\BibitemShut {NoStop}%
\bibitem [{\citenamefont {Elder}\ \emph {et~al.}(2019)\citenamefont {Elder}, \citenamefont {Zaccone},\ and\ \citenamefont {Sirk}}]{Elder}%
  \BibitemOpen
  \bibfield  {author} {\bibinfo {author} {\bibfnamefont {R.~M.}\ \bibnamefont {Elder}}, \bibinfo {author} {\bibfnamefont {A.}~\bibnamefont {Zaccone}}, \ and\ \bibinfo {author} {\bibfnamefont {T.~W.}\ \bibnamefont {Sirk}},\ }\href {\doibase 10.1021/acsmacrolett.9b00505} {\bibfield  {journal} {\bibinfo  {journal} {ACS Macro Letters}\ }\textbf {\bibinfo {volume} {8}},\ \bibinfo {pages} {1160} (\bibinfo {year} {2019})},\ \bibinfo {note} {pMID: 35619458},\ \Eprint {http://arxiv.org/abs/https://doi.org/10.1021/acsmacrolett.9b00505} {https://doi.org/10.1021/acsmacrolett.9b00505} \BibitemShut {NoStop}%
\bibitem [{\citenamefont {Vaibhav}\ \emph {et~al.}(2024)\citenamefont {Vaibhav}, \citenamefont {Sirk},\ and\ \citenamefont {Zaccone}}]{vaibhav}%
  \BibitemOpen
  \bibfield  {author} {\bibinfo {author} {\bibfnamefont {V.}~\bibnamefont {Vaibhav}}, \bibinfo {author} {\bibfnamefont {T.~W.}\ \bibnamefont {Sirk}}, \ and\ \bibinfo {author} {\bibfnamefont {A.}~\bibnamefont {Zaccone}},\ }\href {https://arxiv.org/abs/2406.02113} {\enquote {\bibinfo {title} {Timescale bridging in atomistic simulations of epoxy polymer mechanics using non-affine deformation theory},}\ } (\bibinfo {year} {2024}),\ \Eprint {http://arxiv.org/abs/2406.02113} {arXiv:2406.02113 [cond-mat.soft]} \BibitemShut {NoStop}%
\bibitem [{\citenamefont {Cui}\ \emph {et~al.}(2019{\natexlab{b}})\citenamefont {Cui}, \citenamefont {Zaccone},\ and\ \citenamefont {Rodney}}]{quartz}%
  \BibitemOpen
  \bibfield  {author} {\bibinfo {author} {\bibfnamefont {B.}~\bibnamefont {Cui}}, \bibinfo {author} {\bibfnamefont {A.}~\bibnamefont {Zaccone}}, \ and\ \bibinfo {author} {\bibfnamefont {D.}~\bibnamefont {Rodney}},\ }\href {\doibase 10.1063/1.5129025} {\bibfield  {journal} {\bibinfo  {journal} {The Journal of Chemical Physics}\ }\textbf {\bibinfo {volume} {151}},\ \bibinfo {pages} {224509} (\bibinfo {year} {2019}{\natexlab{b}})}\BibitemShut {NoStop}%
\bibitem [{\citenamefont {Falk}\ and\ \citenamefont {Langer}(1998)}]{Falk}%
  \BibitemOpen
  \bibfield  {author} {\bibinfo {author} {\bibfnamefont {M.~L.}\ \bibnamefont {Falk}}\ and\ \bibinfo {author} {\bibfnamefont {J.~S.}\ \bibnamefont {Langer}},\ }\href {\doibase 10.1103/PhysRevE.57.7192} {\bibfield  {journal} {\bibinfo  {journal} {Phys. Rev. E}\ }\textbf {\bibinfo {volume} {57}},\ \bibinfo {pages} {7192} (\bibinfo {year} {1998})}\BibitemShut {NoStop}%
\bibitem [{\citenamefont {Keim}\ and\ \citenamefont {Arratia}(2014)}]{Arratia}%
  \BibitemOpen
  \bibfield  {author} {\bibinfo {author} {\bibfnamefont {N.~C.}\ \bibnamefont {Keim}}\ and\ \bibinfo {author} {\bibfnamefont {P.~E.}\ \bibnamefont {Arratia}},\ }\href {\doibase 10.1103/PhysRevLett.112.028302} {\bibfield  {journal} {\bibinfo  {journal} {Phys. Rev. Lett.}\ }\textbf {\bibinfo {volume} {112}},\ \bibinfo {pages} {028302} (\bibinfo {year} {2014})}\BibitemShut {NoStop}%
\bibitem [{\citenamefont {Baggioli}\ \emph {et~al.}(2021)\citenamefont {Baggioli}, \citenamefont {Kriuchevskyi}, \citenamefont {Sirk},\ and\ \citenamefont {Zaccone}}]{Baggioli}%
  \BibitemOpen
  \bibfield  {author} {\bibinfo {author} {\bibfnamefont {M.}~\bibnamefont {Baggioli}}, \bibinfo {author} {\bibfnamefont {I.}~\bibnamefont {Kriuchevskyi}}, \bibinfo {author} {\bibfnamefont {T.~W.}\ \bibnamefont {Sirk}}, \ and\ \bibinfo {author} {\bibfnamefont {A.}~\bibnamefont {Zaccone}},\ }\href {\doibase 10.1103/PhysRevLett.127.015501} {\bibfield  {journal} {\bibinfo  {journal} {Phys. Rev. Lett.}\ }\textbf {\bibinfo {volume} {127}},\ \bibinfo {pages} {015501} (\bibinfo {year} {2021})}\BibitemShut {NoStop}%
\bibitem [{\citenamefont {Baggioli}\ \emph {et~al.}(2022)\citenamefont {Baggioli}, \citenamefont {Landry},\ and\ \citenamefont {Zaccone}}]{Landry}%
  \BibitemOpen
  \bibfield  {author} {\bibinfo {author} {\bibfnamefont {M.}~\bibnamefont {Baggioli}}, \bibinfo {author} {\bibfnamefont {M.}~\bibnamefont {Landry}}, \ and\ \bibinfo {author} {\bibfnamefont {A.}~\bibnamefont {Zaccone}},\ }\href {\doibase 10.1103/PhysRevE.105.024602} {\bibfield  {journal} {\bibinfo  {journal} {Phys. Rev. E}\ }\textbf {\bibinfo {volume} {105}},\ \bibinfo {pages} {024602} (\bibinfo {year} {2022})}\BibitemShut {NoStop}%
\bibitem [{\citenamefont {Wu}\ \emph {et~al.}(2023)\citenamefont {Wu}, \citenamefont {Chen}, \citenamefont {Wang}, \citenamefont {Kob},\ and\ \citenamefont {Xu}}]{Wu2023}%
  \BibitemOpen
  \bibfield  {author} {\bibinfo {author} {\bibfnamefont {Z.~W.}\ \bibnamefont {Wu}}, \bibinfo {author} {\bibfnamefont {Y.}~\bibnamefont {Chen}}, \bibinfo {author} {\bibfnamefont {W.-H.}\ \bibnamefont {Wang}}, \bibinfo {author} {\bibfnamefont {W.}~\bibnamefont {Kob}}, \ and\ \bibinfo {author} {\bibfnamefont {L.}~\bibnamefont {Xu}},\ }\href {\doibase 10.1038/s41467-023-38547-w} {\bibfield  {journal} {\bibinfo  {journal} {Nature Communications}\ }\textbf {\bibinfo {volume} {14}},\ \bibinfo {pages} {2955} (\bibinfo {year} {2023})}\BibitemShut {NoStop}%
\bibitem [{\citenamefont {Desmarchelier}\ \emph {et~al.}(2024)\citenamefont {Desmarchelier}, \citenamefont {Fajardo},\ and\ \citenamefont {Falk}}]{desmarchelier}%
  \BibitemOpen
  \bibfield  {author} {\bibinfo {author} {\bibfnamefont {P.}~\bibnamefont {Desmarchelier}}, \bibinfo {author} {\bibfnamefont {S.}~\bibnamefont {Fajardo}}, \ and\ \bibinfo {author} {\bibfnamefont {M.~L.}\ \bibnamefont {Falk}},\ }\href {\doibase 10.1103/PhysRevE.109.L053002} {\bibfield  {journal} {\bibinfo  {journal} {Phys. Rev. E}\ }\textbf {\bibinfo {volume} {109}},\ \bibinfo {pages} {L053002} (\bibinfo {year} {2024})}\BibitemShut {NoStop}%
\bibitem [{\citenamefont {Bera}\ \emph {et~al.}(2024)\citenamefont {Bera}, \citenamefont {Baggioli}, \citenamefont {Petersen}, \citenamefont {Sirk}, \citenamefont {Liu},\ and\ \citenamefont {Zaccone}}]{nexus}%
  \BibitemOpen
  \bibfield  {author} {\bibinfo {author} {\bibfnamefont {A.}~\bibnamefont {Bera}}, \bibinfo {author} {\bibfnamefont {M.}~\bibnamefont {Baggioli}}, \bibinfo {author} {\bibfnamefont {T.~C.}\ \bibnamefont {Petersen}}, \bibinfo {author} {\bibfnamefont {T.~W.}\ \bibnamefont {Sirk}}, \bibinfo {author} {\bibfnamefont {A.~C.~Y.}\ \bibnamefont {Liu}}, \ and\ \bibinfo {author} {\bibfnamefont {A.}~\bibnamefont {Zaccone}},\ }\href {\doibase 10.1093/pnasnexus/pgae315} {\bibfield  {journal} {\bibinfo  {journal} {PNAS Nexus}\ }\textbf {\bibinfo {volume} {3}},\ \bibinfo {pages} {pgae315} (\bibinfo {year} {2024})},\ \Eprint {http://arxiv.org/abs/https://academic.oup.com/pnasnexus/article-pdf/3/9/pgae315/59559307/pgae315.pdf} {https://academic.oup.com/pnasnexus/article-pdf/3/9/pgae315/59559307/pgae315.pdf} \BibitemShut {NoStop}%
\bibitem [{\citenamefont {Cohen}\ \emph {et~al.}(2024)\citenamefont {Cohen}, \citenamefont {Schiller}, \citenamefont {Wang}, \citenamefont {Dijksman},\ and\ \citenamefont {Moshe}}]{moshe}%
  \BibitemOpen
  \bibfield  {author} {\bibinfo {author} {\bibfnamefont {Y.}~\bibnamefont {Cohen}}, \bibinfo {author} {\bibfnamefont {A.}~\bibnamefont {Schiller}}, \bibinfo {author} {\bibfnamefont {D.}~\bibnamefont {Wang}}, \bibinfo {author} {\bibfnamefont {J.}~\bibnamefont {Dijksman}}, \ and\ \bibinfo {author} {\bibfnamefont {M.}~\bibnamefont {Moshe}},\ }\href {https://arxiv.org/abs/2310.09942} {\enquote {\bibinfo {title} {Odd dipole screening in disordered matter},}\ } (\bibinfo {year} {2024}),\ \Eprint {http://arxiv.org/abs/2310.09942} {arXiv:2310.09942 [cond-mat.soft]} \BibitemShut {NoStop}%
\bibitem [{\citenamefont {Mondal}\ \emph {et~al.}(2022)\citenamefont {Mondal}, \citenamefont {Moshe}, \citenamefont {Procaccia}, \citenamefont {Roy}, \citenamefont {Shang},\ and\ \citenamefont {Zhang}}]{procaccia}%
  \BibitemOpen
  \bibfield  {author} {\bibinfo {author} {\bibfnamefont {C.}~\bibnamefont {Mondal}}, \bibinfo {author} {\bibfnamefont {M.}~\bibnamefont {Moshe}}, \bibinfo {author} {\bibfnamefont {I.}~\bibnamefont {Procaccia}}, \bibinfo {author} {\bibfnamefont {S.}~\bibnamefont {Roy}}, \bibinfo {author} {\bibfnamefont {J.}~\bibnamefont {Shang}}, \ and\ \bibinfo {author} {\bibfnamefont {J.}~\bibnamefont {Zhang}},\ }\href {\doibase https://doi.org/10.1016/j.chaos.2022.112609} {\bibfield  {journal} {\bibinfo  {journal} {Chaos, Solitons \& Fractals}\ }\textbf {\bibinfo {volume} {164}},\ \bibinfo {pages} {112609} (\bibinfo {year} {2022})}\BibitemShut {NoStop}%
\bibitem [{\citenamefont {Denisov}\ \emph {et~al.}(2015)\citenamefont {Denisov}, \citenamefont {Dang}, \citenamefont {Struth}, \citenamefont {Zaccone}, \citenamefont {Wegdam},\ and\ \citenamefont {Schall}}]{Denisov2015}%
  \BibitemOpen
  \bibfield  {author} {\bibinfo {author} {\bibfnamefont {D.~V.}\ \bibnamefont {Denisov}}, \bibinfo {author} {\bibfnamefont {M.~T.}\ \bibnamefont {Dang}}, \bibinfo {author} {\bibfnamefont {B.}~\bibnamefont {Struth}}, \bibinfo {author} {\bibfnamefont {A.}~\bibnamefont {Zaccone}}, \bibinfo {author} {\bibfnamefont {G.~H.}\ \bibnamefont {Wegdam}}, \ and\ \bibinfo {author} {\bibfnamefont {P.}~\bibnamefont {Schall}},\ }\href {\doibase 10.1038/srep14359} {\bibfield  {journal} {\bibinfo  {journal} {Scientific Reports}\ }\textbf {\bibinfo {volume} {5}},\ \bibinfo {pages} {14359} (\bibinfo {year} {2015})}\BibitemShut {NoStop}%
\bibitem [{\citenamefont {Regev}\ \emph {et~al.}(2013)\citenamefont {Regev}, \citenamefont {Lookman},\ and\ \citenamefont {Reichhardt}}]{Ido}%
  \BibitemOpen
  \bibfield  {author} {\bibinfo {author} {\bibfnamefont {I.}~\bibnamefont {Regev}}, \bibinfo {author} {\bibfnamefont {T.}~\bibnamefont {Lookman}}, \ and\ \bibinfo {author} {\bibfnamefont {C.}~\bibnamefont {Reichhardt}},\ }\href {\doibase 10.1103/PhysRevE.88.062401} {\bibfield  {journal} {\bibinfo  {journal} {Phys. Rev. E}\ }\textbf {\bibinfo {volume} {88}},\ \bibinfo {pages} {062401} (\bibinfo {year} {2013})}\BibitemShut {NoStop}%
\bibitem [{\citenamefont {Aime}\ \emph {et~al.}(2023)\citenamefont {Aime}, \citenamefont {Truzzolillo}, \citenamefont {Pine}, \citenamefont {Ramos},\ and\ \citenamefont {Cipelletti}}]{Aime2023}%
  \BibitemOpen
  \bibfield  {author} {\bibinfo {author} {\bibfnamefont {S.}~\bibnamefont {Aime}}, \bibinfo {author} {\bibfnamefont {D.}~\bibnamefont {Truzzolillo}}, \bibinfo {author} {\bibfnamefont {D.~J.}\ \bibnamefont {Pine}}, \bibinfo {author} {\bibfnamefont {L.}~\bibnamefont {Ramos}}, \ and\ \bibinfo {author} {\bibfnamefont {L.}~\bibnamefont {Cipelletti}},\ }\href {\doibase 10.1038/s41567-023-02153-w} {\bibfield  {journal} {\bibinfo  {journal} {Nature Physics}\ }\textbf {\bibinfo {volume} {19}},\ \bibinfo {pages} {1673} (\bibinfo {year} {2023})}\BibitemShut {NoStop}%
\bibitem [{\citenamefont {Reichhardt}\ \emph {et~al.}(2023)\citenamefont {Reichhardt}, \citenamefont {Regev}, \citenamefont {Dahmen}, \citenamefont {Okuma},\ and\ \citenamefont {Reichhardt}}]{Ido2}%
  \BibitemOpen
  \bibfield  {author} {\bibinfo {author} {\bibfnamefont {C.}~\bibnamefont {Reichhardt}}, \bibinfo {author} {\bibfnamefont {I.}~\bibnamefont {Regev}}, \bibinfo {author} {\bibfnamefont {K.}~\bibnamefont {Dahmen}}, \bibinfo {author} {\bibfnamefont {S.}~\bibnamefont {Okuma}}, \ and\ \bibinfo {author} {\bibfnamefont {C.~J.~O.}\ \bibnamefont {Reichhardt}},\ }\href {\doibase 10.1103/PhysRevResearch.5.021001} {\bibfield  {journal} {\bibinfo  {journal} {Phys. Rev. Res.}\ }\textbf {\bibinfo {volume} {5}},\ \bibinfo {pages} {021001} (\bibinfo {year} {2023})}\BibitemShut {NoStop}%
\bibitem [{\citenamefont {Schreck}\ \emph {et~al.}(2010)\citenamefont {Schreck}, \citenamefont {Xu},\ and\ \citenamefont {O{'}Hern}}]{Corey}%
  \BibitemOpen
  \bibfield  {author} {\bibinfo {author} {\bibfnamefont {C.~F.}\ \bibnamefont {Schreck}}, \bibinfo {author} {\bibfnamefont {N.}~\bibnamefont {Xu}}, \ and\ \bibinfo {author} {\bibfnamefont {C.~S.}\ \bibnamefont {O{'}Hern}},\ }\href {\doibase 10.1039/C001085E} {\bibfield  {journal} {\bibinfo  {journal} {Soft Matter}\ }\textbf {\bibinfo {volume} {6}},\ \bibinfo {pages} {2960} (\bibinfo {year} {2010})}\BibitemShut {NoStop}%
\bibitem [{\citenamefont {Donev}\ \emph {et~al.}(2004)\citenamefont {Donev}, \citenamefont {Cisse}, \citenamefont {Sachs}, \citenamefont {Variano}, \citenamefont {Stillinger}, \citenamefont {Connelly}, \citenamefont {Torquato},\ and\ \citenamefont {Chaikin}}]{Donev}%
  \BibitemOpen
  \bibfield  {author} {\bibinfo {author} {\bibfnamefont {A.}~\bibnamefont {Donev}}, \bibinfo {author} {\bibfnamefont {I.}~\bibnamefont {Cisse}}, \bibinfo {author} {\bibfnamefont {D.}~\bibnamefont {Sachs}}, \bibinfo {author} {\bibfnamefont {E.~A.}\ \bibnamefont {Variano}}, \bibinfo {author} {\bibfnamefont {F.~H.}\ \bibnamefont {Stillinger}}, \bibinfo {author} {\bibfnamefont {R.}~\bibnamefont {Connelly}}, \bibinfo {author} {\bibfnamefont {S.}~\bibnamefont {Torquato}}, \ and\ \bibinfo {author} {\bibfnamefont {P.~M.}\ \bibnamefont {Chaikin}},\ }\href {\doibase 10.1126/science.1093010} {\bibfield  {journal} {\bibinfo  {journal} {Science}\ }\textbf {\bibinfo {volume} {303}},\ \bibinfo {pages} {990} (\bibinfo {year} {2004})},\ \Eprint {http://arxiv.org/abs/https://www.science.org/doi/pdf/10.1126/science.1093010} {https://www.science.org/doi/pdf/10.1126/science.1093010} \BibitemShut {NoStop}%
\bibitem [{\citenamefont {Rocks}\ and\ \citenamefont {Hoy}(2023)}]{hoy2}%
  \BibitemOpen
  \bibfield  {author} {\bibinfo {author} {\bibfnamefont {S.}~\bibnamefont {Rocks}}\ and\ \bibinfo {author} {\bibfnamefont {R.~S.}\ \bibnamefont {Hoy}},\ }\href {\doibase 10.1039/D3SM00705G} {\bibfield  {journal} {\bibinfo  {journal} {Soft Matter}\ }\textbf {\bibinfo {volume} {19}},\ \bibinfo {pages} {5701} (\bibinfo {year} {2023})}\BibitemShut {NoStop}%
\bibitem [{\citenamefont {Anzivino}\ \emph {et~al.}(2024)\citenamefont {Anzivino}, \citenamefont {Ness}, \citenamefont {Moussa},\ and\ \citenamefont {Zaccone}}]{moussa}%
  \BibitemOpen
  \bibfield  {author} {\bibinfo {author} {\bibfnamefont {C.}~\bibnamefont {Anzivino}}, \bibinfo {author} {\bibfnamefont {C.}~\bibnamefont {Ness}}, \bibinfo {author} {\bibfnamefont {A.~S.}\ \bibnamefont {Moussa}}, \ and\ \bibinfo {author} {\bibfnamefont {A.}~\bibnamefont {Zaccone}},\ }\href {\doibase 10.1103/PhysRevE.109.L042601} {\bibfield  {journal} {\bibinfo  {journal} {Phys. Rev. E}\ }\textbf {\bibinfo {volume} {109}},\ \bibinfo {pages} {L042601} (\bibinfo {year} {2024})}\BibitemShut {NoStop}%
\bibitem [{\citenamefont {Jaeger}(2015)}]{Jaeger_shape}%
  \BibitemOpen
  \bibfield  {author} {\bibinfo {author} {\bibfnamefont {H.~M.}\ \bibnamefont {Jaeger}},\ }\href {\doibase 10.1039/C4SM01923G} {\bibfield  {journal} {\bibinfo  {journal} {Soft Matter}\ }\textbf {\bibinfo {volume} {11}},\ \bibinfo {pages} {12} (\bibinfo {year} {2015})}\BibitemShut {NoStop}%
\bibitem [{\citenamefont {Jain}\ \emph {et~al.}(2014)\citenamefont {Jain}, \citenamefont {Bollinger},\ and\ \citenamefont {Truskett}}]{Truskett_inverse}%
  \BibitemOpen
  \bibfield  {author} {\bibinfo {author} {\bibfnamefont {A.}~\bibnamefont {Jain}}, \bibinfo {author} {\bibfnamefont {J.~A.}\ \bibnamefont {Bollinger}}, \ and\ \bibinfo {author} {\bibfnamefont {T.~M.}\ \bibnamefont {Truskett}},\ }\href {\doibase https://doi.org/10.1002/aic.14491} {\bibfield  {journal} {\bibinfo  {journal} {AIChE Journal}\ }\textbf {\bibinfo {volume} {60}},\ \bibinfo {pages} {2732} (\bibinfo {year} {2014})},\ \Eprint {http://arxiv.org/abs/https://aiche.onlinelibrary.wiley.com/doi/pdf/10.1002/aic.14491} {https://aiche.onlinelibrary.wiley.com/doi/pdf/10.1002/aic.14491} \BibitemShut {NoStop}%
\bibitem [{\citenamefont {Liu}\ \emph {et~al.}(2019)\citenamefont {Liu}, \citenamefont {Henkes},\ and\ \citenamefont {Schwarz}}]{Schwartz_2}%
  \BibitemOpen
  \bibfield  {author} {\bibinfo {author} {\bibfnamefont {K.}~\bibnamefont {Liu}}, \bibinfo {author} {\bibfnamefont {S.}~\bibnamefont {Henkes}}, \ and\ \bibinfo {author} {\bibfnamefont {J.~M.}\ \bibnamefont {Schwarz}},\ }\href {\doibase 10.1103/PhysRevX.9.021006} {\bibfield  {journal} {\bibinfo  {journal} {Phys. Rev. X}\ }\textbf {\bibinfo {volume} {9}},\ \bibinfo {pages} {021006} (\bibinfo {year} {2019})}\BibitemShut {NoStop}%
\bibitem [{\citenamefont {Zeng}\ \emph {et~al.}(2022)\citenamefont {Zeng}, \citenamefont {Zhang}, \citenamefont {Zheng}, \citenamefont {Xia}, \citenamefont {Kob}, \citenamefont {Yuan},\ and\ \citenamefont {Wang}}]{Yujie}%
  \BibitemOpen
  \bibfield  {author} {\bibinfo {author} {\bibfnamefont {Z.}~\bibnamefont {Zeng}}, \bibinfo {author} {\bibfnamefont {S.}~\bibnamefont {Zhang}}, \bibinfo {author} {\bibfnamefont {X.}~\bibnamefont {Zheng}}, \bibinfo {author} {\bibfnamefont {C.}~\bibnamefont {Xia}}, \bibinfo {author} {\bibfnamefont {W.}~\bibnamefont {Kob}}, \bibinfo {author} {\bibfnamefont {Y.}~\bibnamefont {Yuan}}, \ and\ \bibinfo {author} {\bibfnamefont {Y.}~\bibnamefont {Wang}},\ }\href {\doibase 10.1103/PhysRevLett.129.228004} {\bibfield  {journal} {\bibinfo  {journal} {Phys. Rev. Lett.}\ }\textbf {\bibinfo {volume} {129}},\ \bibinfo {pages} {228004} (\bibinfo {year} {2022})}\BibitemShut {NoStop}%
\bibitem [{\citenamefont {Candelier}\ and\ \citenamefont {Dauchot}(2009)}]{Dauchot}%
  \BibitemOpen
  \bibfield  {author} {\bibinfo {author} {\bibfnamefont {R.}~\bibnamefont {Candelier}}\ and\ \bibinfo {author} {\bibfnamefont {O.}~\bibnamefont {Dauchot}},\ }\href {\doibase 10.1103/PhysRevLett.103.128001} {\bibfield  {journal} {\bibinfo  {journal} {Phys. Rev. Lett.}\ }\textbf {\bibinfo {volume} {103}},\ \bibinfo {pages} {128001} (\bibinfo {year} {2009})}\BibitemShut {NoStop}%
\bibitem [{\citenamefont {Henkes}\ \emph {et~al.}(2016)\citenamefont {Henkes}, \citenamefont {Quint}, \citenamefont {Fily},\ and\ \citenamefont {Schwarz}}]{Henkes}%
  \BibitemOpen
  \bibfield  {author} {\bibinfo {author} {\bibfnamefont {S.}~\bibnamefont {Henkes}}, \bibinfo {author} {\bibfnamefont {D.~A.}\ \bibnamefont {Quint}}, \bibinfo {author} {\bibfnamefont {Y.}~\bibnamefont {Fily}}, \ and\ \bibinfo {author} {\bibfnamefont {J.~M.}\ \bibnamefont {Schwarz}},\ }\href {\doibase 10.1103/PhysRevLett.116.028301} {\bibfield  {journal} {\bibinfo  {journal} {Phys. Rev. Lett.}\ }\textbf {\bibinfo {volume} {116}},\ \bibinfo {pages} {028301} (\bibinfo {year} {2016})}\BibitemShut {NoStop}%
\bibitem [{\citenamefont {Papadopoulos}\ \emph {et~al.}(2018)\citenamefont {Papadopoulos}, \citenamefont {Porter}, \citenamefont {Daniels},\ and\ \citenamefont {Bassett}}]{Daniels}%
  \BibitemOpen
  \bibfield  {author} {\bibinfo {author} {\bibfnamefont {L.}~\bibnamefont {Papadopoulos}}, \bibinfo {author} {\bibfnamefont {M.~A.}\ \bibnamefont {Porter}}, \bibinfo {author} {\bibfnamefont {K.~E.}\ \bibnamefont {Daniels}}, \ and\ \bibinfo {author} {\bibfnamefont {D.~S.}\ \bibnamefont {Bassett}},\ }\href {\doibase 10.1093/comnet/cny005} {\bibfield  {journal} {\bibinfo  {journal} {Journal of Complex Networks}\ }\textbf {\bibinfo {volume} {6}},\ \bibinfo {pages} {485} (\bibinfo {year} {2018})},\ \Eprint {http://arxiv.org/abs/https://academic.oup.com/comnet/article-pdf/6/4/485/25451976/cny005.pdf} {https://academic.oup.com/comnet/article-pdf/6/4/485/25451976/cny005.pdf} \BibitemShut {NoStop}%
\bibitem [{\citenamefont {Az\'ema}\ and\ \citenamefont {Radja\"{\i}}(2012)}]{Azema}%
  \BibitemOpen
  \bibfield  {author} {\bibinfo {author} {\bibfnamefont {E.}~\bibnamefont {Az\'ema}}\ and\ \bibinfo {author} {\bibfnamefont {F.}~\bibnamefont {Radja\"{\i}}},\ }\href {\doibase 10.1103/PhysRevE.85.031303} {\bibfield  {journal} {\bibinfo  {journal} {Phys. Rev. E}\ }\textbf {\bibinfo {volume} {85}},\ \bibinfo {pages} {031303} (\bibinfo {year} {2012})}\BibitemShut {NoStop}%
\bibitem [{\citenamefont {Huang}\ \emph {et~al.}(2005)\citenamefont {Huang}, \citenamefont {Ovarlez}, \citenamefont {Bertrand}, \citenamefont {Rodts}, \citenamefont {Coussot},\ and\ \citenamefont {Bonn}}]{bonn}%
  \BibitemOpen
  \bibfield  {author} {\bibinfo {author} {\bibfnamefont {N.}~\bibnamefont {Huang}}, \bibinfo {author} {\bibfnamefont {G.}~\bibnamefont {Ovarlez}}, \bibinfo {author} {\bibfnamefont {F.}~\bibnamefont {Bertrand}}, \bibinfo {author} {\bibfnamefont {S.}~\bibnamefont {Rodts}}, \bibinfo {author} {\bibfnamefont {P.}~\bibnamefont {Coussot}}, \ and\ \bibinfo {author} {\bibfnamefont {D.}~\bibnamefont {Bonn}},\ }\href {\doibase 10.1103/PhysRevLett.94.028301} {\bibfield  {journal} {\bibinfo  {journal} {Phys. Rev. Lett.}\ }\textbf {\bibinfo {volume} {94}},\ \bibinfo {pages} {028301} (\bibinfo {year} {2005})}\BibitemShut {NoStop}%
\bibitem [{\citenamefont {Mermet-Guyennet}\ \emph {et~al.}(2015)\citenamefont {Mermet-Guyennet}, \citenamefont {{Gianfelice de Castro}}, \citenamefont {Varol}, \citenamefont {Habibi}, \citenamefont {Hosseinkhani}, \citenamefont {Martzel}, \citenamefont {Sprik}, \citenamefont {Denn}, \citenamefont {Zaccone}, \citenamefont {Parekh},\ and\ \citenamefont {Bonn}}]{bonn_composite}%
  \BibitemOpen
  \bibfield  {author} {\bibinfo {author} {\bibfnamefont {M.}~\bibnamefont {Mermet-Guyennet}}, \bibinfo {author} {\bibfnamefont {J.}~\bibnamefont {{Gianfelice de Castro}}}, \bibinfo {author} {\bibfnamefont {H.}~\bibnamefont {Varol}}, \bibinfo {author} {\bibfnamefont {M.}~\bibnamefont {Habibi}}, \bibinfo {author} {\bibfnamefont {B.}~\bibnamefont {Hosseinkhani}}, \bibinfo {author} {\bibfnamefont {N.}~\bibnamefont {Martzel}}, \bibinfo {author} {\bibfnamefont {R.}~\bibnamefont {Sprik}}, \bibinfo {author} {\bibfnamefont {M.}~\bibnamefont {Denn}}, \bibinfo {author} {\bibfnamefont {A.}~\bibnamefont {Zaccone}}, \bibinfo {author} {\bibfnamefont {S.}~\bibnamefont {Parekh}}, \ and\ \bibinfo {author} {\bibfnamefont {D.}~\bibnamefont {Bonn}},\ }\href {\doibase https://doi.org/10.1016/j.polymer.2015.07.041} {\bibfield  {journal} {\bibinfo  {journal} {Polymer}\ }\textbf {\bibinfo {volume} {73}},\ \bibinfo {pages} {170} (\bibinfo {year} {2015})}\BibitemShut {NoStop}%
\bibitem [{\citenamefont {Kalghatgi}\ \emph {et~al.}(0)\citenamefont {Kalghatgi}, \citenamefont {Kumar},\ and\ \citenamefont {Mani}}]{Kumar}%
  \BibitemOpen
  \bibfield  {author} {\bibinfo {author} {\bibfnamefont {S.~R.}\ \bibnamefont {Kalghatgi}}, \bibinfo {author} {\bibfnamefont {S.~K.}\ \bibnamefont {Kumar}}, \ and\ \bibinfo {author} {\bibfnamefont {E.}~\bibnamefont {Mani}},\ }\href {\doibase 10.1021/acs.macromol.4c01560} {\bibfield  {journal} {\bibinfo  {journal} {Macromolecules}\ }\textbf {\bibinfo {volume} {0}},\ \bibinfo {pages} {null} (\bibinfo {year} {0})},\ \Eprint {http://arxiv.org/abs/https://doi.org/10.1021/acs.macromol.4c01560} {https://doi.org/10.1021/acs.macromol.4c01560} \BibitemShut {NoStop}%
\bibitem [{\citenamefont {Ishima}\ \emph {et~al.}(2023{\natexlab{a}})\citenamefont {Ishima}, \citenamefont {Saitoh}, \citenamefont {Otsuki},\ and\ \citenamefont {Hayakawa}}]{Hayakawa1}%
  \BibitemOpen
  \bibfield  {author} {\bibinfo {author} {\bibfnamefont {D.}~\bibnamefont {Ishima}}, \bibinfo {author} {\bibfnamefont {K.}~\bibnamefont {Saitoh}}, \bibinfo {author} {\bibfnamefont {M.}~\bibnamefont {Otsuki}}, \ and\ \bibinfo {author} {\bibfnamefont {H.}~\bibnamefont {Hayakawa}},\ }\href {\doibase 10.1103/PhysRevE.107.054902} {\bibfield  {journal} {\bibinfo  {journal} {Phys. Rev. E}\ }\textbf {\bibinfo {volume} {107}},\ \bibinfo {pages} {054902} (\bibinfo {year} {2023}{\natexlab{a}})}\BibitemShut {NoStop}%
\bibitem [{\citenamefont {Ishima}\ \emph {et~al.}(2023{\natexlab{b}})\citenamefont {Ishima}, \citenamefont {Saitoh}, \citenamefont {Otsuki},\ and\ \citenamefont {Hayakawa}}]{Hayakawa2}%
  \BibitemOpen
  \bibfield  {author} {\bibinfo {author} {\bibfnamefont {D.}~\bibnamefont {Ishima}}, \bibinfo {author} {\bibfnamefont {K.}~\bibnamefont {Saitoh}}, \bibinfo {author} {\bibfnamefont {M.}~\bibnamefont {Otsuki}}, \ and\ \bibinfo {author} {\bibfnamefont {H.}~\bibnamefont {Hayakawa}},\ }\href {\doibase 10.1103/PhysRevE.107.034904} {\bibfield  {journal} {\bibinfo  {journal} {Phys. Rev. E}\ }\textbf {\bibinfo {volume} {107}},\ \bibinfo {pages} {034904} (\bibinfo {year} {2023}{\natexlab{b}})}\BibitemShut {NoStop}%
\bibitem [{\citenamefont {Wu}\ \emph {et~al.}(2009)\citenamefont {Wu}, \citenamefont {Zaccone}, \citenamefont {Tsoutsoura}, \citenamefont {Lattuada},\ and\ \citenamefont {Morbidelli}}]{Hua}%
  \BibitemOpen
  \bibfield  {author} {\bibinfo {author} {\bibfnamefont {H.}~\bibnamefont {Wu}}, \bibinfo {author} {\bibfnamefont {A.}~\bibnamefont {Zaccone}}, \bibinfo {author} {\bibfnamefont {A.}~\bibnamefont {Tsoutsoura}}, \bibinfo {author} {\bibfnamefont {M.}~\bibnamefont {Lattuada}}, \ and\ \bibinfo {author} {\bibfnamefont {M.}~\bibnamefont {Morbidelli}},\ }\href {\doibase 10.1021/la803789s} {\bibfield  {journal} {\bibinfo  {journal} {Langmuir}\ }\textbf {\bibinfo {volume} {25}},\ \bibinfo {pages} {4715} (\bibinfo {year} {2009})},\ \bibinfo {note} {pMID: 19260654},\ \Eprint {http://arxiv.org/abs/https://doi.org/10.1021/la803789s} {https://doi.org/10.1021/la803789s} \BibitemShut {NoStop}%
\bibitem [{\citenamefont {Vani}\ \emph {et~al.}(2024)\citenamefont {Vani}, \citenamefont {Escudier}, \citenamefont {Jeong},\ and\ \citenamefont {Sauret}}]{Sauret}%
  \BibitemOpen
  \bibfield  {author} {\bibinfo {author} {\bibfnamefont {N.}~\bibnamefont {Vani}}, \bibinfo {author} {\bibfnamefont {S.}~\bibnamefont {Escudier}}, \bibinfo {author} {\bibfnamefont {D.-H.}\ \bibnamefont {Jeong}}, \ and\ \bibinfo {author} {\bibfnamefont {A.}~\bibnamefont {Sauret}},\ }\href {\doibase 10.1103/PhysRevResearch.6.L032060} {\bibfield  {journal} {\bibinfo  {journal} {Phys. Rev. Res.}\ }\textbf {\bibinfo {volume} {6}},\ \bibinfo {pages} {L032060} (\bibinfo {year} {2024})}\BibitemShut {NoStop}%
\bibitem [{\citenamefont {Heussinger}\ and\ \citenamefont {Frey}(2006)}]{frey}%
  \BibitemOpen
  \bibfield  {author} {\bibinfo {author} {\bibfnamefont {C.}~\bibnamefont {Heussinger}}\ and\ \bibinfo {author} {\bibfnamefont {E.}~\bibnamefont {Frey}},\ }\href {\doibase 10.1103/PhysRevLett.97.105501} {\bibfield  {journal} {\bibinfo  {journal} {Phys. Rev. Lett.}\ }\textbf {\bibinfo {volume} {97}},\ \bibinfo {pages} {105501} (\bibinfo {year} {2006})}\BibitemShut {NoStop}%
\bibitem [{\citenamefont {Petridou}\ \emph {et~al.}(2021)\citenamefont {Petridou}, \citenamefont {Corominas-Murtra}, \citenamefont {Heisenberg},\ and\ \citenamefont {Hannezo}}]{Heisenberg}%
  \BibitemOpen
  \bibfield  {author} {\bibinfo {author} {\bibfnamefont {N.~I.}\ \bibnamefont {Petridou}}, \bibinfo {author} {\bibfnamefont {B.}~\bibnamefont {Corominas-Murtra}}, \bibinfo {author} {\bibfnamefont {C.-P.}\ \bibnamefont {Heisenberg}}, \ and\ \bibinfo {author} {\bibfnamefont {E.}~\bibnamefont {Hannezo}},\ }\href {\doibase https://doi.org/10.1016/j.cell.2021.02.017} {\bibfield  {journal} {\bibinfo  {journal} {Cell}\ }\textbf {\bibinfo {volume} {184}},\ \bibinfo {pages} {1914} (\bibinfo {year} {2021})}\BibitemShut {NoStop}%
\bibitem [{\citenamefont {Bi}\ \emph {et~al.}(2016)\citenamefont {Bi}, \citenamefont {Yang}, \citenamefont {Marchetti},\ and\ \citenamefont {Manning}}]{Manning}%
  \BibitemOpen
  \bibfield  {author} {\bibinfo {author} {\bibfnamefont {D.}~\bibnamefont {Bi}}, \bibinfo {author} {\bibfnamefont {X.}~\bibnamefont {Yang}}, \bibinfo {author} {\bibfnamefont {M.~C.}\ \bibnamefont {Marchetti}}, \ and\ \bibinfo {author} {\bibfnamefont {M.~L.}\ \bibnamefont {Manning}},\ }\href {\doibase 10.1103/PhysRevX.6.021011} {\bibfield  {journal} {\bibinfo  {journal} {Phys. Rev. X}\ }\textbf {\bibinfo {volume} {6}},\ \bibinfo {pages} {021011} (\bibinfo {year} {2016})}\BibitemShut {NoStop}%
\bibitem [{\citenamefont {Bi}\ \emph {et~al.}(2015)\citenamefont {Bi}, \citenamefont {Lopez}, \citenamefont {Schwarz},\ and\ \citenamefont {Manning}}]{Schwartz}%
  \BibitemOpen
  \bibfield  {author} {\bibinfo {author} {\bibfnamefont {D.}~\bibnamefont {Bi}}, \bibinfo {author} {\bibfnamefont {J.~H.}\ \bibnamefont {Lopez}}, \bibinfo {author} {\bibfnamefont {J.~M.}\ \bibnamefont {Schwarz}}, \ and\ \bibinfo {author} {\bibfnamefont {M.~L.}\ \bibnamefont {Manning}},\ }\href {\doibase 10.1038/nphys3471} {\bibfield  {journal} {\bibinfo  {journal} {Nature Physics}\ }\textbf {\bibinfo {volume} {11}},\ \bibinfo {pages} {1074} (\bibinfo {year} {2015})}\BibitemShut {NoStop}%
\bibitem [{\citenamefont {Atia}\ and\ \citenamefont {Fredberg}(2023)}]{Fredberg}%
  \BibitemOpen
  \bibfield  {author} {\bibinfo {author} {\bibfnamefont {L.}~\bibnamefont {Atia}}\ and\ \bibinfo {author} {\bibfnamefont {J.~J.}\ \bibnamefont {Fredberg}},\ }\href {\doibase 10.1063/5.0179719} {\bibfield  {journal} {\bibinfo  {journal} {Biophysics Reviews}\ }\textbf {\bibinfo {volume} {4}},\ \bibinfo {pages} {041304} (\bibinfo {year} {2023})},\ \Eprint {http://arxiv.org/abs/https://pubs.aip.org/aip/bpr/article-pdf/doi/10.1063/5.0179719/18270285/041304\_1\_5.0179719.pdf} {https://pubs.aip.org/aip/bpr/article-pdf/doi/10.1063/5.0179719/18270285/041304\_1\_5.0179719.pdf} \BibitemShut {NoStop}%
\bibitem [{\citenamefont {Garcia}\ \emph {et~al.}(2015)\citenamefont {Garcia}, \citenamefont {Hannezo}, \citenamefont {Elgeti}, \citenamefont {Joanny}, \citenamefont {Silberzan},\ and\ \citenamefont {Gov}}]{Hannezo}%
  \BibitemOpen
  \bibfield  {author} {\bibinfo {author} {\bibfnamefont {S.}~\bibnamefont {Garcia}}, \bibinfo {author} {\bibfnamefont {E.}~\bibnamefont {Hannezo}}, \bibinfo {author} {\bibfnamefont {J.}~\bibnamefont {Elgeti}}, \bibinfo {author} {\bibfnamefont {J.-F.}\ \bibnamefont {Joanny}}, \bibinfo {author} {\bibfnamefont {P.}~\bibnamefont {Silberzan}}, \ and\ \bibinfo {author} {\bibfnamefont {N.~S.}\ \bibnamefont {Gov}},\ }\href {\doibase 10.1073/pnas.1510973112} {\bibfield  {journal} {\bibinfo  {journal} {Proceedings of the National Academy of Sciences}\ }\textbf {\bibinfo {volume} {112}},\ \bibinfo {pages} {15314} (\bibinfo {year} {2015})},\ \Eprint {http://arxiv.org/abs/https://www.pnas.org/doi/pdf/10.1073/pnas.1510973112} {https://www.pnas.org/doi/pdf/10.1073/pnas.1510973112} \BibitemShut {NoStop}%
\bibitem [{\citenamefont {Janssen}(2019)}]{Janssen}%
  \BibitemOpen
  \bibfield  {author} {\bibinfo {author} {\bibfnamefont {L.~M.~C.}\ \bibnamefont {Janssen}},\ }\href {\doibase 10.1088/1361-648X/ab3e90} {\bibfield  {journal} {\bibinfo  {journal} {Journal of Physics: Condensed Matter}\ }\textbf {\bibinfo {volume} {31}},\ \bibinfo {pages} {503002} (\bibinfo {year} {2019})}\BibitemShut {NoStop}%
\bibitem [{\citenamefont {Fang}\ \emph {et~al.}(2020)\citenamefont {Fang}, \citenamefont {Han}, \citenamefont {Zhang}, \citenamefont {Jiang}, \citenamefont {Lin}, \citenamefont {Shi}, \citenamefont {Wu}, \citenamefont {Meng}, \citenamefont {Gao}, \citenamefont {Griffin}, \citenamefont {Xiao}, \citenamefont {Tsai}, \citenamefont {Zhou}, \citenamefont {Zuo}, \citenamefont {Zhang}, \citenamefont {Chu}, \citenamefont {Zhang}, \citenamefont {Gao}, \citenamefont {Roth}, \citenamefont {Bleher}, \citenamefont {Ma}, \citenamefont {Jiang}, \citenamefont {Yue}, \citenamefont {Kao}, \citenamefont {Chen}, \citenamefont {Tokmakoff}, \citenamefont {Wang}, \citenamefont {Jaeger},\ and\ \citenamefont {Tian}}]{Fang2020}%
  \BibitemOpen
  \bibfield  {author} {\bibinfo {author} {\bibfnamefont {Y.}~\bibnamefont {Fang}}, \bibinfo {author} {\bibfnamefont {E.}~\bibnamefont {Han}}, \bibinfo {author} {\bibfnamefont {X.-X.}\ \bibnamefont {Zhang}}, \bibinfo {author} {\bibfnamefont {Y.}~\bibnamefont {Jiang}}, \bibinfo {author} {\bibfnamefont {Y.}~\bibnamefont {Lin}}, \bibinfo {author} {\bibfnamefont {J.}~\bibnamefont {Shi}}, \bibinfo {author} {\bibfnamefont {J.}~\bibnamefont {Wu}}, \bibinfo {author} {\bibfnamefont {L.}~\bibnamefont {Meng}}, \bibinfo {author} {\bibfnamefont {X.}~\bibnamefont {Gao}}, \bibinfo {author} {\bibfnamefont {P.~J.}\ \bibnamefont {Griffin}}, \bibinfo {author} {\bibfnamefont {X.}~\bibnamefont {Xiao}}, \bibinfo {author} {\bibfnamefont {H.-M.}\ \bibnamefont {Tsai}}, \bibinfo {author} {\bibfnamefont {H.}~\bibnamefont {Zhou}}, \bibinfo {author} {\bibfnamefont {X.}~\bibnamefont {Zuo}}, \bibinfo {author} {\bibfnamefont {Q.}~\bibnamefont {Zhang}}, \bibinfo {author} {\bibfnamefont {M.}~\bibnamefont {Chu}}, \bibinfo {author}
  {\bibfnamefont {Q.}~\bibnamefont {Zhang}}, \bibinfo {author} {\bibfnamefont {Y.}~\bibnamefont {Gao}}, \bibinfo {author} {\bibfnamefont {L.~K.}\ \bibnamefont {Roth}}, \bibinfo {author} {\bibfnamefont {R.}~\bibnamefont {Bleher}}, \bibinfo {author} {\bibfnamefont {Z.}~\bibnamefont {Ma}}, \bibinfo {author} {\bibfnamefont {Z.}~\bibnamefont {Jiang}}, \bibinfo {author} {\bibfnamefont {J.}~\bibnamefont {Yue}}, \bibinfo {author} {\bibfnamefont {C.-M.}\ \bibnamefont {Kao}}, \bibinfo {author} {\bibfnamefont {C.-T.}\ \bibnamefont {Chen}}, \bibinfo {author} {\bibfnamefont {A.}~\bibnamefont {Tokmakoff}}, \bibinfo {author} {\bibfnamefont {J.}~\bibnamefont {Wang}}, \bibinfo {author} {\bibfnamefont {H.~M.}\ \bibnamefont {Jaeger}}, \ and\ \bibinfo {author} {\bibfnamefont {B.}~\bibnamefont {Tian}},\ }\href {\doibase 10.1016/j.matt.2020.01.008} {\bibfield  {journal} {\bibinfo  {journal} {Matter}\ }\textbf {\bibinfo {volume} {2}},\ \bibinfo {pages} {948} (\bibinfo {year} {2020})}\BibitemShut {NoStop}%
\bibitem [{\citenamefont {van Oosten}\ \emph {et~al.}(2019)\citenamefont {van Oosten}, \citenamefont {Chen}, \citenamefont {Chin}, \citenamefont {Cruz}, \citenamefont {Patteson}, \citenamefont {Pogoda}, \citenamefont {Shenoy},\ and\ \citenamefont {Janmey}}]{Janmey}%
  \BibitemOpen
  \bibfield  {author} {\bibinfo {author} {\bibfnamefont {A.~S.~G.}\ \bibnamefont {van Oosten}}, \bibinfo {author} {\bibfnamefont {X.}~\bibnamefont {Chen}}, \bibinfo {author} {\bibfnamefont {L.}~\bibnamefont {Chin}}, \bibinfo {author} {\bibfnamefont {K.}~\bibnamefont {Cruz}}, \bibinfo {author} {\bibfnamefont {A.~E.}\ \bibnamefont {Patteson}}, \bibinfo {author} {\bibfnamefont {K.}~\bibnamefont {Pogoda}}, \bibinfo {author} {\bibfnamefont {V.~B.}\ \bibnamefont {Shenoy}}, \ and\ \bibinfo {author} {\bibfnamefont {P.~A.}\ \bibnamefont {Janmey}},\ }\href {\doibase 10.1038/s41586-019-1516-5} {\bibfield  {journal} {\bibinfo  {journal} {Nature}\ }\textbf {\bibinfo {volume} {573}},\ \bibinfo {pages} {96} (\bibinfo {year} {2019})}\BibitemShut {NoStop}%
\bibitem [{\citenamefont {Shivers}\ \emph {et~al.}(2020)\citenamefont {Shivers}, \citenamefont {Feng}, \citenamefont {van Oosten}, \citenamefont {Levine}, \citenamefont {Janmey},\ and\ \citenamefont {MacKintosh}}]{Shivers}%
  \BibitemOpen
  \bibfield  {author} {\bibinfo {author} {\bibfnamefont {J.~L.}\ \bibnamefont {Shivers}}, \bibinfo {author} {\bibfnamefont {J.}~\bibnamefont {Feng}}, \bibinfo {author} {\bibfnamefont {A.~S.~G.}\ \bibnamefont {van Oosten}}, \bibinfo {author} {\bibfnamefont {H.}~\bibnamefont {Levine}}, \bibinfo {author} {\bibfnamefont {P.~A.}\ \bibnamefont {Janmey}}, \ and\ \bibinfo {author} {\bibfnamefont {F.~C.}\ \bibnamefont {MacKintosh}},\ }\href {\doibase 10.1073/pnas.2003037117} {\bibfield  {journal} {\bibinfo  {journal} {Proceedings of the National Academy of Sciences}\ }\textbf {\bibinfo {volume} {117}},\ \bibinfo {pages} {21037} (\bibinfo {year} {2020})},\ \Eprint {http://arxiv.org/abs/https://www.pnas.org/doi/pdf/10.1073/pnas.2003037117} {https://www.pnas.org/doi/pdf/10.1073/pnas.2003037117} \BibitemShut {NoStop}%
\bibitem [{\citenamefont {Zhao}\ \emph {et~al.}(2024)\citenamefont {Zhao}, \citenamefont {Hu}, \citenamefont {Huang}, \citenamefont {Liu}, \citenamefont {Yan}, \citenamefont {Xu}, \citenamefont {Zhang}, \citenamefont {Wang},\ and\ \citenamefont {Xu}}]{Zhao2024}%
  \BibitemOpen
  \bibfield  {author} {\bibinfo {author} {\bibfnamefont {Y.}~\bibnamefont {Zhao}}, \bibinfo {author} {\bibfnamefont {H.}~\bibnamefont {Hu}}, \bibinfo {author} {\bibfnamefont {Y.}~\bibnamefont {Huang}}, \bibinfo {author} {\bibfnamefont {H.}~\bibnamefont {Liu}}, \bibinfo {author} {\bibfnamefont {C.}~\bibnamefont {Yan}}, \bibinfo {author} {\bibfnamefont {C.}~\bibnamefont {Xu}}, \bibinfo {author} {\bibfnamefont {R.}~\bibnamefont {Zhang}}, \bibinfo {author} {\bibfnamefont {Y.}~\bibnamefont {Wang}}, \ and\ \bibinfo {author} {\bibfnamefont {Q.}~\bibnamefont {Xu}},\ }\href {\doibase 10.1038/s41467-024-45964-y} {\bibfield  {journal} {\bibinfo  {journal} {Nature Communications}\ }\textbf {\bibinfo {volume} {15}},\ \bibinfo {pages} {1691} (\bibinfo {year} {2024})}\BibitemShut {NoStop}%
\end{thebibliography}%

\end{document}